\documentclass[usenatbib]{mnras}
\usepackage{natbib,amssymb,amsmath,times,graphicx,url,aas_macros}
\usepackage{epstopdf}       % to convert eps images to pdf images
\usepackage{verbatim}       % to allow multiline comments
\usepackage{bm}		  % allow shorthand to bold text
\usepackage{fleqn}		  % flush left equations
\usepackage{Wurster_macros}
\defcitealias{WursterBatePrice2018hd}{Paper: Isolated}
\defcitealias{WursterBatePrice2019}{Paper: Cluster}

\title[Do we need non-ideal MHD to model discs?]{Do we need non-ideal magnetohydrodynamics to model protostellar discs?}

\author[Wurster]{James Wurster$^{1}$\thanks{jhw5@st-andrews.ac.uk} \\
$^{1}$Scottish Universities Physics Alliance (SUPA), School of Physics and Astronomy, University of St. Andrews, North Haugh, St Andrews, Fife KY16 9SS, UK
}
\date{Submitted: Revised: Accepted: }
\pagerange{\pageref{firstpage}--\pageref{lastpage}} \pubyear{2020}

\begin{document}
\label{firstpage}
\bibliographystyle{mnras}
\maketitle
%231 of a 250 limit
\begin{abstract}
We investigate and discuss protostellar discs in terms of where the various non-ideal magnetohydrodynamics (MHD) processes are important.  We find that the traditional picture of a magnetised disc (where Ohmic resistivity is dominant near the mid-plane, surrounded by a region dominated by the Hall effect, with the remainder of the disc dominated by ambipolar diffusion) is a great oversimplification.  In simple parameterised discs, we find that the Hall effect is typically the dominant term throughout the majority of the disc.  More importantly, we find that in much of our parameterised discs, at least two non-ideal processes have coefficients within a factor of 10 of one another, indicating that both are important and that naming a dominant term underplays the importance of the other terms.  Discs that were self-consistently formed in our previous studies are also dominated by the Hall effect, and the ratio of ambipolar diffusion and Hall coefficients is typically less than 10, suggesting that both terms are equally important and listing a dominant term is misleading.  These conclusions become more robust once the magnetic field geometry is taken into account.  In agreement with the literature we review, we conclude that non-ideal MHD processes are important for the formation and evolution of protostellar discs.  Ignoring any of the non-ideal processes, especially ambipolar diffusion and the Hall effect, yields an incorrect description of disc evolution.  
\end{abstract}

\begin{keywords}
protoplanetary discs --- magnetic fields --- (magnetohydrodynamics) MHD --- dust, extinction --- methods: numerical
\end{keywords}

%----------------------------------------------------------------------------------------------------------------
\section{Introduction}
\label{sec:intro}
Protostellar discs are interesting and important objects, most notably since they are where planets form.  They are incredibly complex and often highly structured objects, and are host to many components and processes: gas, dust, chemical processes, radiative processes and non-ideal magnetohydrodynamic (MHD) processes.  

Many theoretical studies have shown the effect that some or all of the non-ideal MHD effects have on star formation and the subsequent formation of discs, and many additional studies have discussed non-ideal MHD in terms of the evolution of the discs themselves.  All of these studies conclude that non-ideal MHD affects the system and thus demonstrate its necessity.  Despite this, a complete description of non-ideal MHD is still frequently neglected throughout the literature when discussing star and disc formation and disc evolution.  

In this paper, we discuss non-ideal MHD with the aim of reinforcing its necessity in protostellar disc simulations (formation and evolution).  We start in \secref{sec:review} with a review of the observational motivation and the theoretical work to date.  In \secref{sec:id}, we discuss the non-ideal MHD coefficients and relative importance in an idealised, parameterised disc; this allows us to investigate the validity of the traditional picture of where the non-ideal effects are important (see \figref{fig:cartoon} and associated text below).  We further discuss the complexity of the geometry of the non-ideal processes and show that a direct comparison of the non-ideal coefficients must be treated with caution.  In \secref{sec:rd}, we show and discuss the magnetic field structure and the relative importance of the non-ideal terms in discs that were self-consistently formed in our previous studies \citep{\wbp2018hd,\wbp2019}.  In \secref{sec:disc}, we discuss the implications of these results on the long term evolution of the disc, and we conclude in \secref{sec:conc}.  In Appendix~\ref{app:nicil}, we introduce version 2.1 of the \textsc{nicil} library.  

%----
\section{Review of previous studies}
\label{sec:review}
\subsection{Observational Motivation}
Over the past few years, there has been a plethora of surveys aimed at obs-eps-converted-to.pdferving protostellar discs \citepeg{Cox+2015,Tobin+2015,Pascucci+2016,Barenfeld+2016,Barenfeld+2017,Ansdell+2016,Ansdell+2017,Ansdell+2018,Tazzari+2017,Tychoniec+2018,Andrews+2018,Eisner+2018,Sadavoy+2018,Andersen+2019,Loomis+2020,Tobin+2020,Villenave+2020}, with considerable focus on the dust components.  Trends have been discovered (e.g. the relationship between the size of the dust disc and their millimetre luminosity), however, these trends vary between regions as do the dust discs sizes themselves \citepeg{Ansdell+2016,Hendler+2020}.  Therefore, discs and disc properties appear to be dependent on their host environment.

In the interstellar medium (ISM), the gas-to-dust ratio is \sm0.01 \citep{BohlinSavageDrake1978} where the dust size distribution can be modelled using an MRN (Mathis, Rumpl \& Nordsieck) power-law with an exponent of $-3.5$ \citep{MathisRumplNordsieck1977}.  However, this approximation translates poorly to protostellar discs \citepeg{Pinte+2016,Ansdell+2016,Birnstiel+2018}, indicating a departure between ISM and disc properties.  In their survey of discs in Lupus, \citet{Ansdell+2016} found that the dust-to-gas ratios varied from \sm0.005 to \sm0.5 throughout their sample with no clustering around any specific value.

The dust-to-gas ratio is also a function of position within the disc since the dust tends to settle to the mid-plane \citepeg{Pinte+2008,KwonLooneyMundy2011,Kwon+2015,Pinte+2016,Birnstiel+2018,Huang+2018,Lee+2018young}, although this settling can be counteracted by stirring mechanisms \citepeg{GaraudLin2004,DullemondDominik2005,FromangPapaloizou2006}.  In their detailed study of HL Tau, \citet{Pinte+2016} showed that the dust settling -- and even the dust structure -- is not consistent throughout the disc.  The ratio of the gas and dust disc scale heights varies largely throughout the disc, and there are more large grains in the inner disc than the outer disc; this is either due to faster grain growth in the inner disc or to more efficient inwards migration of the larger dust grains.  As a result, an MRN slope of $-4.5$ is required to fit the dust profile in the outer disc, while the fiducial slope of $-3.5$ fits the dust in the inner disc.

Dust has been suggested as a tracer of magnetic fields \citepeg{Lazarian2007}; specifically, grains will align with the magnetic field, thus the polarisation vector will be 90$^\circ$ to the magnetic field vector.  While the typical `hour-glass' magnetic field structure has been inferred on large scales \citepeg{GirartRaoMarrone2006,Hull+2014,Stephens+2013,Maury+2018,Kwon+2019}, the magnetic structure within the disc is less well-known.  In HL Tau, \citet{Stephens+2014} concluded that the vertical component of the magnetic field could not be the dominant component.  Although a toroidal field was a better fit to the data, there was still a high degree of uncertainty, especially in the outer disc.   Similarly, \citet{Rao+2014} and \citet{Seguracox+2015} suggested toroidal fields reasonably fit the discs in IRAS 16293-2422 B and L1527, respectively.  Radial components of the magnetic field are typically not considered when modelling the dust profile since it is expected that this component will be sheared into the toroidal component on a short timescale due to differential rotation \citepeg{Stephens+2014}.

Unfortunately, magnetic fields are not the only process to cause the polarisation of dust \citepeg{Kataoka+2015,Kataoka+2017}.  It is likely that the polarisation of small dust grains is caused by radiation fields rather than magnetic fields \citepeg{TazakiLazarianNomura2017}.  Therefore, determining the structure of magnetic fields in discs is challenging, since polarisation at many wavelengths is consistent with self-scattering \citepeg{Stephens+2017,Lee+2018,Harris+2018}; while this may rule out certain magnetic field configurations for various discs (both vertical and toroidal configurations), it cannot conclusively confirm the remaining possibilities.  Therefore, further observational work is required to better understand the magnetic field structure in protostellar discs.

%----
\subsection{Theoretical background}
From the observations, it is clear that gas, dust and magnetic fields are important when modelling protostellar discs (and their formation).  The discs are ionised by cosmic rays \citepeg{SpitzerTomasko1968,UmebayashiNakano1981}, UV and X-rays from the host star \citepeg{IgeaGlassgold1999,TurnerSano2008,GortiHollenbach2009} and radio-nuclide decay \citepeg{UmebayashiNakano2009,Zhao+2018}.  Realistically, these processes do not ionise the disc evenly since the ionising particles are attenuated as they pass through the gas towards the disc mid-plane.   As the ionisation fraction is reduced, the coupling between the gas and the magnetic fields weakens.  In the mid-plane where the ionising particles may not reach, the coupling between the gas and magnetic field may be removed altogether,  leading to the formation of a `dead' zone where there is negligible magnetic activity \citep{Gammie1996,Wardle1997}.  Dead zones are unlikely to form in geometrically thin discs where ionising particles can reach the mid-plane \citep{GlassgoldLizanoGalli2017}.

In discs, it is not just gas that can become ionised; dust grains can absorb electrons and ions to suppress the coupling of the magnetic field to the gas, where their ability to do this depends on their size, with smaller grains better reducing the coupling between the gas and the magnetic field \citep{Wardle2007,Bai2011grain}.  Thus, grain dynamics (distribution, kinematics, growth and destruction) are very important in determining the ionisation fraction throughout a disc \citepeg{Bai2011grain,Marchand+2016,Wurster2016,Zhao+2016,ZhaoCaselliLi2018,Tsukamoto+2020}.

\begin{figure}
\centering
\includegraphics[trim=0 25 0 0,clip,width=\columnwidth]{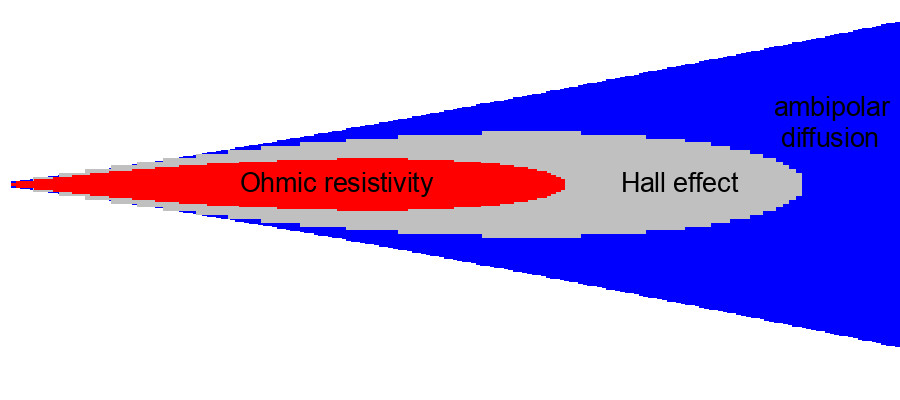}
\caption{A cartoon of a typical protostellar disc, where the regions are marked by what is generally assumed to be the dominant non-ideal MHD term in each region.  Although this is the traditional picture, analytical, semi-analytical and numerical studies show that this picture is an oversimplification.}
\label{fig:cartoon}
\end{figure}

In partially ionised discs, non-ideal MHD processes become important to account for the interaction between the charged particles (ions, electrons and grains), the neutral gas and the magnetic field.  The three processes important for disc formation and evolution are Ohmic resistivity, ambipolar diffusion (ion-neutral drift) and the Hall effect (ion-electron drift); while the first two processes diffuse the magnetic field, the Hall effect is a dispersive term that evolves the magnetic field vector \citepeg{Wardle2004}.  Each process represents a different coupling of the charged particles with the magnetic field, thus it is reasonable to assume that each processes is dominant in a different regime of the magnetic field strength-density phase space \citepeg{Wardle2007}.  
In terms of protostellar discs, the traditional picture is that Ohmic resistivity is dominant in the mid-plane near the star, ambipolar diffusion is dominant near the surface and in outer disc, and the Hall effect is dominant between these two regions \citepeg{DeschMouschovias2001,Wardle2004,Wardle2007,Bai2014,Bai2015,Bai2017,Simon+2018}, as shown in the cartoon sketch in \figref{fig:cartoon}.

There has been considerable research into these terms with respect to protostellar discs, including analytical/semi-analytical studies \citepeg{Wardle2004,BraidingWardle2012sf,BraidingWardle2012acc}, shearing-box simulations \citepeg{SanoStone2002b,LesurKunzFromang2014,Bai2015,RiolsLesur2018}, 2D simulations \citepeg{KrasnopolskyLiShang2011,LiKrasnopolskyShang2011,Zhao+2016,Zhao+2020,Bai2017,Simon+2018,WangBaiGoodman2019}, and 3D disc formation simulations \citepeg{Tsukamoto+2015oa,Tsukamoto+2015hall,Tsukamoto+2017,\wpb2016,\wbp2018hd,Zhao+2018,Vaytet+2018}.  The results from all of these studies indicate that the sketch in \figref{fig:cartoon} is a great oversimplification.  

In discs, Ohmic resistivity is often claimed to be unimportant, and some parameter studies neglect it altogether \citepeg{Bai2011grain,LiKrasnopolskyShang2011,Zhao+2016}.  This is justified since its effect is much weaker than the Hall effect and ambipolar diffusion, and given the traditional view of the disc, the region where Ohmic resistivity should be important coincides with the dead zone.   However,  \citet{LesurKunzFromang2014} concluded that the strength of Ohmic resistivity was comparable to the other terms in the inner disc and could not be ignored.  Therefore, the relevance of Ohmic resistivity may be dependent on the properties of the disc itself.

Ambipolar diffusion dissipates the magnetic field in the regime with low density and strong magnetic fields; this is traditionally the outer disc and surrounding environment.  Therefore, it is expected that ambipolar diffusion is more important in the disc formation rather than the disc evolution phase \citepeg{MellonLi2009,LiKrasnopolskyShang2011,Hennebelle+2016}.  This conclusion is dependent on grain population, since at these lower densities, removing the smaller grains increases the effect of ambipolar diffusion \citep{Zhao+2016,ZhaoCaselliLi2018,Tsukamoto+2020}.  Assuming $B \propto \rho^{1/2}$ (where $B$ is the magnetic field strength and $\rho$ is the gas density), \citet{ZhaoCaselliLi2018} find that removing the small grains at \emph{all} densities increases the effect of ambipolar diffusion.  However, at high densities and weak magnetic fields, \citet{Tsukamoto+2020} finds that removing small grains \emph{decreases} the effect of ambipolar diffusion. Therefore, the importance of ambipolar diffusion is dependent on all the environmental properties, especially density, grain populations and magnetic field strength.  

The Hall effect has been increasingly studied over the past several years in connection with formation and evolution of protostellar discs.  Given its vector evolution of the magnetic field, the Hall effect directly affects the angular momentum budget in a disc.  Due to this vector evolution, the relative angle $\theta$ between the magnetic field vector and the rotation vector of the disc becomes important.  Moreover, the sign of the Hall coefficient can be either positive or negative, depending on local microphysics \citep[for a study on where this term changes sign, see][]{XuBai2016}.  Therefore, the evolution of the disc will depend on the sign of the coefficient (i.e. the local microphysics) and on $\theta$.  

If the gas is initially not rotating, then the Hall effect can induce a rotation where the direction of the rotation is dependent on coefficient's sign and the relative angle \citepeg{KrasnopolskyLiShang2011,LiKrasnopolskyShang2011,BraidingWardle2012sf,Tsukamoto+2015hall,\wpb2016,Marchand+2019}.  If the gas is already rotating, then it will either be spun up or spun down, depending on the local parameters, causing a bi-modality in disc sizes \citepeg{KrasnopolskyLiShang2011,BraidingWardle2012sf,Tsukamoto+2015hall,\wpb2016}; simulations have shown that discs of \sm30~au form when the magnetic field and rotation vectors are anti-aligned ($\theta = 180^\circ$) and small \sm5~au discs form when the vectors are aligned ($\theta = 0^\circ$).  

However, \citet{Zhao+2020} recently found that the bi-modality no longer holds since discs of 10-20~au form independent of the initial orientation of the magnetic field.   They found that in the anti-aligned configuration, discs of \sm30–50 au formed, but only the inner region remained long-lived, resulting in the disc shrinking to $\lesssim$10–20 au.  In their aligned configuration, a counter-rotating disc formed later in the evolution with a radius of \sm20-40~au that subsequently shrank to \sm10~au.  These simulations were evolved over a longer period of time than \citet{Tsukamoto+2015hall} and \citet{\wpb2016} who initially suggested the bi-modality; all the discs in these earlier studies were rotating in the same direction as the initial collapsing gas.

The Hall effect is expected to become important once the grains have settled and are no longer the dominant charge carrier \citepeg{Wardle2007}.  \citet{KrasnopolskyLiShang2011} found that the Hall effect is stronger for an MRN dust grain distribution than a single dust grain species; \citet{Zhao+2020} concluded the Hall effect was only important when the small grains %(1e-6cm)
have been removed, with its effect strongest when the minimum grain size was $\sim0.04\mu$m \citep{ZhaoCaselliLi2018}.  \citet{LesurKunzFromang2014} even found that the Hall effect can `revive' dead zones.  In general, it has been found that the Hall effect is important over a wide range of conditions, and can often be the dominant term or at least comparable to the strength of ambipolar diffusion in a large portion of the disc \citepeg{LiKrasnopolskyShang2011,Wardle2004,Wardle2007,Bai2014,Bai2015}.  Even when its value is smaller compared to Ohmic resistivity or ambipolar diffusion, its effect is very evident \citep{BraidingWardle2012acc}.  

%----------------------------------------------------------------------------------------------------------------
\section{Structure of idealised discs}
\label{sec:id}
To better understand where the various non-ideal effects are important in a disc, we first consider an idealised disc.  For simplicity, we consider a 2D slice of a disc assuming azimuthal symmetry, and parameterise it similar to that commonly found in the literature.  

The density profile of the disc is \citepeg{Pringle1981} %eqn 3.14 (also 298 in the Phantom paper)
\begin{equation}
\label{eq:discrho}
\rho(r,z) = \rho_0 \left(\frac{r}{r_\text{in}}\right)^{-p} \exp\left(\frac{-z^2}{2H^2(r)}\right),
\end{equation}
where $r_\text{in}$ is the inner edge of the disc, $\rho_0$ is the density at $r = r_\text{in}$, and $H(r) = c_\text{s}(r)/\Omega(r)$ is the scale height.  The rotation is given by $\Omega = \sqrt{GM_*/r^3}$, where $M_*$ is the mass of the central star, and the sound speed is $c_\text{s} = c_\text{s,in}(r/r_\text{in})^{-q}$ \citep{Dipierro+2015}.  For this study, we set $r_\text{in} = 1$~au, $\rho_0 = 10^{-9}$~\gpercc{},  $M_*=$~0.8\Msun{}, $c_\text{s,in} = 2\times10^5$~\cms{} $p=3/2$ and $q = 1/4$.  Although $c_\text{s,in}$ may seem high, the resulting profile approximately matches the temperature in the newly formed disc presented in \secref{sec:rd:iso}.
%To match DS Tau, we set $M_*=$~0.8\Msun{} and $q = 0.25$ \citep{Veronesi+2020}.  \citep{LodatoPringle2007} suggests $p = 3/2$.

It is commonly assumed that the magnetic field strength scales as $B \propto \rho^{1/2}$ \citepeg{MyersGoodman1988,Wardle2007,LiKrasnopolskyShang2011,ZhaoCaselliLi2018}, however a shallower relationship of $B \propto \rho^{1/4}$ may be more reasonable at higher densities \citepeg{WardleNg1999}; an analysis of the magnetic field strength in the discs in \citet{\wbp2019} agrees with this shallower slope, thus our parametrised magnetic field strength is given by
\begin{equation}
\label{eq:discB}
B = B_0\left(\frac{\rho}{\rho_0}\right)^{1/4},
\end{equation}
where we test $B_0 = 0.001$, $0.01$, $0.1$ and $1$~G.

Despite the complex dust structure discussed in \secref{sec:review}, we assume a constant dust-to-gas ratio of 0.01 and simplified grain populations.  Although not realistic, these assumptions are typically used in disc formation simulations where the dust is only required for the non-ideal MHD coefficients and is not self-consistently modelled \citepeg{Tsukamoto+2015hall,\wpb2016,\wbp2018hd,Vaytet+2018,Marchand+2019}. 

In this study, we explore three grain distributions: %, all assuming a global dust-to-gas ratio of 0.01 and a bulk density of 3~\gpercc{} \citep{Pollack+1994}:
\begin{enumerate}
\item Single grain sizes of $a_\text{g} = 3\times10^{-6}$, $10^{-5}$ and $3\times10^{-5}$~cm,
\item An MRN grain size distribution with $10^{-6} < a_\text{g}/\text{cm} < 0.1$, and 
\item A settled distribution from an initial MRN distribution with $10^{-6} < a_\text{g}/\text{cm} < 0.1$; see Appendix~\ref{app:settled}.
\end{enumerate}

Our typical ionisation source is cosmic rays with the constant rate of \zetatwoeq{1.2}{-17} \citep{Mcelroy+2013}.  We also consider one model where we approximate the attenuation of the cosmic rays as they pass through the gas  \citepeg{NakanoNishiUmebayashi2002,UmebayashiNakano2009,Zhao+2016}.  See Appendix~\ref{app:nicil} for details.

%----
\subsection{The non-ideal coefficients}
\begin{figure}
\includegraphics[width=0.95\columnwidth]{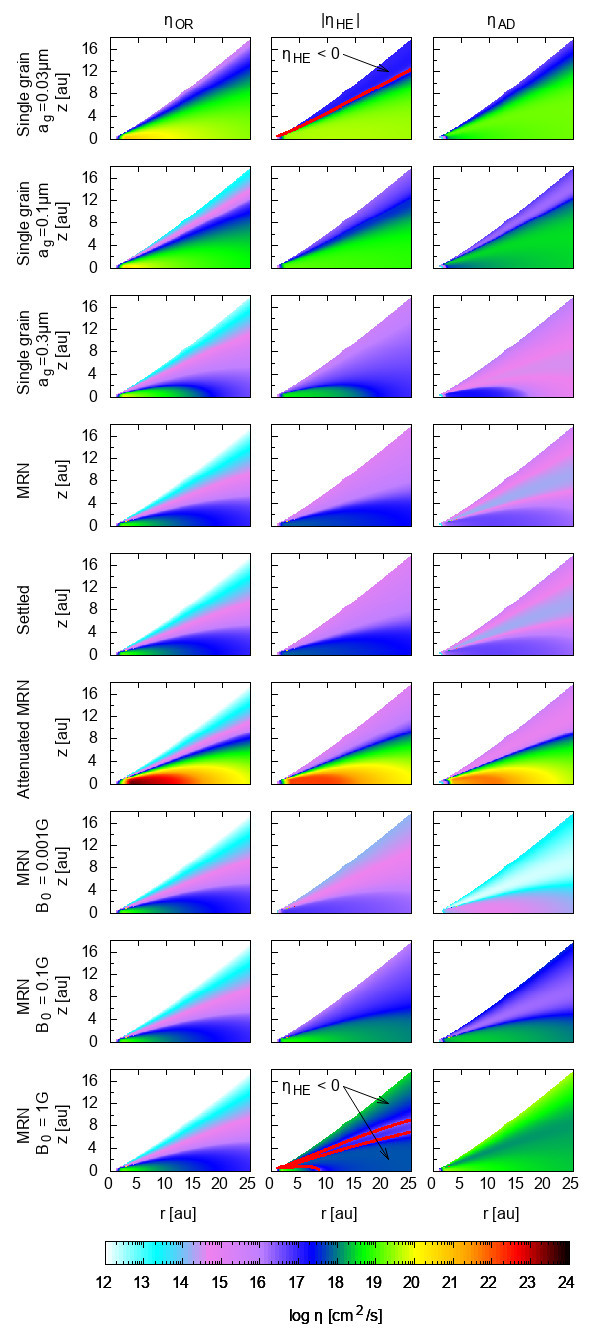}  %Made on Mythos
\caption{The strength of the non-ideal coefficients for nine idealised discs.  Unless otherwise stated, cosmic rays are unattenuated, $B_0 = 0.01$~G, and the values are only shown for \rhoge{-15}.   In the middle column, the red contours represent $\eta_\text{HE} = 0$, and we explicitly point to the region where $\eta_\text{HE} < 0$; unidentified regions are $\eta_\text{HE} > 0$.  Increasing the grain size or switching from a single grain to an MRN distribution reduces the strength of the coefficients, while including cosmic ray attenuation increases their strength near the mid-plane.  Except for the single grain disc with $a_\text{g} = 0.03$~\mum{} and the MRN disc with $B_0 = 1$~G, $\eta_\text{HE} > 0$.  }
\label{fig:idealised:eta}
\end{figure}

\figref{fig:idealised:eta} shows the non-ideal coefficients for nine idealised discs, where the coefficients are calculated using v2.1 of the \textsc{Nicil} library.  In most of our discs, $\eta_\text{HE} > 0$.  In the single grain disc with $a_\text{g} = 0.03$~\mum{}, $\eta_\text{HE} < 0$ at the surface of the disc, while in the MRN disc with $B_0 = 1$~G, $\eta_\text{HE} < 0$ except for a small ray.  The sign of $\eta_\text{HE}$ is briefly discussed in Appendix~\ref{app:signeta}.  The value of the coefficients is important in determining their influence on a system.  In the collapse to stellar densities simulations of \citet{\wbp2018sd}, we found that weak coefficients of $|\eta | \lesssim 10^{15}$~\ueta{} had negligible effect on the evolution.  Therefore, in the majority of the idealised discs in \figref{fig:idealised:eta}, the non-ideal effects \emph{will} influence their evolution since $|\eta | >  10^{15}$~\ueta{}, where the effect will naturally be stronger near the mid-plane and in the models with stronger magnetic field strengths.  

\begin{figure}
\includegraphics[width=0.95\columnwidth]{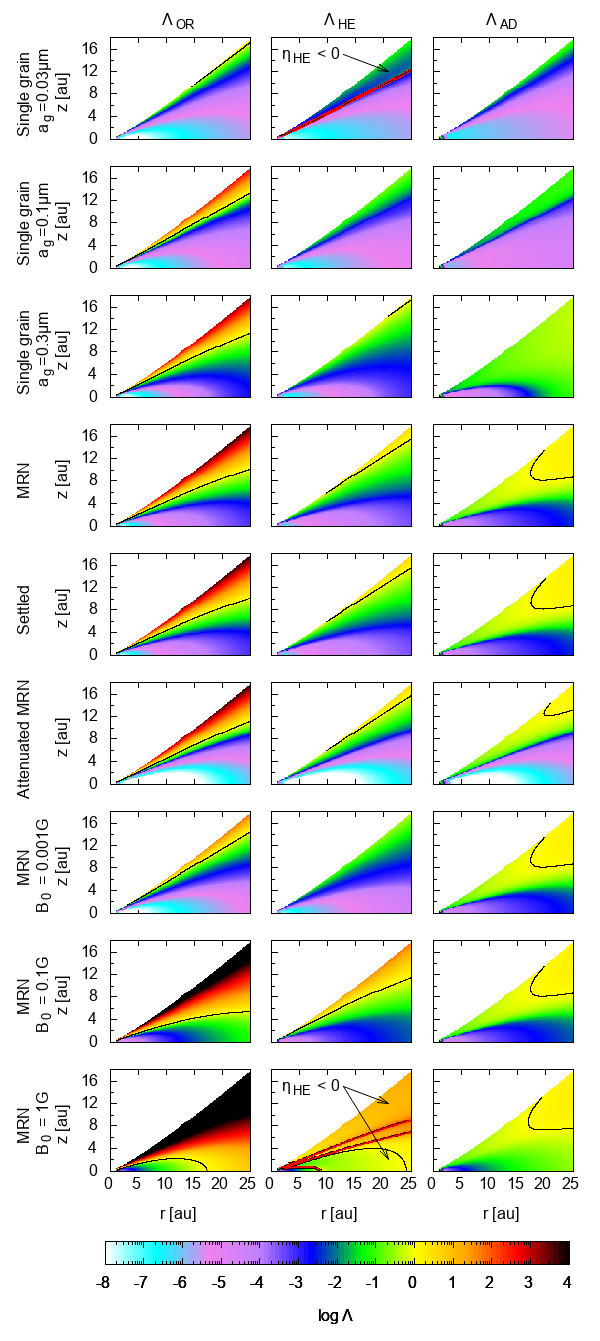}  %Made on Mythos
\caption{The Elsasser number for the nine discs in \figref{fig:idealised:eta}.  Black contour lines represent $\Lambda = 1$ and red contours represent $\eta_\text{HE} = 0$.   As expected, the coupling between the neutral fluid and the magnetic fields decreases towards the mid-plane, indicating that non-ideal MHD is more important in the mid-plane of the disc rather than near its surface. In many discs, $\Lambda < 1$ throughout much of the disc, indicating that non-ideal MHD will always play a role in the evolution of the disc.}
\label{fig:idealised:elsasser}
\end{figure}
\begin{figure}
\includegraphics[width=0.95\columnwidth]{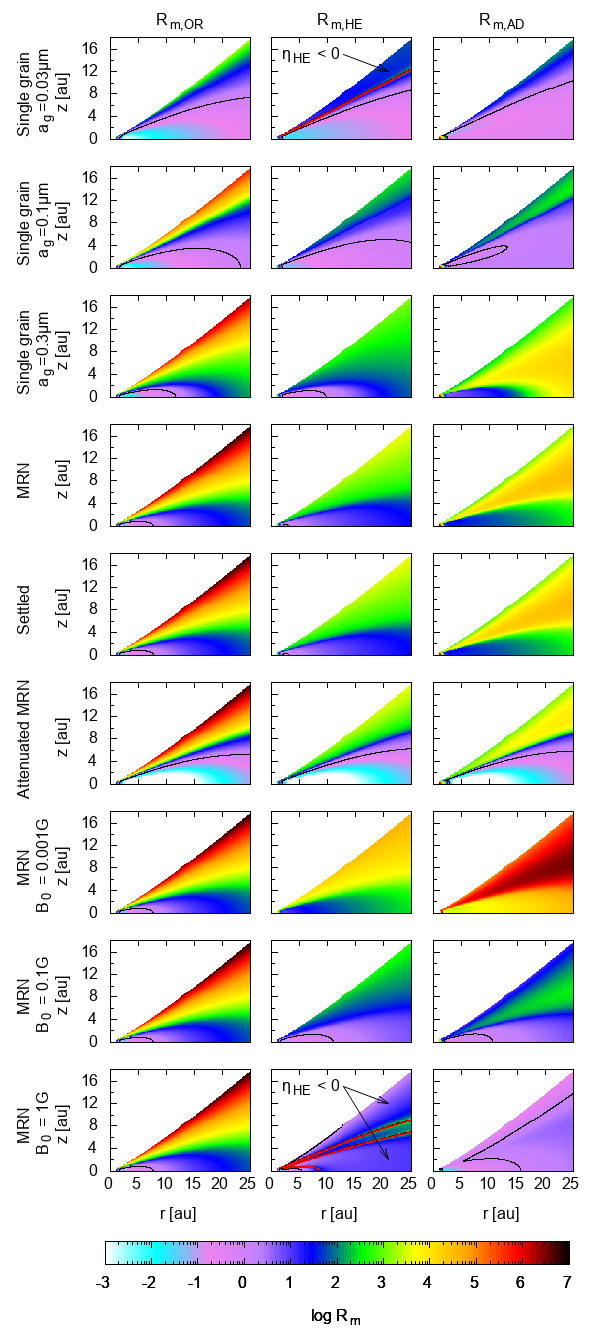}  %Made on Mythos
\caption{The magnetic Reynolds number for the nine discs in \figref{fig:idealised:eta}.  Black contour lines represent $R_\text{m} = 1$ and red contours represent $\eta_\text{HE} = 0$.  Aside from the single grain models with  $a_\text{g} \le 0.1$~\mum{}, $R_\text{m} > 1$, suggesting that non-ideal MHD may be less important than other processes within the disc. }
\label{fig:idealised:reynold}
\end{figure}
\begin{figure}
\includegraphics[width=0.95\columnwidth]{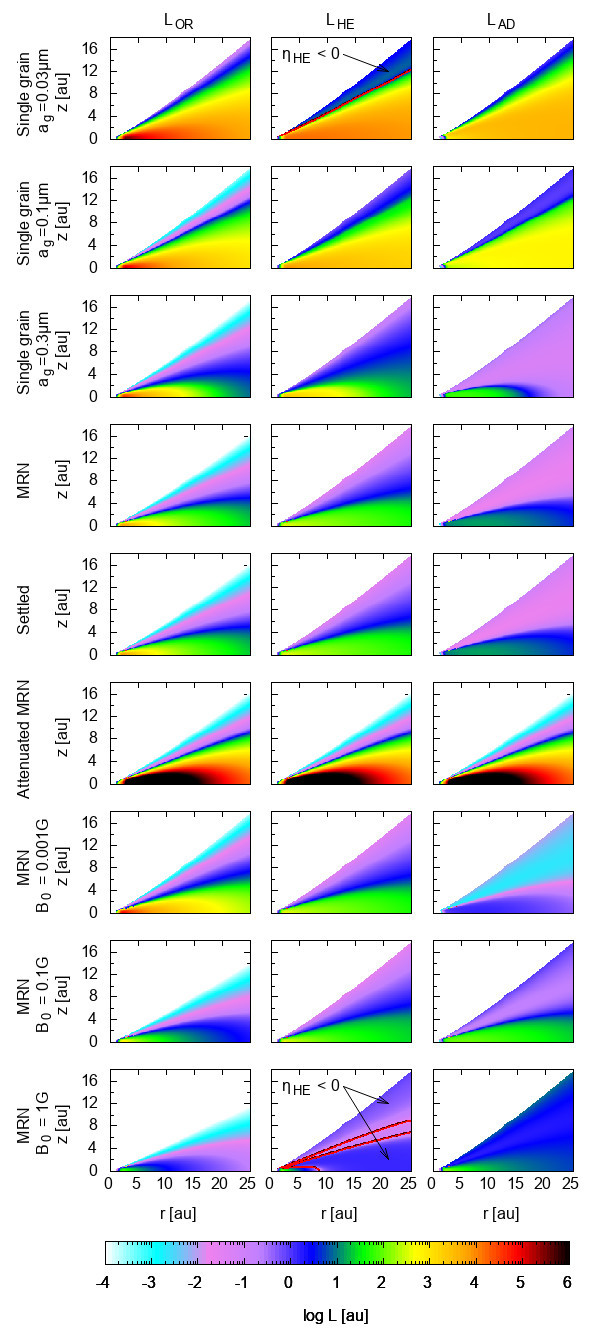}  %Made on Mythos
\caption{The characteristic scale length of the non-ideal processes for the nine discs in \figref{fig:idealised:eta}.  Red contours represent $\eta_\text{HE} = 0$. The large scale lengths in the mid-plane indicate regions where non-ideal MHD is important.}
\label{fig:idealised:scale}
\end{figure}

Given the well-parametrised discs, we can relate the non-ideal coefficients to the disc properties to determine their relative importance within the disc.  \figref{fig:idealised:elsasser} shows the dimensionless Elsasser numbers, 
\begin{equation}
\label{eq:Elsasser}
\Lambda = \frac{v_\text{A}^2}{\Omega |\eta|},
\end{equation}
where $v_\text{A}$ is the \alfven{} velocity, which compares the magnetic forces to the Coriolis forces.    
\figref{fig:idealised:reynold} shows the magnetic Reynolds numbers, 
\begin{equation}
\label{eq:reynold}
R_\text{m} = \frac{c_\text{s}H}{|\eta|},
\end{equation}
which compares the timescales of advection compared to diffusion of the magnetic field; following \citet{FlockHenningKlahr2012}, we have chosen the local sound speed and disc scale height for the characteristic velocity and scale length, respectively.  
\figref{fig:idealised:scale} shows the characteristic scale length of the non-ideal processes, 
\begin{equation}
\label{eq:scale}
L = \frac{|\eta|}{v_\text{A}}.
\end{equation}
Both dimensionless values give an indication of the importance of the magnetic field and its diffusion.  Independently, values of $\Lambda \gg 1$ and  $R_\text{m} \gg 1$ represent strong coupling between the neutral gas and the magnetic field, suggesting that these regions are well-described by ideal MHD \citepeg{WardleSalmeron2012,Tomida+2013,Lin2014}.

In all discs, $|\eta|$, $\Lambda^{-1}$, $R_\text{m}^{-1}$ and $L$ smoothly increase from the surface of the disc to the mid-plane; the exception to this trend is where the sign of $\eta_\text{HE}$ changes.  There is a similar increase for decreasing radius, but this is not as pronounced.  Therefore, as expected and previously discussed in the literature, non-ideal MHD is more important in the mid-plane than near the surface.

The single grain models (top three rows in \figrref{fig:idealised:eta}{fig:idealised:scale}) are the least realistic in terms of grain model, however, they best match the sub-grid grain profile used in the non-ideal MHD disc formation studies in the literature (e.g. \citealp{Tsukamoto+2015oa,Tsukamoto+2015hall} used $a_\text{g} = 0.035$~\mum{} and \citealp{\wpb2016,\wbp2018hd} used $a_\text{g} = 0.1$~\mum{}).  The discs with $a_\text{g} \le 0.1$~\mum{} are dominated by large coefficients ($|\eta| \gtrsim 10^{18}$~\ueta{}).  This yields reasonably large scale lengths and low Elsasser numbers throughout the discs, indicating the importance of non-ideal MHD.  For increasing $a_\text{g}$, the region of $R_\text{m} < 1$ decreases, indicating that magnetic diffusion has less of an effect on the evolution of the magnetic field than advection for discs with larger grains.  These values suggest that non-ideal MHD is important throughout the discs with $a_\text{g} \le 0.1$~\mum{} , and will affect their evolution.  The increase in the non-ideal coefficients for decreasing grain size is consistent with \citet{Tsukamoto+2020} who used magnetic fields of comparable strength, but opposite of \citet{ZhaoCaselliLi2018} whose magnetic fields were much stronger.

The MRN models intentionally include a large range of grain sizes to represent all the grain populations in a disc \citepeg{Dipierro+2015}.  Although there are considerably more grains of smaller sizes, their total mass is considerably less than the total mass of the larger grain species.  This corresponds to a depletion of the smaller grain sizes, which is analogous to increasing the grain size in single grain models.  Thus, given our magnetic field strengths, switching from single grains to the MRN distribution decreases the non-ideal coefficients by a few orders of magnitude.  $\Lambda < 1$ throughout much of the disc while $R_\text{m} > 1$; this suggests that there is at least some decoupling between the neutrals and the magnetic field, indicating that even with lower values of the non-ideal coefficients, the non-ideal processes are important for the evolution of the disc.

Our model where the grain distribution is determined from a dust settling simulation (fifth row; see Appendix~\ref{app:settled}) yields nearly identical results to that of the MRN model.  At our magnetic field strengths, the coefficients are predominantly affected by the small grains, and our five smallest grain sizes are nearly perfectly coupled to the gas in the dust settling simulation; thus, the grain profile of these small grains is the same in both the MRN and settled models.  The larger grains are concentrated towards the mid-plane, however this enhancement over the MRN distribution trivially affects the non-ideal coefficients.  

The unattenuated cosmic ray ionisation rate is $\zeta_\text{cr} = 1.2\times10^{-17}$~\pers{}, yet in the mid-plane the rate decreases to $\zeta_\text{cr} \sim 10^{-20}$~\pers{} for $r > 15$~au and to the imposed floor of $\zeta_\text{min} = 1.1\times10^{-22}$~\pers{} closer to the star\footnote{The ionisation rate becomes unreasonably low if $\zeta_\text{min}$ is not imposed.}.  While the outer regions of the disc are similar to the unattenuated MRN disc, the gas in the mid-plane of the attenuated MRN disc (sixth row) is mostly neutral and very poorly coupled to the magnetic field ($\Lambda \ll 1$ and $R_\text{m}  \ll 1$).  This suggests the existence of a magnetic dead zone.  Although attenuation is realistic and a parameterised attenuation rate is computationally efficient, low ionisation rates yield large coefficients which yield very small numerical timesteps, making numerical simulations slow or even prohibitively expensive to run \citep{\wbp2018ion}.  Even if the mid-plane is essentially neutral, it cannot be modelled using pure hydrodynamics since the regions around the dead zone are weakly ionised thus are somewhat influenced by the magnetic field.  Thus, new and innovative techniques\footnote{This is beyond the scope of this study.} must be derived if the formation and evolution of the dead zone is to be included in simulations of disc formation and early evolution.

As well known, the Hall and ambipolar diffusion coefficients are dependent on the magnetic field strength (bottom three rows in  \figrref{fig:idealised:eta}{fig:idealised:scale}), leading to weak coefficients for $B_0 = 0.001$~G and strong coefficients for $B_0 = 1$~G.  Only with strong magnetic fields of $B_0 = 1$~G do we recover $\eta_\text{HE} < 0$ in the majority of the disc, although there remains a ray of $\eta_\text{HE} > 0$.  Ambipolar diffusion is strongest in this disc, second only to the mid-plane values in the attenuated MRN model.  As the magnetic field strength is increased, the region where the neutrals are decoupled from the magnetic field ($\Lambda < 1$) decreases for Ohmic resistivity and the Hall effect.  Despite the increasing field strength and increasing value of $|\eta_\text{HE}|$, this suggests a weakening influence of these two processes throughout the disc, although they remain important in the mid-plane.  Ambipolar diffusion remains important throughout most of the disc ($\Lambda < 1$), independent of the field strength.  This relationship is reasonable given that $\Lambda_\text{OR} \propto B^2$,  $\Lambda_\text{HE} \propto B$,  and  $\Lambda_\text{AD} \propto B^0$.  When considering the magnetic Reynolds number, $R_\text{m} > 1$ throughout most of the discs, suggesting that the magnetic field is advected rather than diffused and that the non-ideal MHD processes have a weak or negligible influence on these discs.  This is corroborated by the small scale lengths (except near the mid-plane), suggesting only a small region of influence.  Therefore, the non-ideal coefficients increase for the increasing field strength will influence the evolution of the disc; however, their influence may be secondary to other processes within the discs. 

Aside from the discs with a single grain population of small grains, $R_\text{m} > 1$ in much of the discs.  This suggests that in these idealised discs advection is typically more important than diffusion; these are also the regions with weaker non-ideal coefficients ($|\eta| \lesssim 10^{18}$~\ueta{}; in agreement with \citealp{Tomida+2013}).  Large fractions of many of the discs include $\Lambda < 1$ indicating decoupling of the neutral fluid and the magnetic field.  It is indisputable that non-ideal MHD is important in all of these discs near the mid-plane, and the importance throughout the rest of the disc is dependent on the grain properties and magnetic field strength.  These results strongly suggest that non-ideal MHD cannot be ignored.

%----
\subsection{Relative importance of the coefficients}
\begin{figure}
\includegraphics[width=0.95\columnwidth]{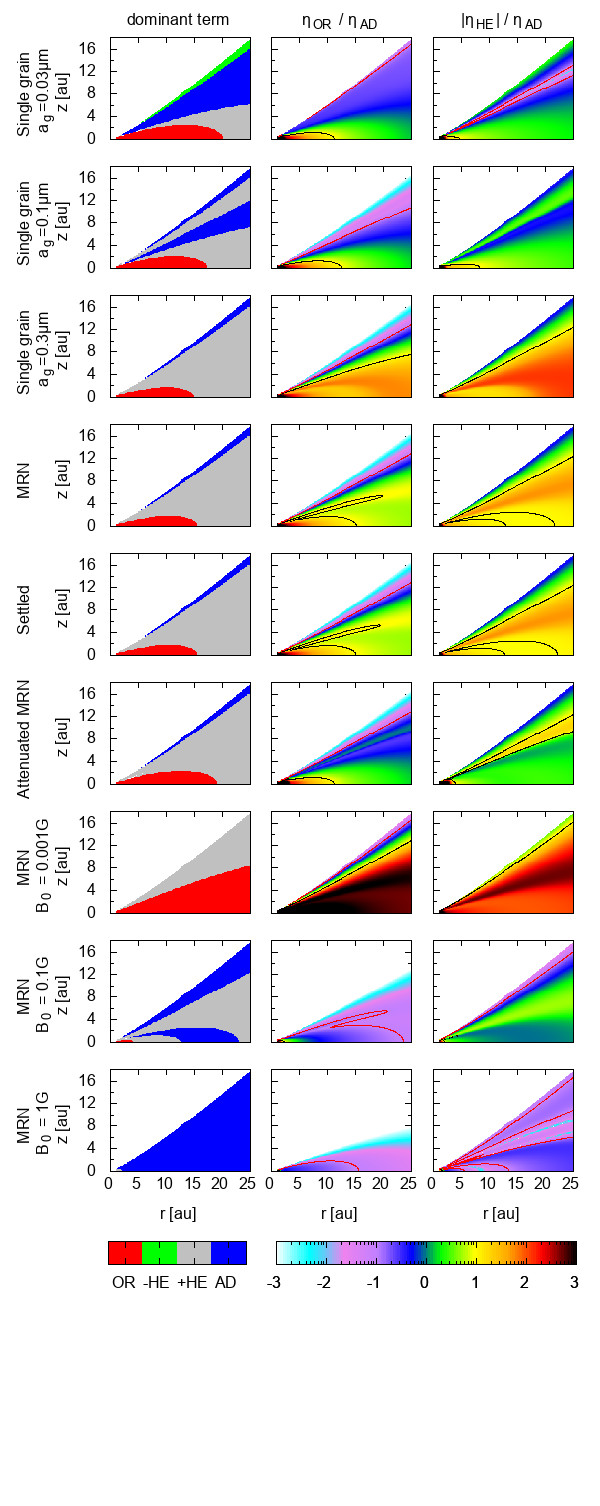}  %Made on Mythos
\caption{The dominant non-ideal term (left-hand column), the ratio of $\eta_\text{OR}/\eta_\text{AD}$ (centre) and  $|\eta_\text{HE}|/\eta_\text{AD}$ (right-hand column) for the nine idealised discs in \figref{fig:idealised:eta}.  Black contours are ratios of 10 and red contours are ratios of 0.1; for clarity, we have not included contours of $\eta_\text{HE} = 0$.  For $B_0 = 0.01$~G, the Hall effect is predominantly the dominant term, but ambipolar diffusion become more important for stronger magnetic field strengths.  Throughout many of the discs,  $0.1 <  \{\eta_\text{OR}/\eta_\text{AD}, |\eta_\text{HE}|/\eta_\text{AD}\} < 10$ indicating the importance of multiple processes in many regions of the discs.}
\label{fig:idealised}
\end{figure}

Once the non-ideal coefficients are calculated, we can easily determine where each term is dominant, which is shown in the first column of \figref{fig:idealised}.  By design, the discs with the MRN grain distribution and the fiducial magnetic field strength of $B_0 = 0.01$~G have the similar nested structure to the cartoon in \figref{fig:cartoon}.  This suggests that, for these initial conditions, the Hall effect is the most important term.  Unlike in  \figref{fig:cartoon}, the Hall effect is dominant in the majority of the disc rather than just a region near the mid-plane.  With these discs (and several others that we tested), the traditional cartoon is difficult to recover, indicating that it is far too idealised, even when we can fine tune all the parameters.

Although each region has a dominant term, the other two terms cannot simply be ignored, as often alluded to when qualitatively describing the disc and showing images similar to \figref{fig:cartoon}.  The second and third columns of \figref{fig:idealised} show the ratios of $\eta_\text{OR}/\eta_\text{AD}$ and $|\eta_\text{HE}|/\eta_\text{AD}$, respectively; the contour lines are at ratios of 0.1 and 10.

With the exceptions of the three MRN discs with $B_0 \ne 0.01$~G, the ratio $0.1 < \eta_\text{OR}/\eta_\text{AD} < 10$ is maintained for \sm30-90 per cent of the disc, showing that both terms are equally important \citep[in agreement with][]{LesurKunzFromang2014}.  Although this contradicts what is generally expected for where Ohmic resistivity is important, we caution that these discs are threaded with a moderate magnetic field strength.  Therefore, even if not the dominant term, Ohmic resistivity will affect these idealised discs.

With the exceptions of the MRN disc with $B_0 = 0.001 $~G, the ratio $0.1 < |\eta_\text{HE}|/\eta_\text{AD} < 10$ is maintained for \sm30-99 per cent of the disc.  This indicates that ambipolar diffusion and the Hall effect are both equally important in determining the evolution of the disc -- at least in terms of the relative strength of their coefficients -- and that neither can be ignored.  This reinforces many similar warnings in the literature \citepeg{SanoStone2002a,SanoStone2002b,Wardle2004,Wardle2007,BraidingWardle2012sf,BraidingWardle2012acc,Bai2014,Bai2015}.

Therefore, independent of which parameter is dominant, all three non-ideal effects must be accounted for when modelling the disc since their coefficients typically differ by less than a factor of 10 in at least a small region of the disc.  

Additional idealised discs can be generated and investigated using the \textsc{Nicil} library's disc generation program, as described in Appendix~\ref{app:nicil:discs}.

%----
\subsection{Contribution to the evolution of the magnetic field}
\label{sec:id:comp}
The evolution of the magnetic field is governed by the induction equation,
\begin{eqnarray}
\label{eq:ideal}
\left.\frac{\text{d} \bm{B}}{\text{d} t}\right|_\text{ideal} =\bm{\nabla} \times \left(  \bm{v} \times \bm{B}\right),
\end{eqnarray}
where $\bm{v}$ is velocity, and the contribution from the non-ideal MHD processes,
\begin{eqnarray}
\label{eq:nonideal}
\left.\frac{\text{d} \bm{B}}{\text{d} t}\right|_\text{non-ideal} =& -&\bm{\nabla} \times \left[  \eta_\text{OR}      \left(\bm{\nabla}\times\bm{B}\right)\right] \notag\\
 &-&\bm{\nabla} \times \left[  \eta_\text{HE}       \left(\bm{\nabla}\times\bm{B}\right)\times\bm{\hat{B}}\right] \notag \\
 &+&\bm{\nabla} \times \left\{ \eta_\text{AD}\left[\left(\bm{\nabla}\times\bm{B}\right)\times\bm{\hat{B}}\right]\times\bm{\hat{B}}\right\}. 
\end{eqnarray}
Therefore, understanding the relative importance of the coefficients is only part of the picture:   We must also understand the relative contribution to each component of the magnetic field.

In keeping with the spirit of this section, we make some simple assumptions about the structure of the disc to better understand the vector evolution of the magnetic field.  We assume cylindrical coordinates, that the gas velocity is purely rotational (i.e. $\bm{v} = v_\phi\hat{\bm{\phi}}$) and that the disc is azimuthally symmetric (i.e. $\partial_\phi = 0$).

For a purely vertical magnetic field, $\bm{B} = B_\text{z}(r,\phi,z)\hat{\bm{z}}$, the total contribution to the magnetic field is
\begin{flalign}
\label{eq:contribute:Bz}
\frac{\text{d} \bm{B}}{\text{d} t} =\bm{\nabla} &\times \left[  -v_\phi B_z \hat{\bm{r}} - \eta_\text{HE}   \partial_\text{r}B_\text{z} \hat{\bm{r}}  - \left(\eta_\text{OR} + \eta_\text{AD}\right)  \partial_\text{r}B_\text{z}\hat{\bm{\phi}} \right] \notag \\
= -&\left(\eta_\text{OR} + \eta_\text{AD}\right) \partial_\text{z} \partial_\text{r} B_\text{z} \hat{\bm{r}} + \left(\eta_\text{OR} + \eta_\text{AD}\right)\partial^2_\text{r} B_\text{z}\hat{\bm{z}} \notag \\
+& \left[\eta_\text{HE} \partial_\text{z}\partial_\text{r} B_\text{z} + \partial_\text{z} \left(v_\phi B_\text{z}\right)\right] \hat{\bm{\phi}}.
\end{flalign}
For a purely radial magnetic field, $\bm{B} = B_\text{r}(r,\phi,z)\hat{\bm{r}}$, the total contribution to the magnetic field is
\begin{flalign}
\frac{\text{d} \bm{B}}{\text{d} t} = & \bm{\nabla} \times \left[  -v_\phi B_r \hat{\bm{z}}  -\eta_\text{HE}  \partial_\text{z}B_\text{r} \hat{\bm{z}} - \left(\eta_\text{OR} + \eta_\text{AD}\right)  \partial_\text{z}B_\text{r} \hat{\bm{\phi}} \right] \notag \\
= &\left(\eta_\text{OR} + \eta_\text{AD}\right) \partial^2_\text{z} B_\text{r} \hat{\bm{r}} - \left(\eta_\text{OR} + \eta_\text{AD}\right)\partial_\text{r} \partial_\text{z} B_\text{r}\hat{\bm{z}} \notag \\
+& \left[\eta_\text{HE} \partial_\text{r}\partial_\text{z} B_\text{r} + \partial_\text{r} (v_\phi B_\text{r})\right] \hat{\bm{\phi}}.
\end{flalign}
Finally, for a purely toroidal magnetic field, $\bm{B} = B_{\phi}(r,\phi,z)\hat{\bm{\phi}}$, the total contribution to the magnetic field is
\begin{flalign}
\frac{\text{d} \bm{B}}{\text{d} t} &= \bm{\nabla} \times \left[\left(\eta_\text{OR} + \eta_\text{AD}\right) \left( \partial_\text{z}B_\phi  \hat{\bm{r}}  - \partial_\text{r}B_\phi  \hat{\bm{z}}\right)\right] \notag \\
&= \left(\eta_\text{OR} + \eta_\text{AD}\right) \left(\partial^2_\text{z} B_\phi + \partial^2_\text{r} B_\phi\right)\hat{\bm{\phi}}.
\end{flalign}

From these simple examples, it is clear that the relative contribution from Ohmic resistivity and ambipolar diffusion is dependent on the strengths of their coefficients since they contribute similarly to each component.  For the evolution of the poloidal magnetic field (i.e. the $\hat{\bm{r}}$ and $\hat{\bm{z}}$ components), both Ohmic resistivity and ambipolar diffusion diffuse the poloidal component while the Hall effect (and induction equation) generate a toroidal component.  In this case, it is inappropriate to compare the relative strengths of Ohmic resistivity and ambipolar diffusion versus the Hall effect since they affect different components.  Although the dispersive and dissipative processes affect different magnetic field components, their relative effect \emph{is} dependent on the strength of the coefficients (e.g. comparing $\eta_\text{AD} \partial_\text{z} \partial_\text{r} B_\text{z} \hat{\bm{r}}$ and $\eta_\text{HE} \partial_\text{z}\partial_\text{r} B_\text{z}  \hat{\bm{\phi}}$ from \eqnref{eq:contribute:Bz}, where both terms include $\partial_\text{z} \partial_\text{r}B_\text{z}$).

When we consider a seed toroidal magnetic field, Ohmic resistivity and ambipolar diffusion diffuse the toroidal field while the Hall effect (and the induction equation) do not contribute to the evolution of the magnetic field at all!  Thus, in this specific case only, it is safe to neglect the Hall effect.

From these calculations, the different behaviour of the dissipative (ambipolar diffusion and Ohmic resistivity) and dispersive (Hall effect) terms is clear.  It also suggests that the ratio of $\eta_\text{OR}/\eta_\text{AD}$ must always be considered since both terms affect the magnetic field in a qualitatively similar way \citepeg{Bai2011grain,XuBai2016} whereas the importance of $\eta_\text{HE}/\eta_\text{AD}$ is dependent on the magnetic field geometry.    Therefore, when determining the relative importance of the non-ideal effects, the magnetic field geometry must also be taken into account.  This further shows that the sketch in \figref{fig:cartoon} is an oversimplification.

As will be shown in \secref{sec:rd:iso}, discs are not axi-symmetric, nor is there only a single component to the magnetic field.  Therefore, in reality, each process will contribute to each magnetic field component, and their relative importance (even when comparing Ohmic resistivity to ambipolar diffusion) will not be a direct ratio of the coefficients, highlighting the complexity of non-ideal MHD in realistic discs.

%----------------------------------------------------------------------------------------------------------------
\section{Structure of self-consistently formed discs}
\label{sec:rd}

In previous studies, we have self-consistently formed protostellar discs both during the formation of an isolated star (\citealp{\wbp2018hd}; hereafter \citetalias{\wbp2018hd}) and during the formation and evolution of a stellar cluster (\citealp{\wbp2019}; hereafter \citetalias{\wbp2019}).  The former disc has the flared shape as expected, whereas the discs in the latter study are less well-defined due to the presence of multiple systems and dynamical interactions.   Since we remove the requirement of choosing the parameters for the disc, analysing the discs in these studies will yield a better understanding of the importance of the non-ideal processes in the discs and how they compare to the idealised discs in \secref{sec:id}.  

Both studies used the 3D smoothed particle hydrodynamics (SPH) code \textsc{sphNG} to solve the self-gravitating, radiation non-ideal magnetohydrodynamics equations.  This code originated from \citet{Benz1990}, but has since been heavily modified to improve both the physical and numerical algorithms \citep{BateBonnellPrice1995,BorveOmangTrulsen2001,WhitehouseBateMonaghan2005,WhitehouseBate2006,PriceMonaghan2007,Price2012,TriccoPrice2012,WursterPriceAyliffe2014}.  Both studies used version 1.2.1 of the \textsc{Nicil} library \citep{Wurster2016} with a single dust grain size of $a_g = 0.1$~\mum{}, and the non-ideal processes were always included in the calculations of the magnetic field.  Due to the long runtime of both studies and due to the different goals, there are small differences in the \textsc{sphNG} versions between the two studies.  We summarise the differences below, however, they are not expected to affect our conclusions:
\begin{enumerate}
\item \citetalias{\wbp2018hd} was initialised with a 1~\Msun{} sphere of gas of uniform density that was undergoing solid body rotation; it was threaded with a magnetic field that was anti-aligned with the rotation axis and had a strength of 5 times the critical mass-to-flux ratio \citepeg{Mestel1999,MaclowKlessen2004}.  \citetalias{\wbp2019} was initialised with a 50~\Msun{} sphere of gas of uniform density that was seeded with a turbulent velocity field; we modelled four different initial magnetic field strengths.
\item \citetalias{\wbp2018hd} self-consistently modelled the stellar core without a sink particle \citep{BateBonnellPrice1995}.  \citetalias{\wbp2019} used 0.5~au sink particles, where one sink particle represented one star.
\item \citetalias{\wbp2018hd} used the radiative transfer method from \citet{WhitehouseBateMonaghan2005} and \citet{WhitehouseBate2006}.  \citetalias{\wbp2019} used the same method in the dense regions, but the method from  \citet{BateKeto2015} to model the diffuse ISM.
\item \citetalias{\wbp2018hd} used the artificial resistivity algorithm from \citet{Phantom2018}.  \citetalias{\wbp2019} used the more resistive artificial resistivity algorithm from \citet{TriccoPriceBate2016}.  See \citet{Wurster+2017} for a discussion of the resistivities.
\item \citetalias{\wbp2018hd} used a resolution of $3.33\times 10^{-7}$~\Msun{} per particle.  \citetalias{\wbp2019} used a lower resolution of $10^{-5}$~\Msun{} per particle given the additional mass in the simulation.  Discs are resolved in both simulations.
\end{enumerate}

%----
\subsection{Structure of discs formed in isolation}
\label{sec:rd:iso}

\begin{figure*}
\centering
\includegraphics[width=0.24\textwidth]{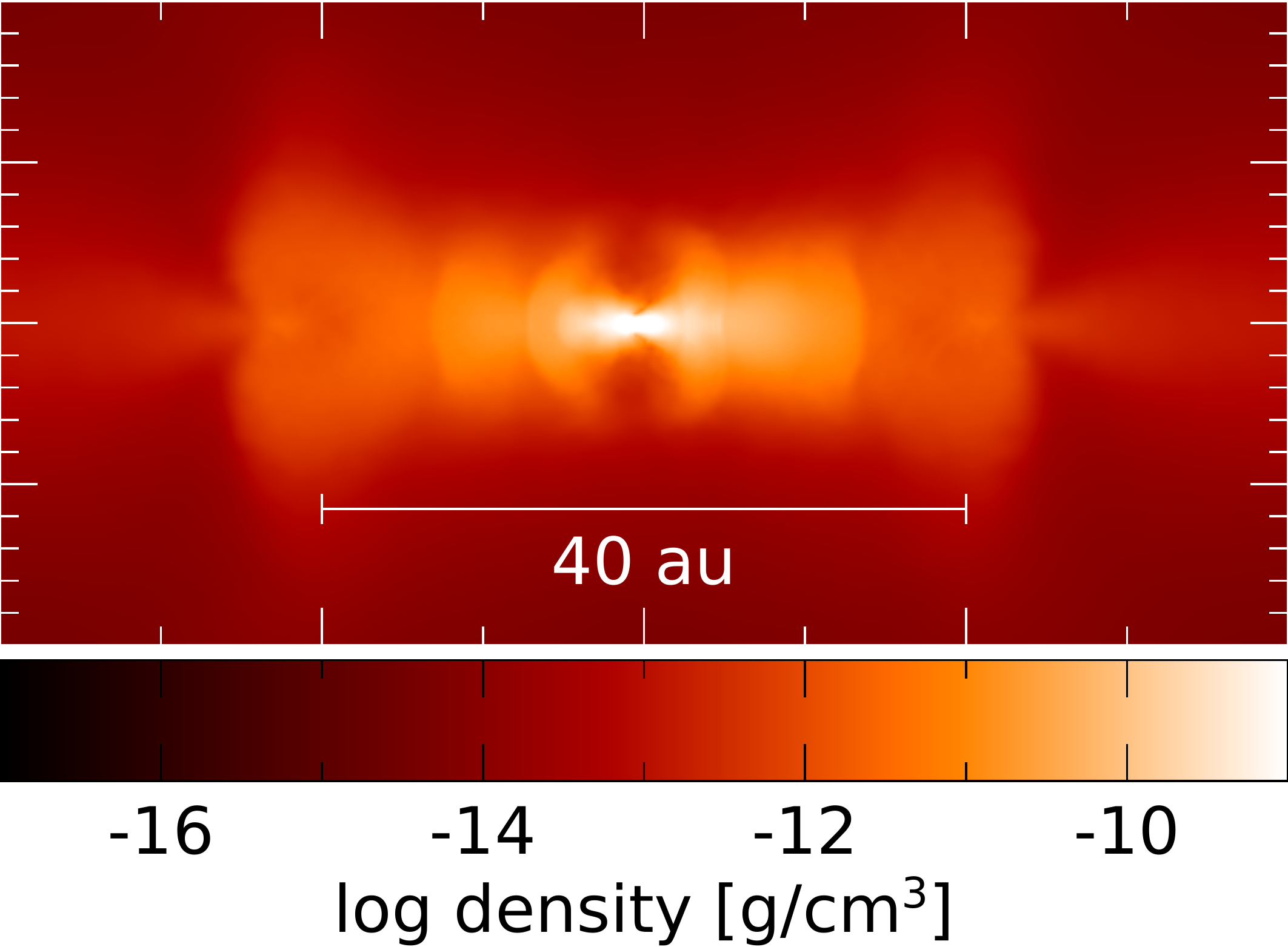}  %Made on DiAL
\includegraphics[width=0.24\textwidth]{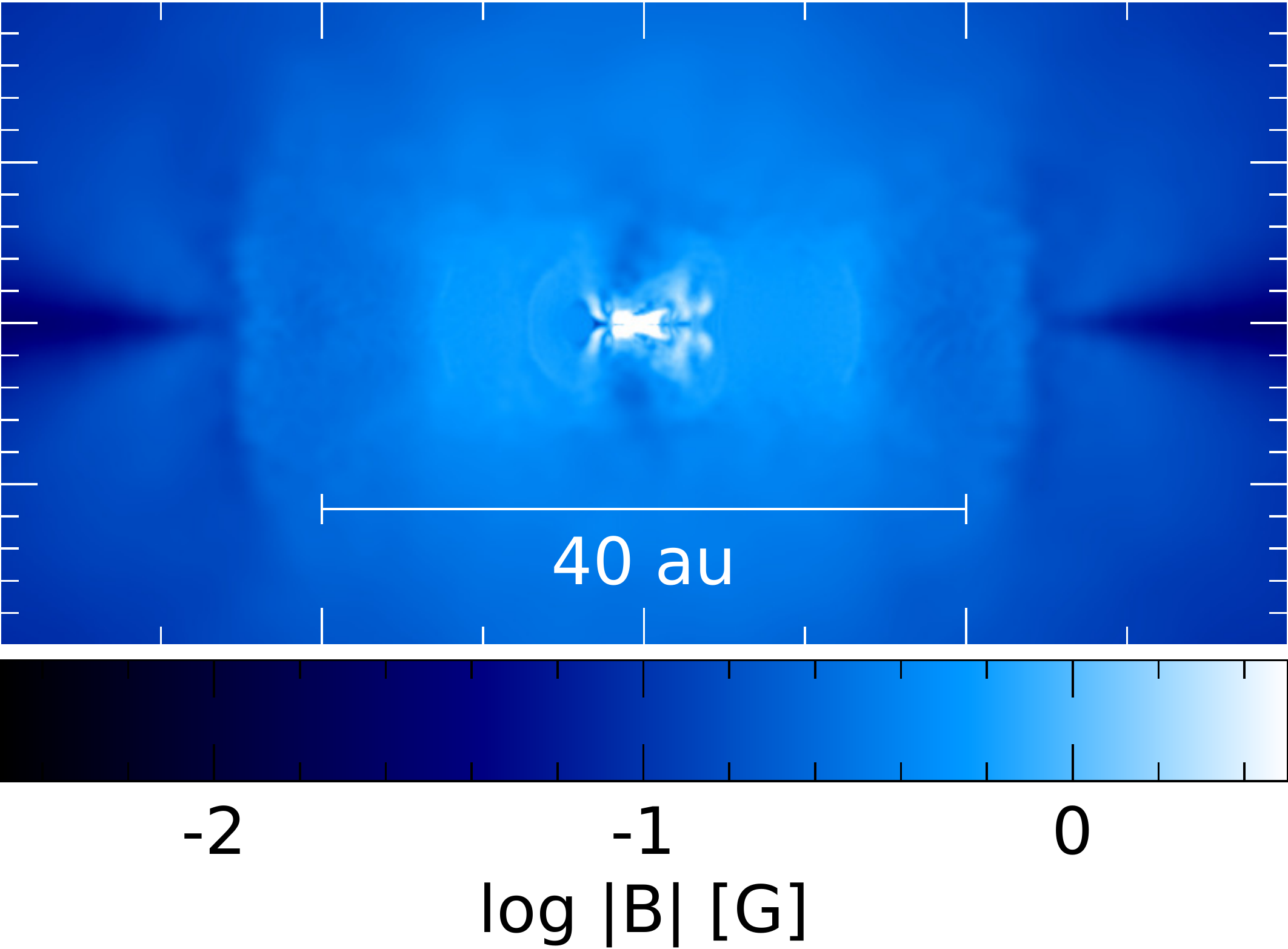}  %Made on DiAL
\includegraphics[width=0.24\textwidth]{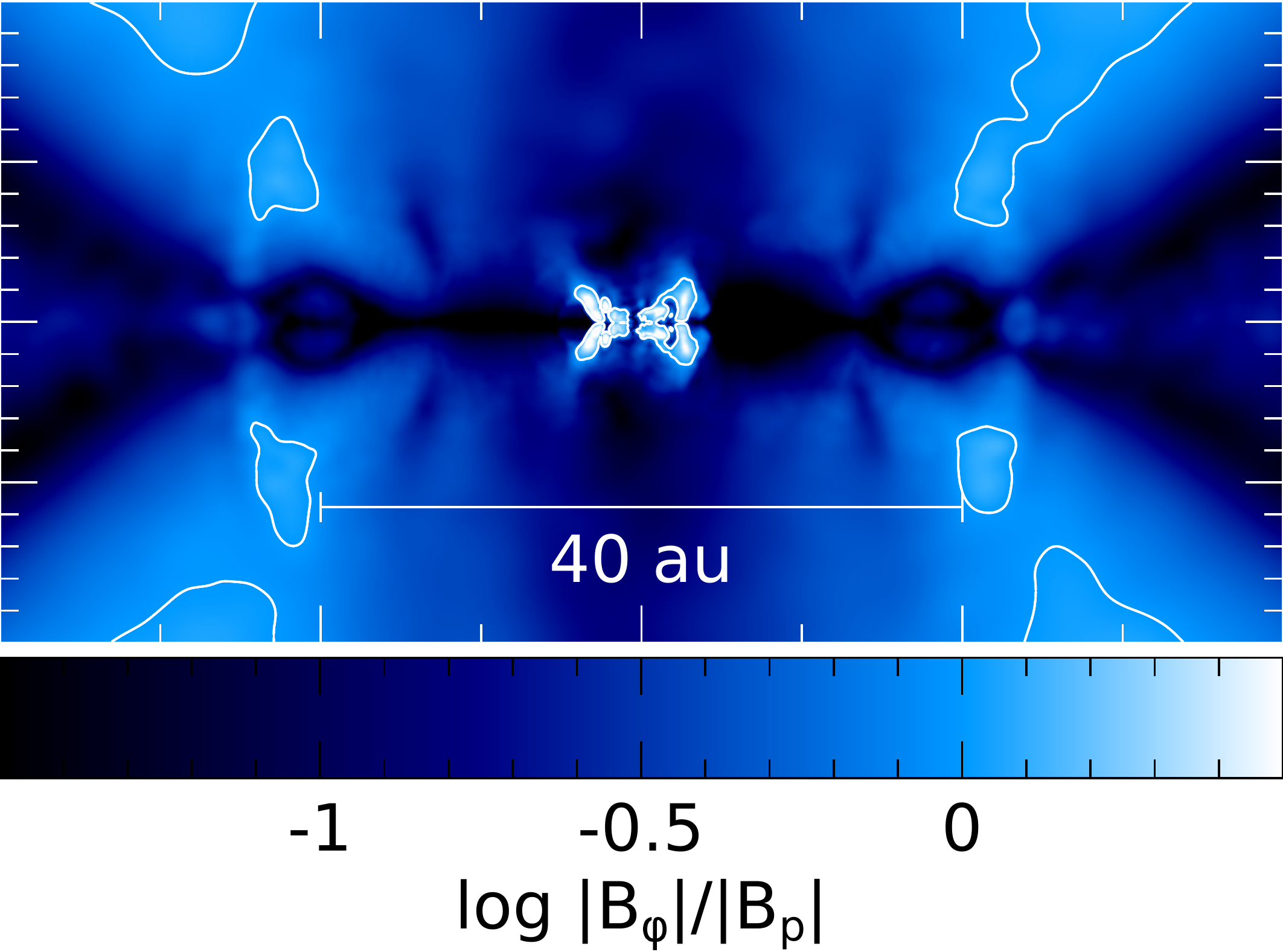}  %Made on DiAL
\includegraphics[width=0.24\textwidth]{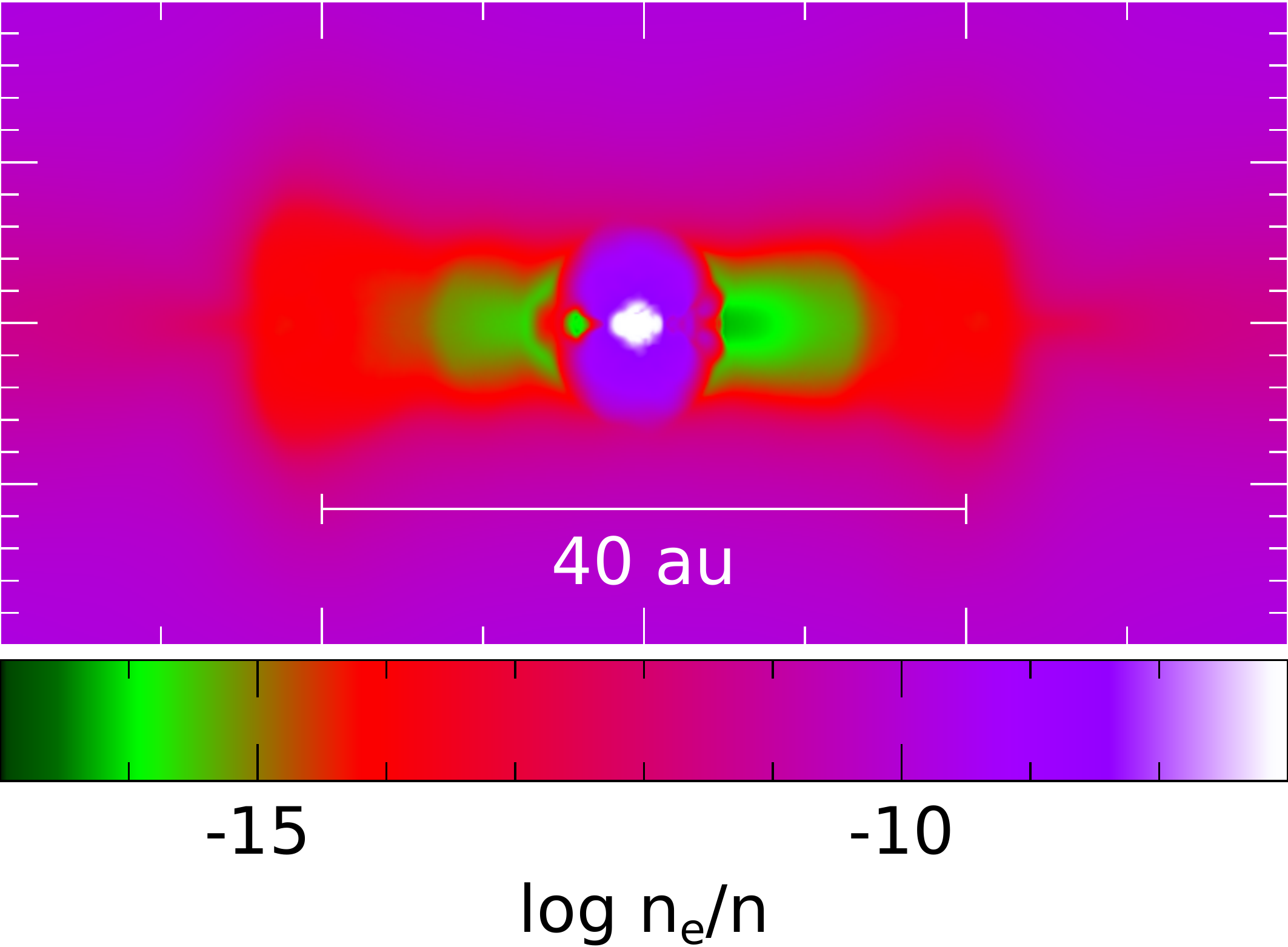}  %Made on DiAL
\includegraphics[width=0.24\textwidth]{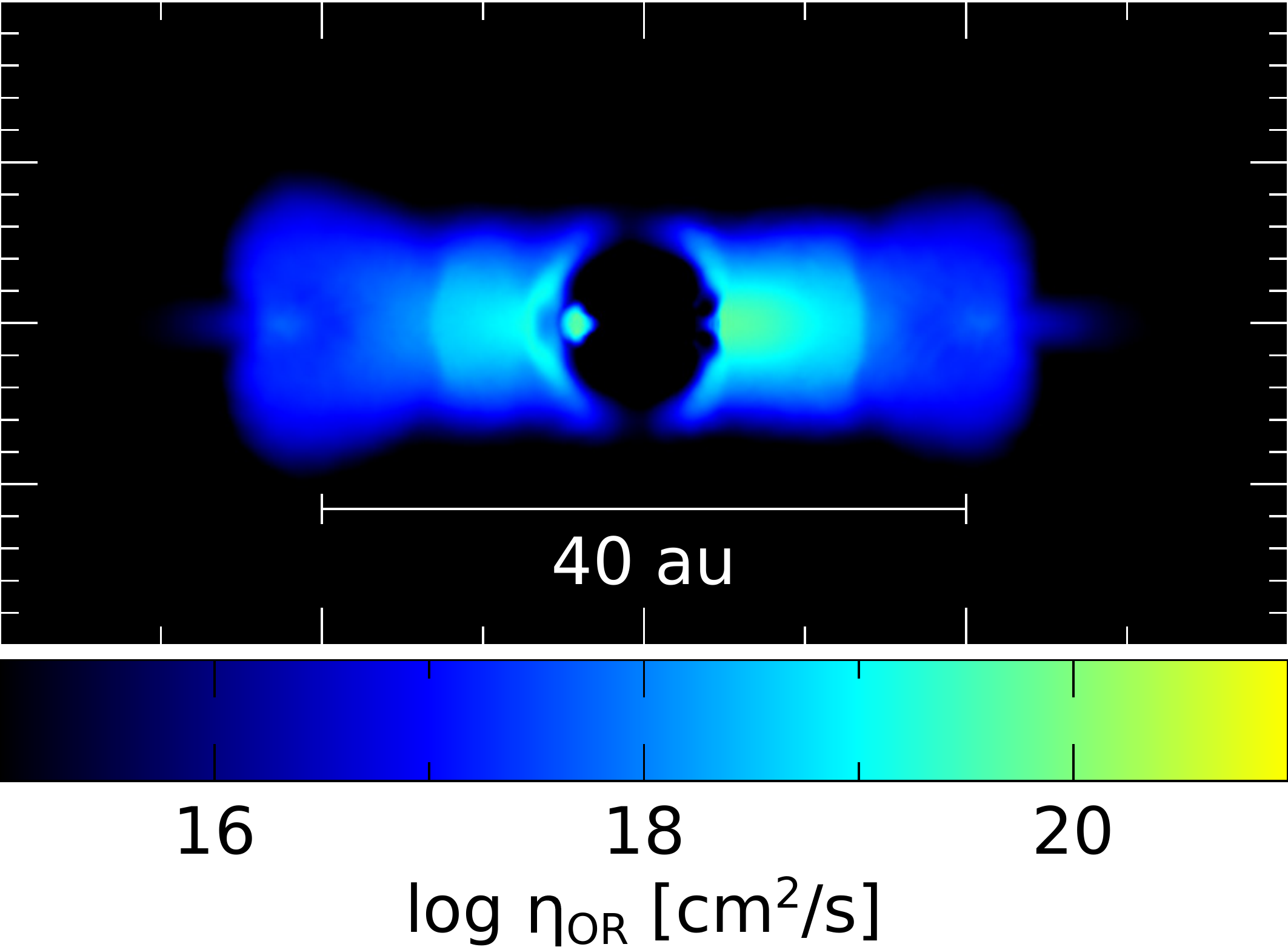}  %Made on DiAL
\includegraphics[width=0.24\textwidth]{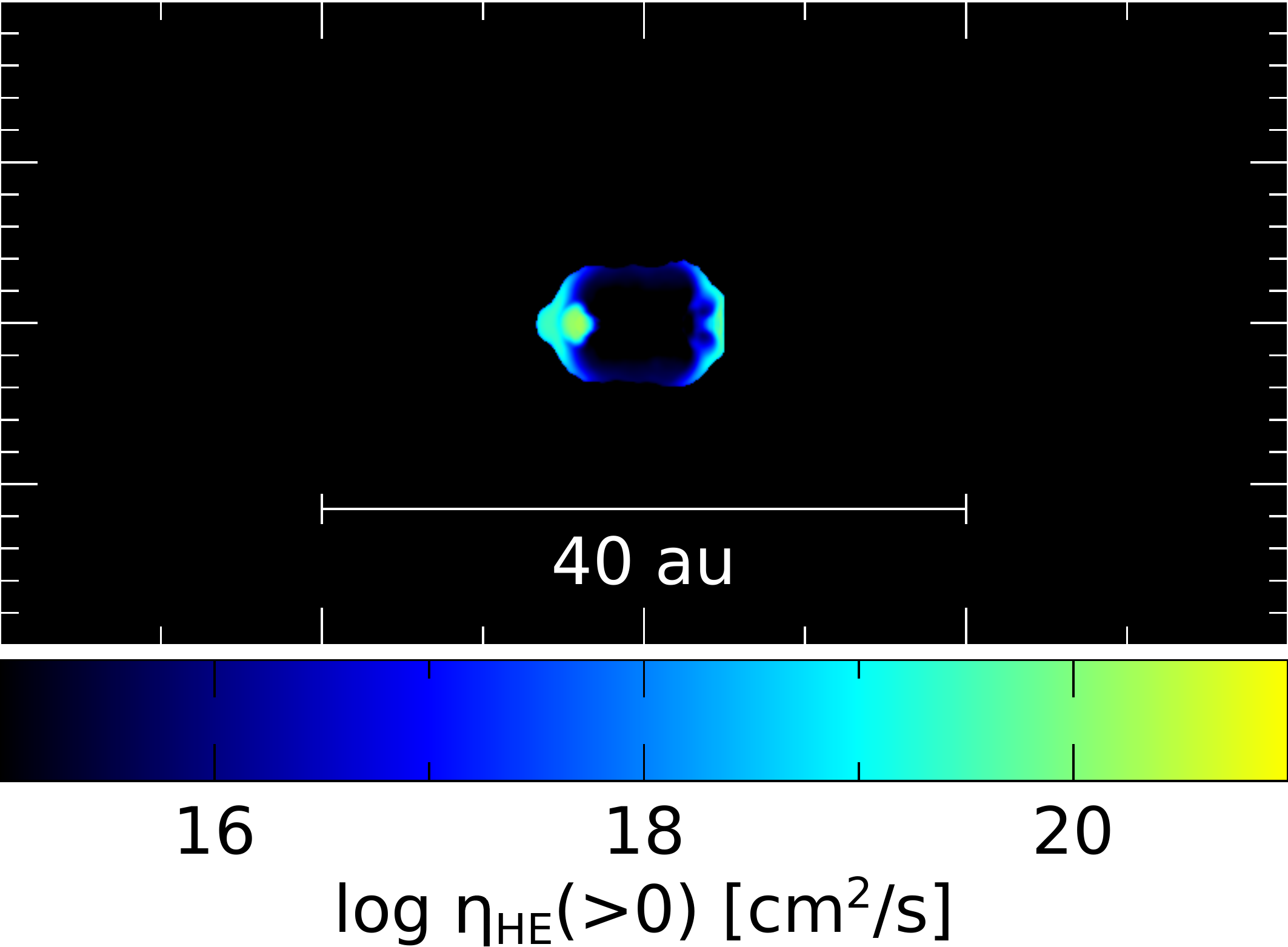}  %Made on DiAL
\includegraphics[width=0.24\textwidth]{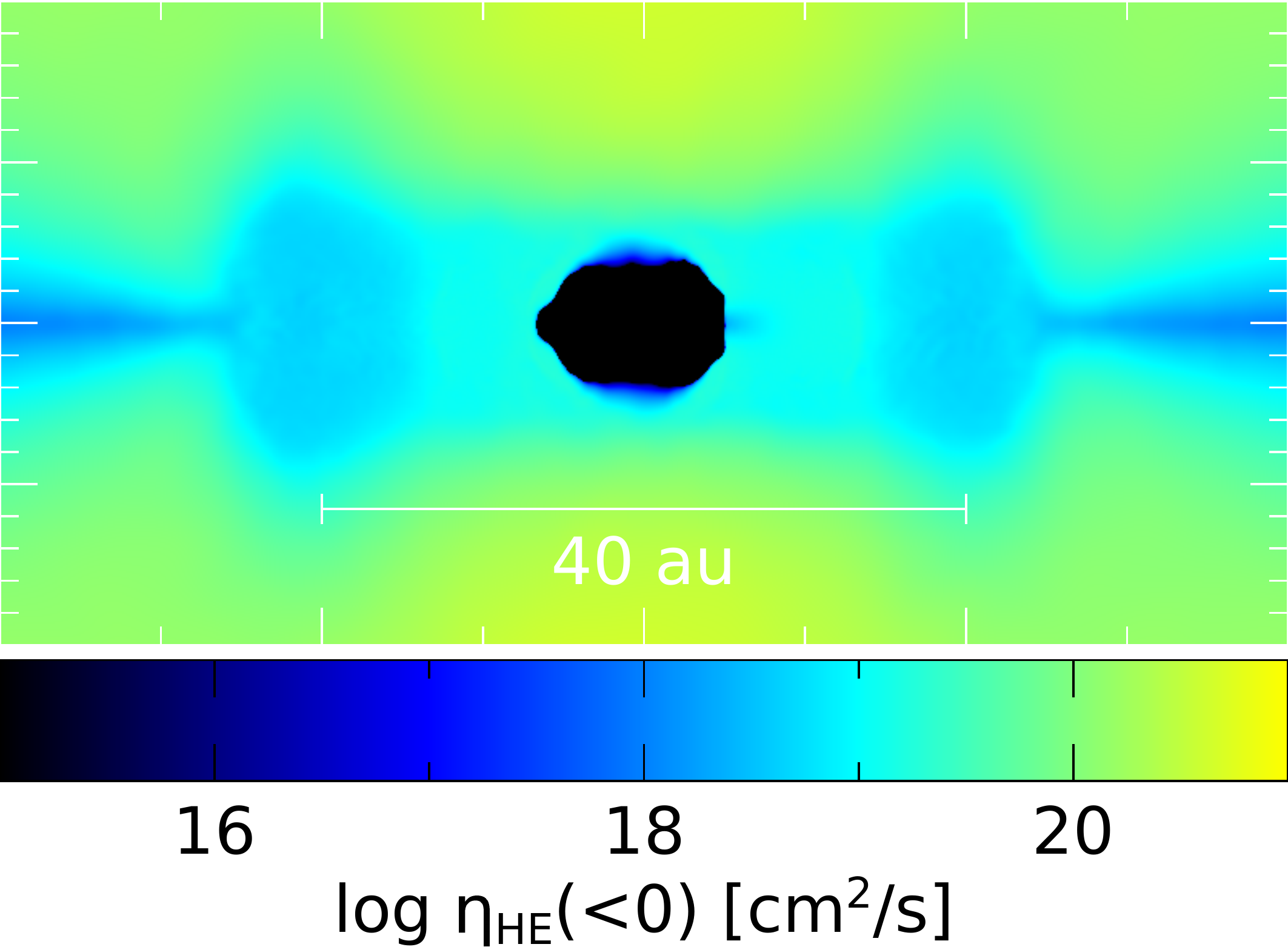}  %Made on DiAL
\includegraphics[width=0.24\textwidth]{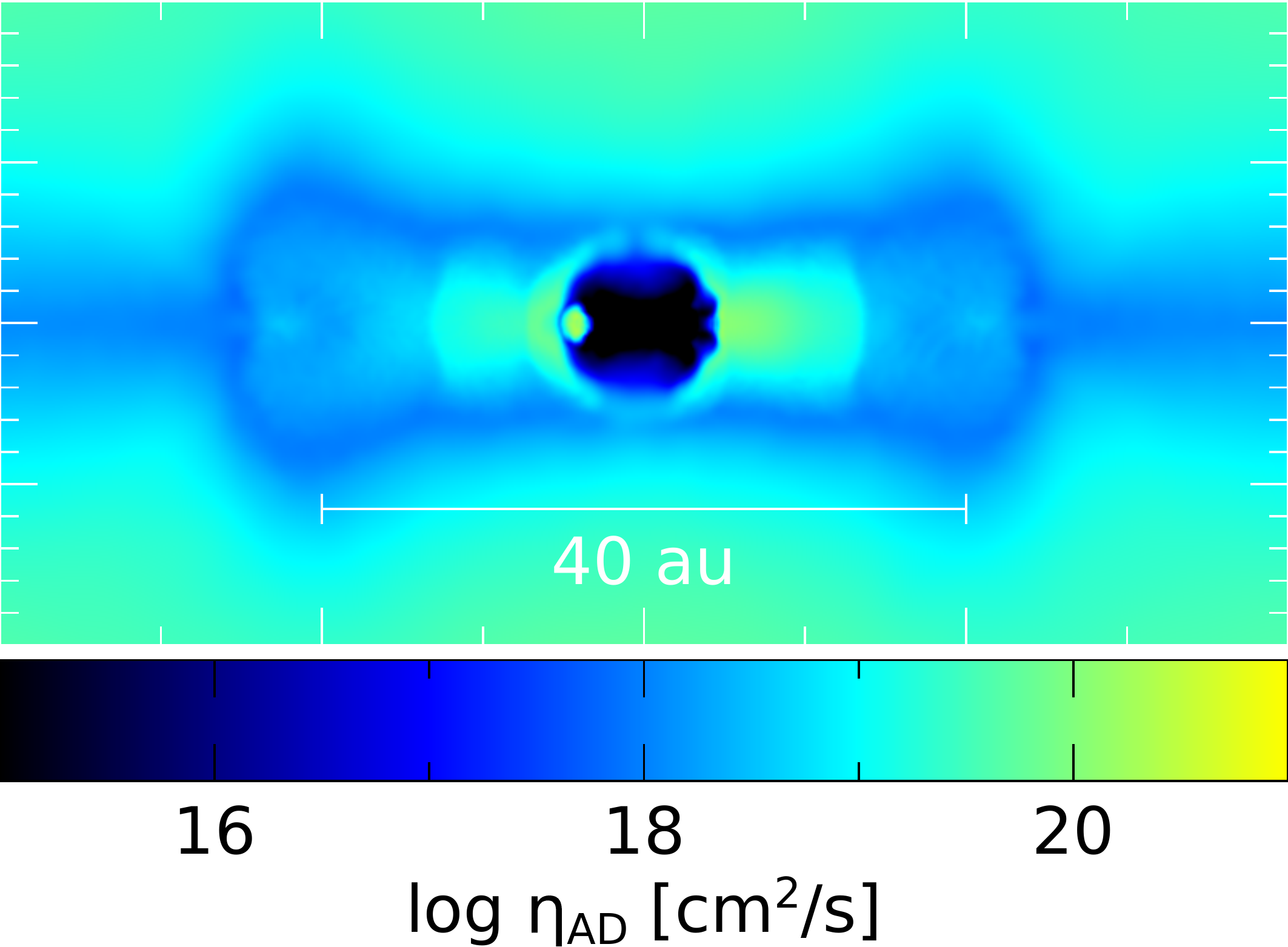}  %Made on DiAL
\includegraphics[width=0.24\textwidth]{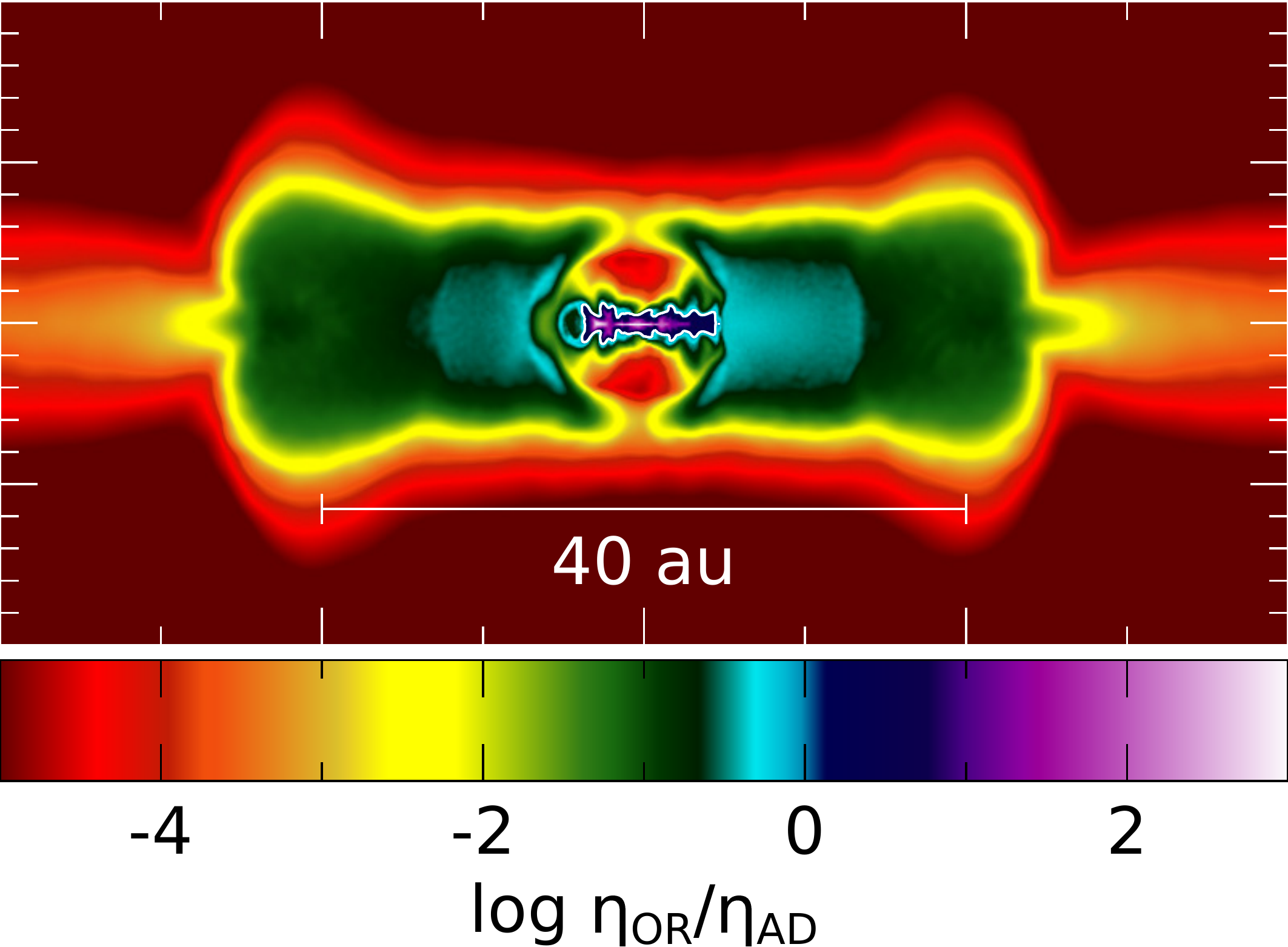}  %Made on DiAL
\includegraphics[width=0.24\textwidth]{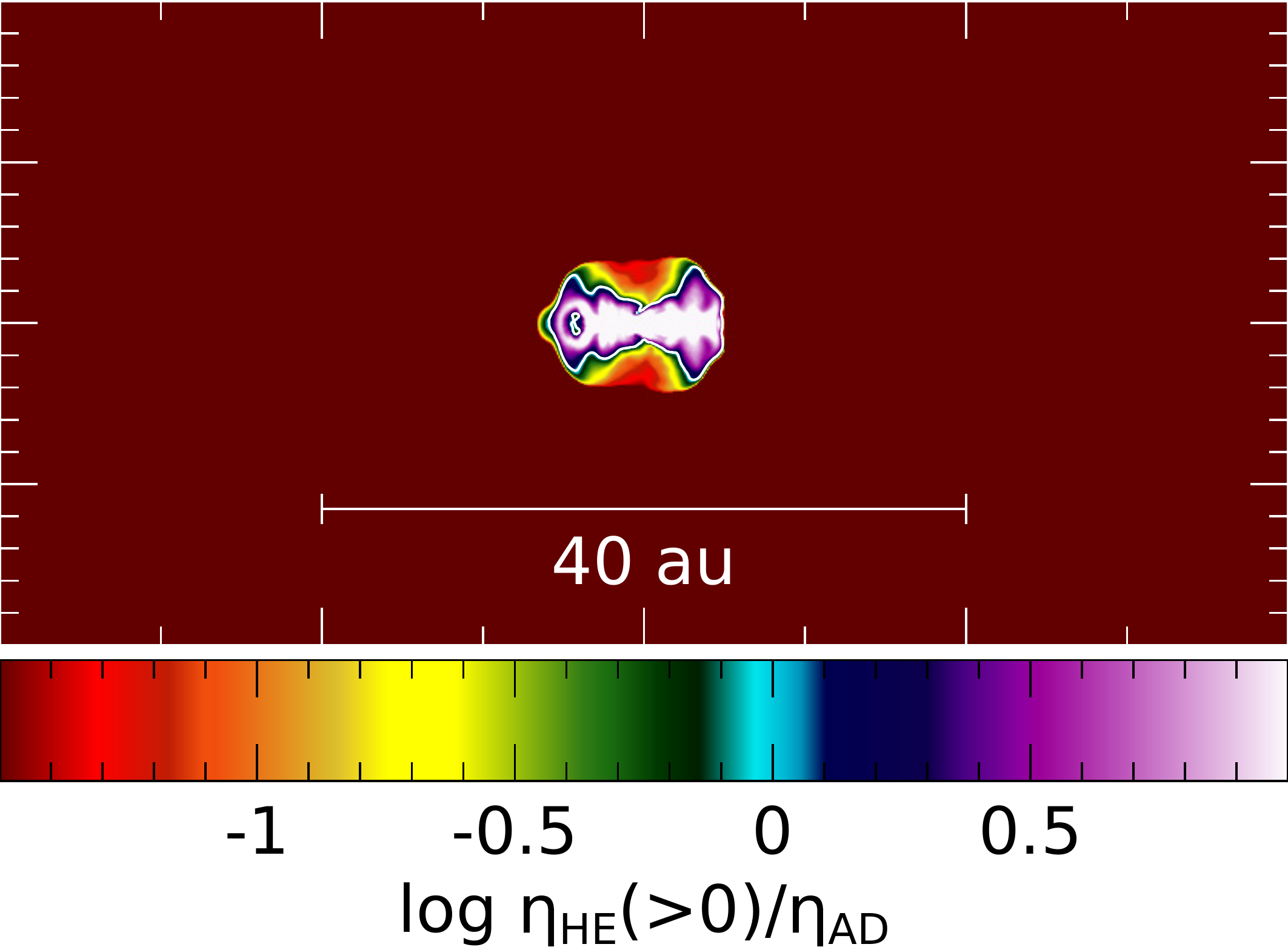}  %Made on DiAL
\includegraphics[width=0.24\textwidth]{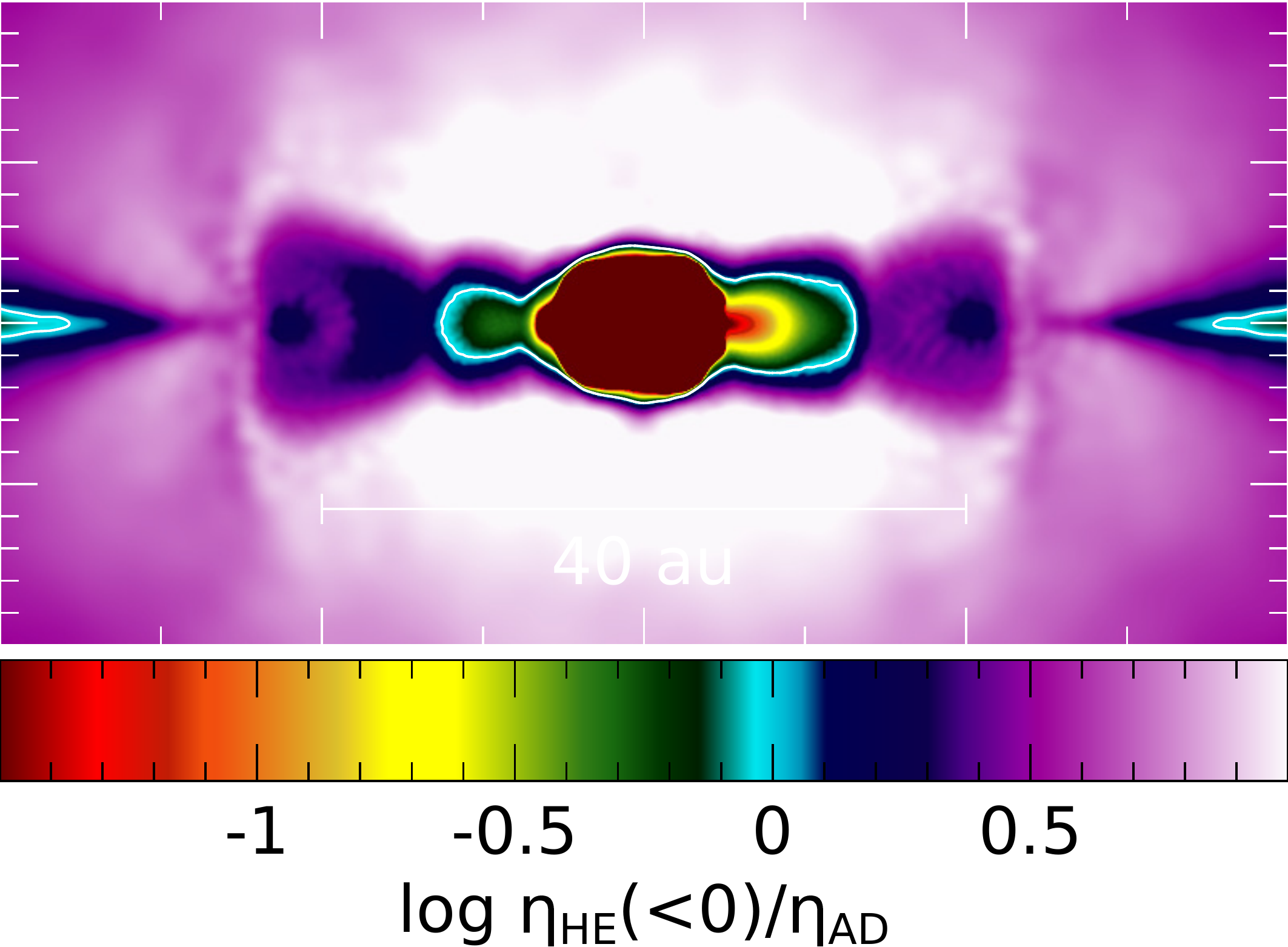}  %Made on DiAL
\caption{Various properties of the isolated disc from \citet[][herein \citetalias{\wbp2018hd}]{\wbp2018hd}.  The images are in a slice through the centre of the core in the $yz$-plane, taken \sm60~yr after their formation which is 9.5~yr after the formation of the stellar core.  Each frame measures 80 x 40~au.  In the ratio plots, the contour is at 1.  In the inner disc, the ambipolar diffusion coefficient is stronger than the Hall coefficient, while in the outer disc and surrounding environment, the Hall coefficient is stronger; this is opposite to the generally accepted structure of the non-ideal MHD coefficients.}
\label{fig:isolated}
\end{figure*}

In \citetalias{\wbp2018hd}, we self-consistently formed a protostellar disc by following the gravitational collapse of a cloud core.  Thus, the disc in this section is a more realistic representation of a disc than those discussed in \secref{sec:id}.  \figref{fig:isolated} shows the gas density, magnetic field strength, ratio of the poloidal field to toroidal magnetic field, ionisation fraction, the non-ideal MHD coefficients and ratios of the coefficients relative to $\eta_\text{AD}$.  The images are taken 9.5~yr after the formation of the stellar core.
\begin{figure*}
\centering
\includegraphics[width=0.24\textwidth]{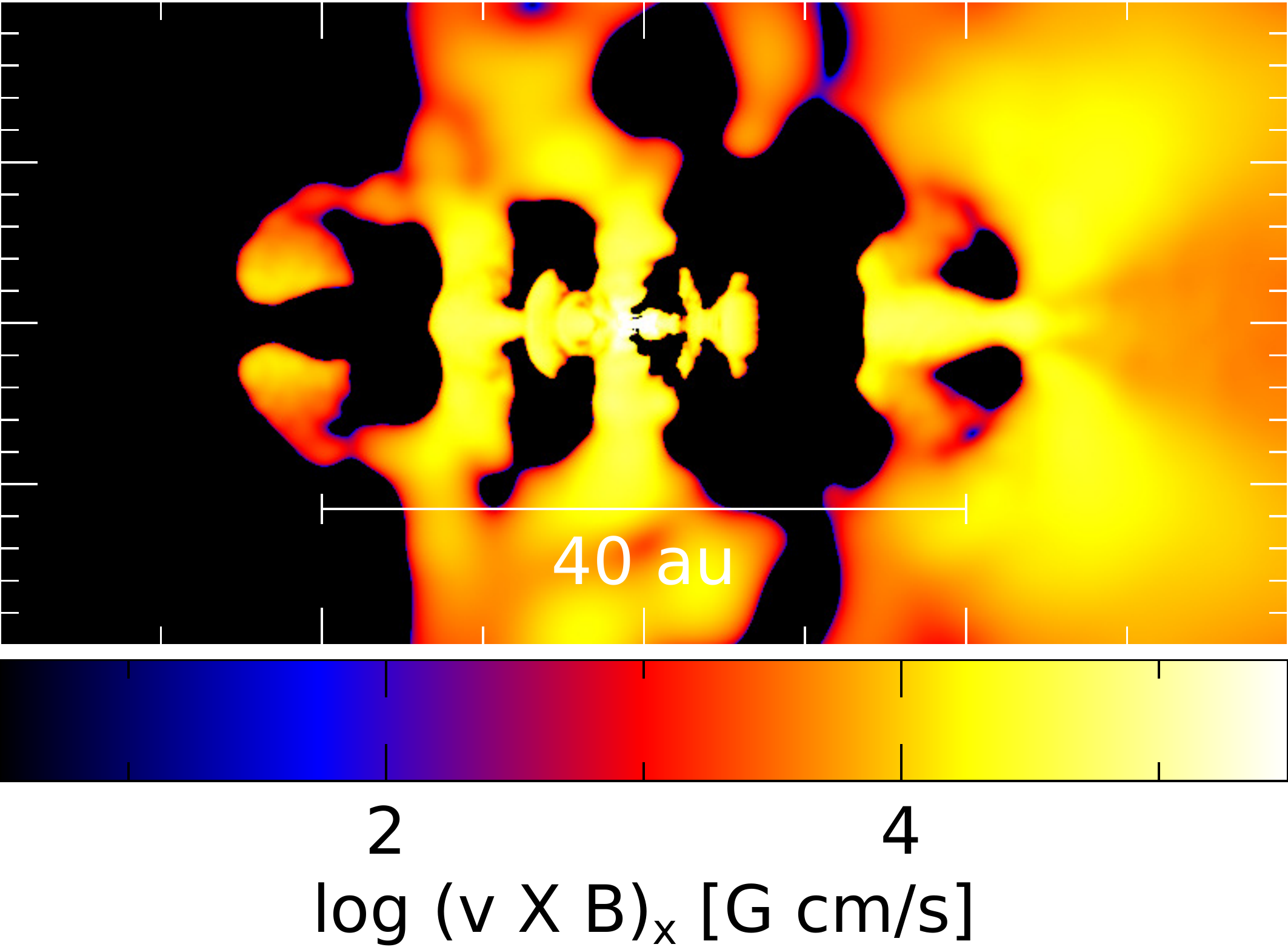}  %Made on DiAL
\includegraphics[width=0.24\textwidth]{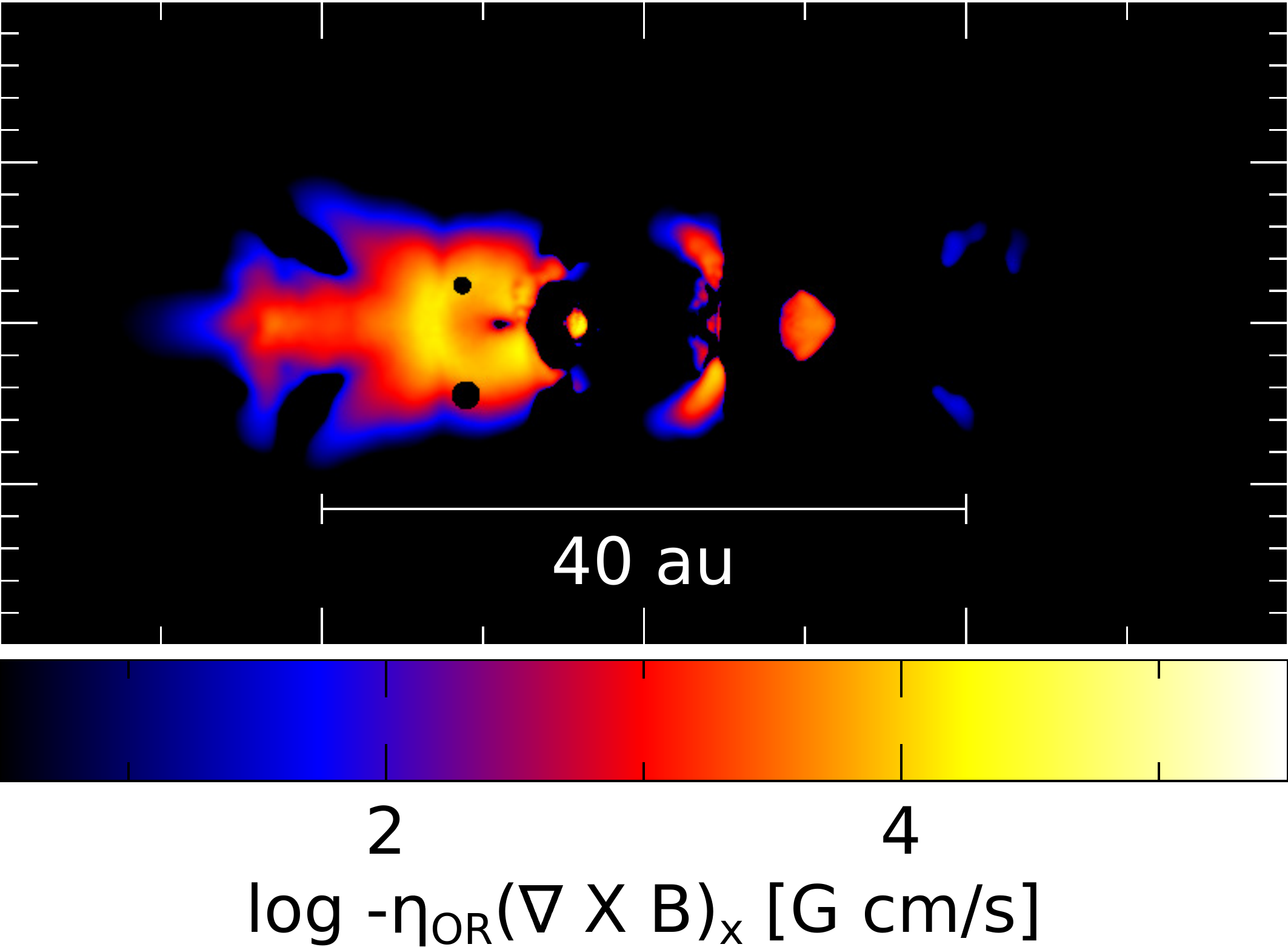}  %Made on DiAL
\includegraphics[width=0.24\textwidth]{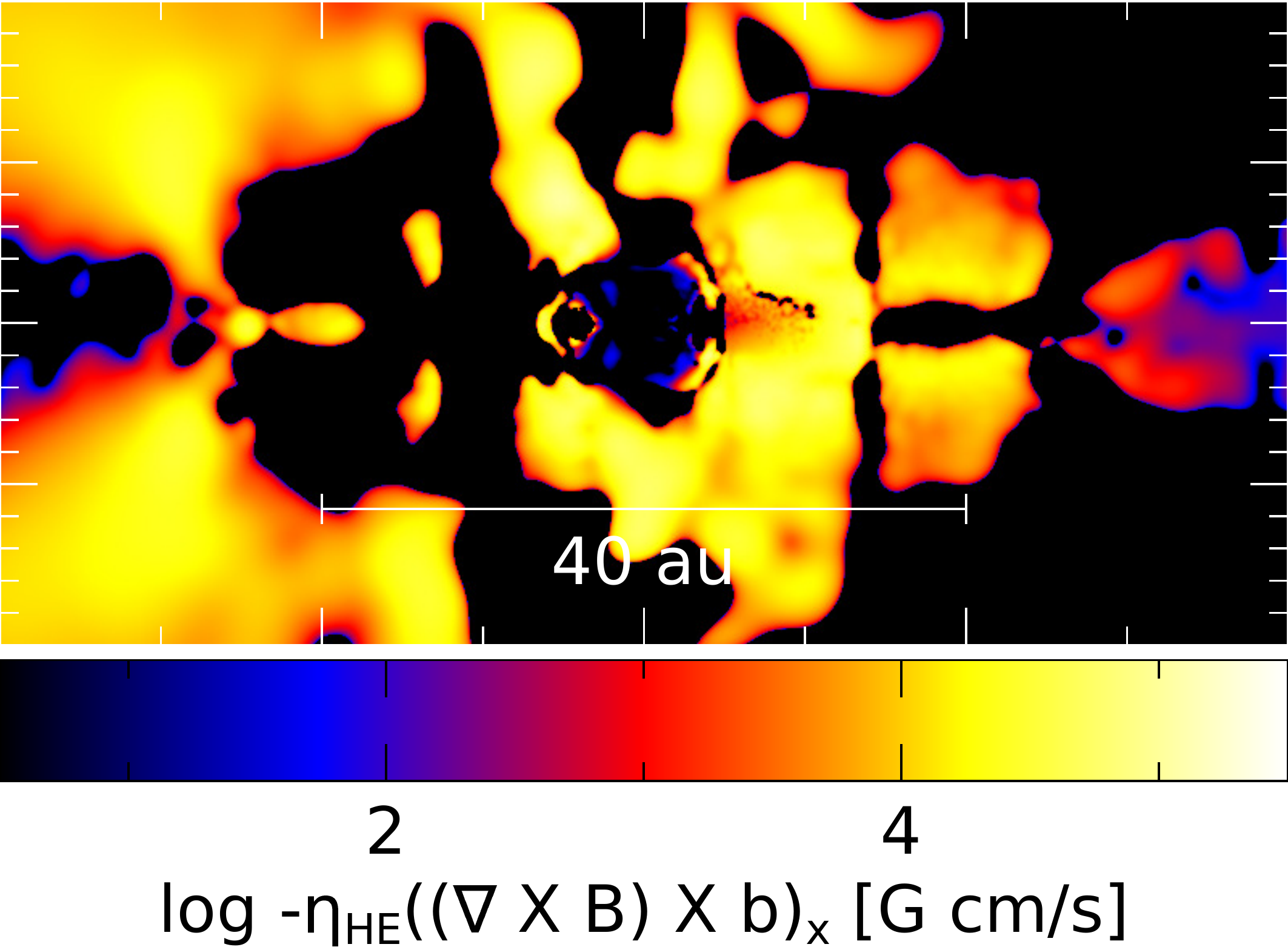}  %Made on DiAL
\includegraphics[width=0.24\textwidth]{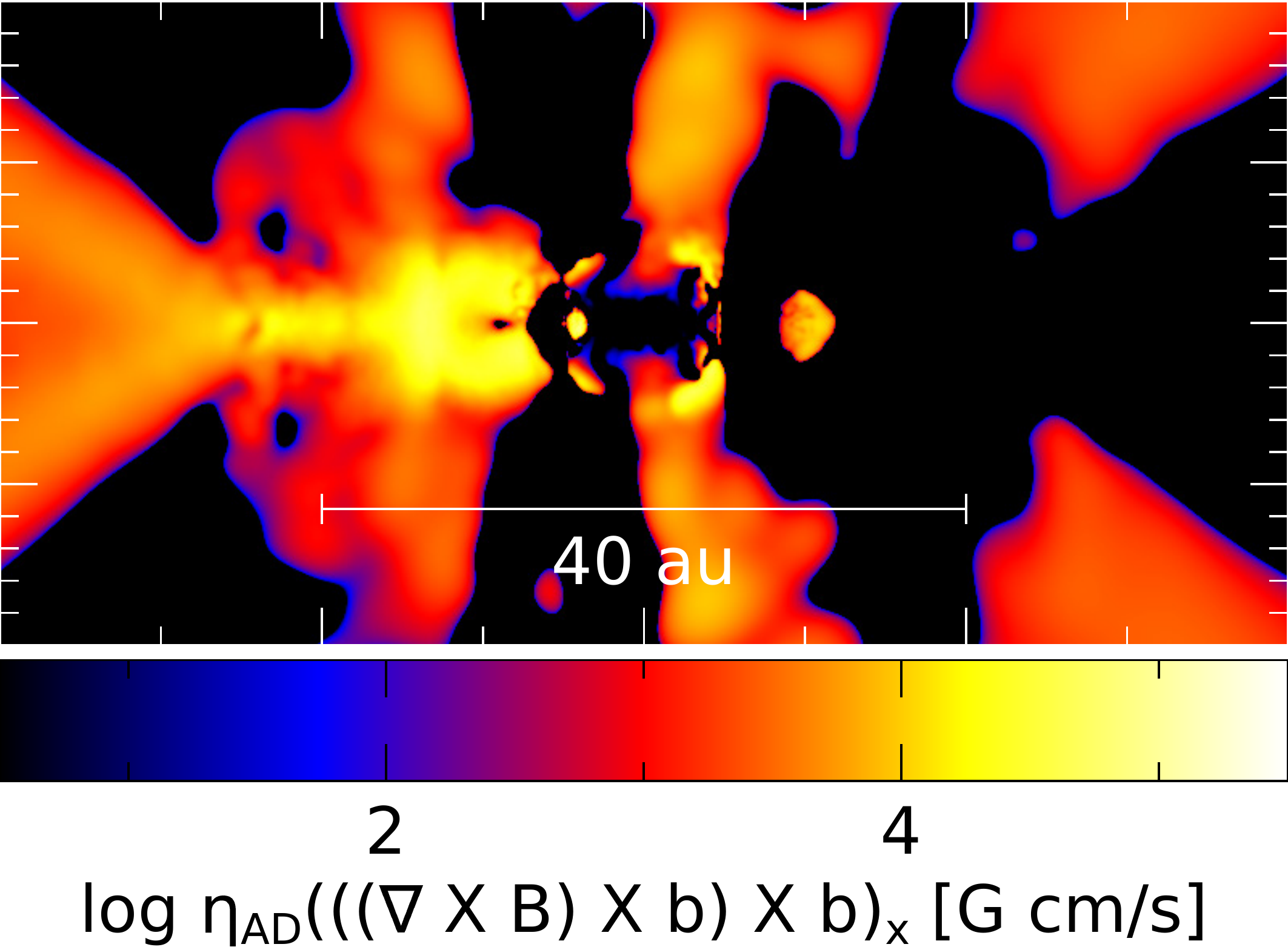}  %Made on DiAL
\includegraphics[width=0.24\textwidth]{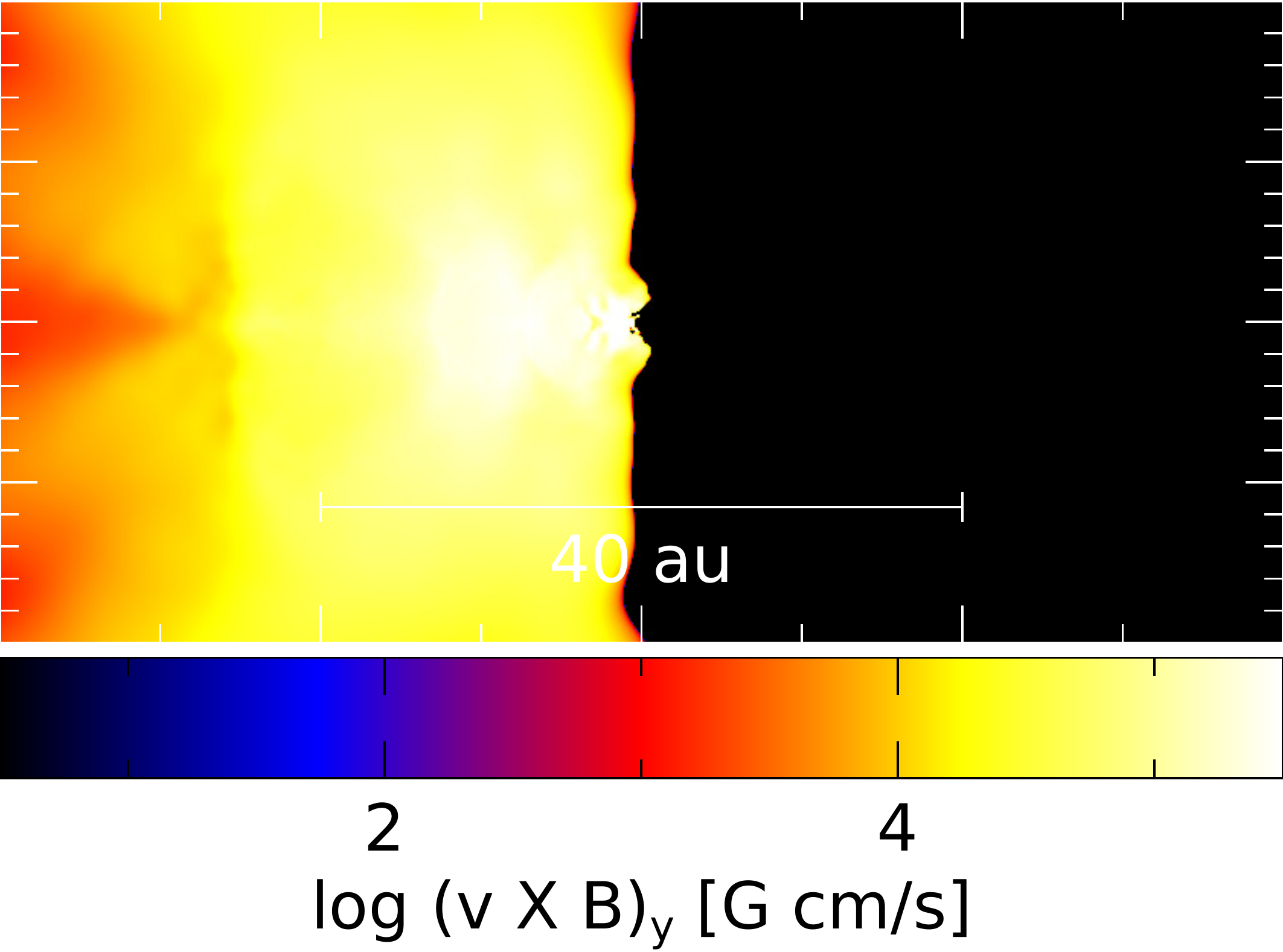}  %Made on DiAL
\includegraphics[width=0.24\textwidth]{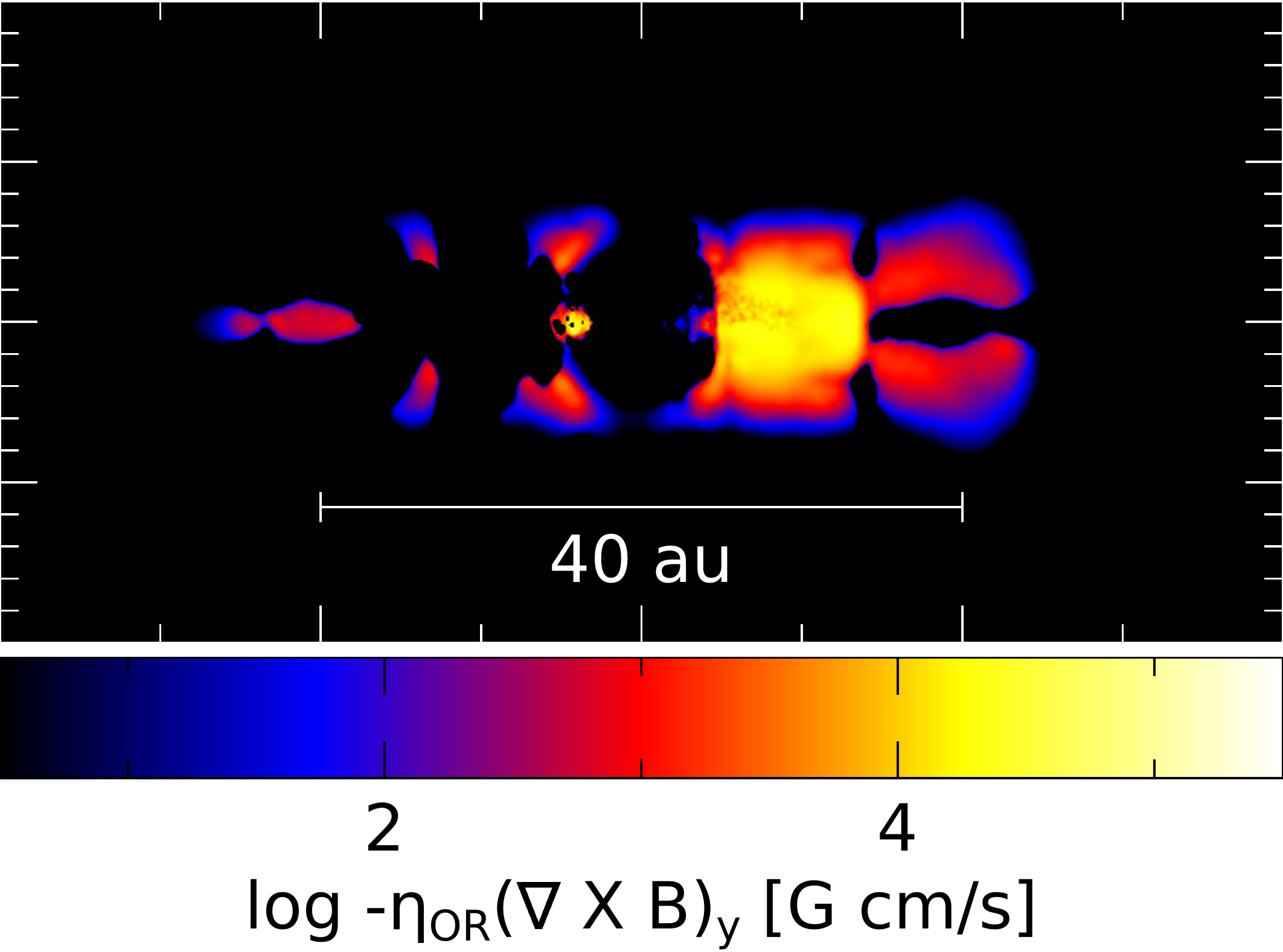}  %Made on DiAL
\includegraphics[width=0.24\textwidth]{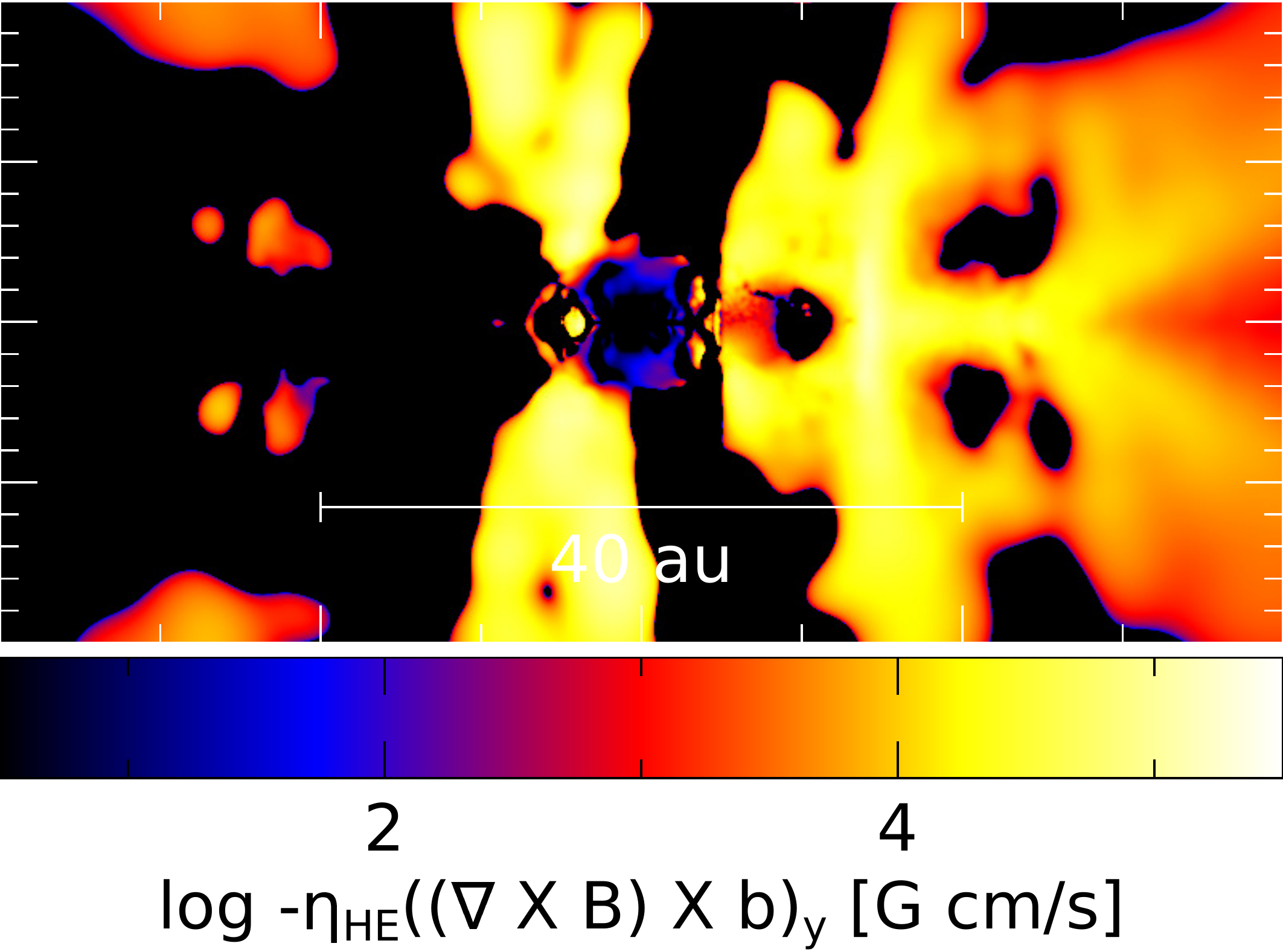}  %Made on DiAL
\includegraphics[width=0.24\textwidth]{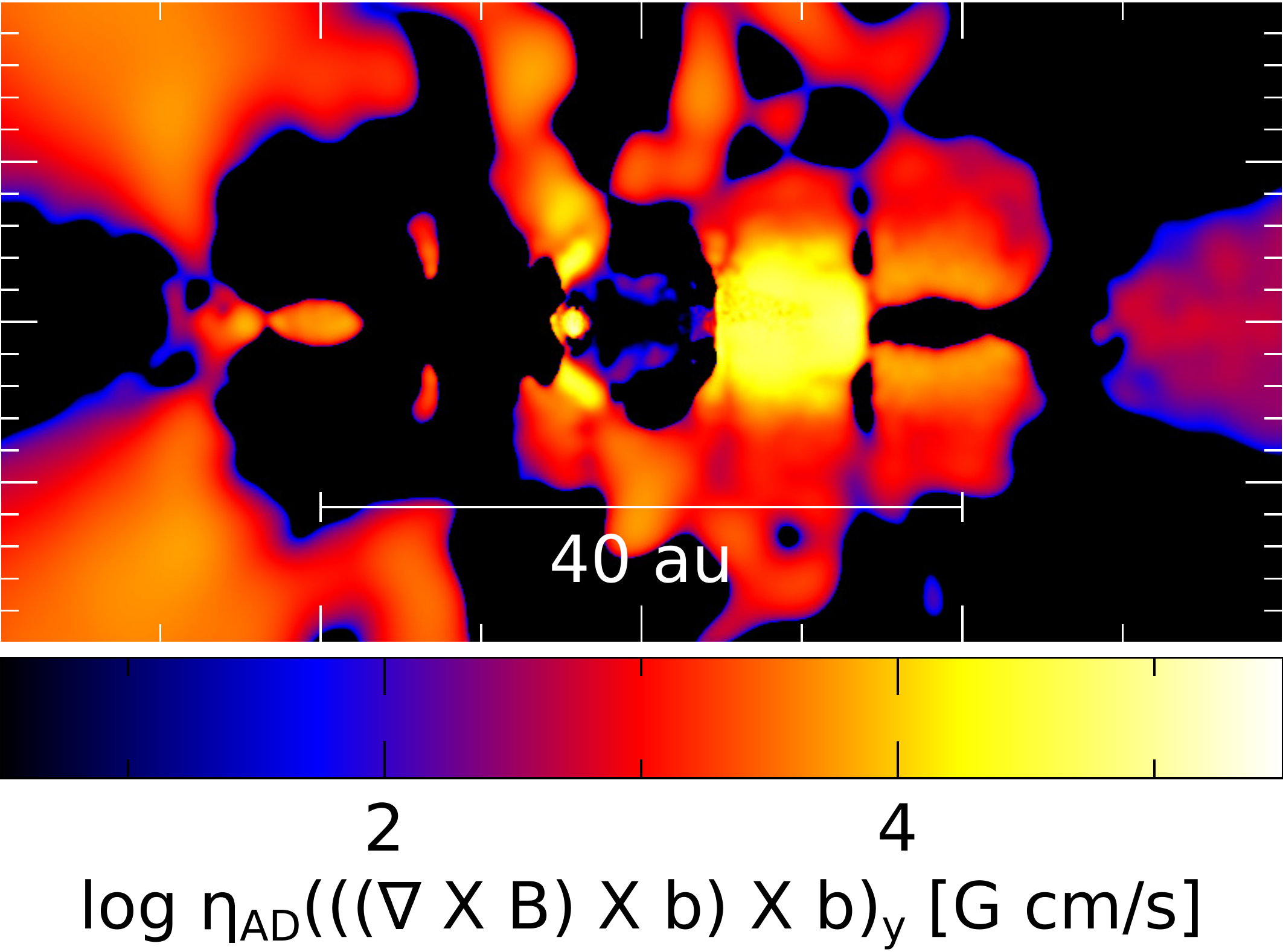}  %Made on DiAL
\includegraphics[width=0.24\textwidth]{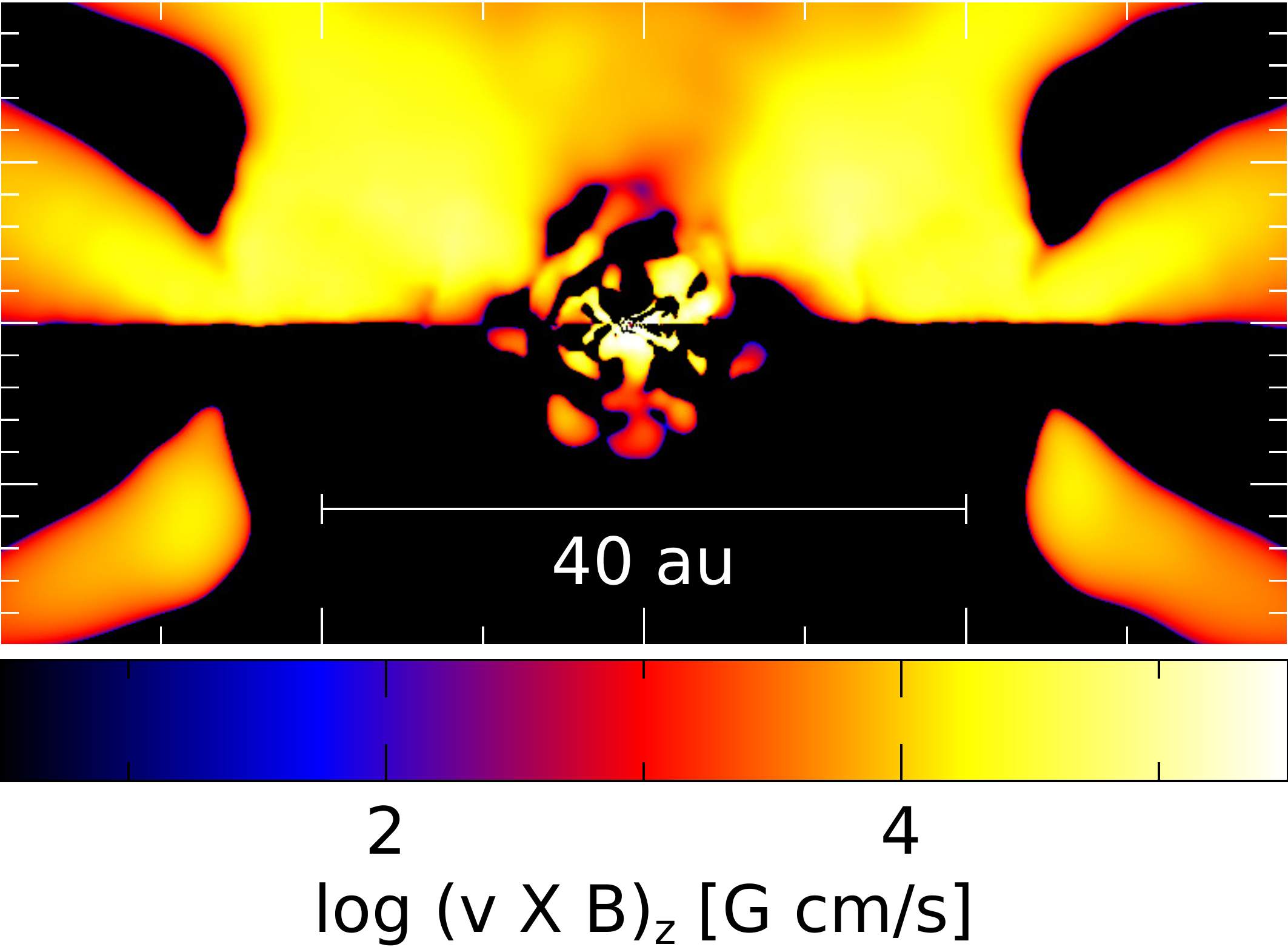}  %Made on DiAL
\includegraphics[width=0.24\textwidth]{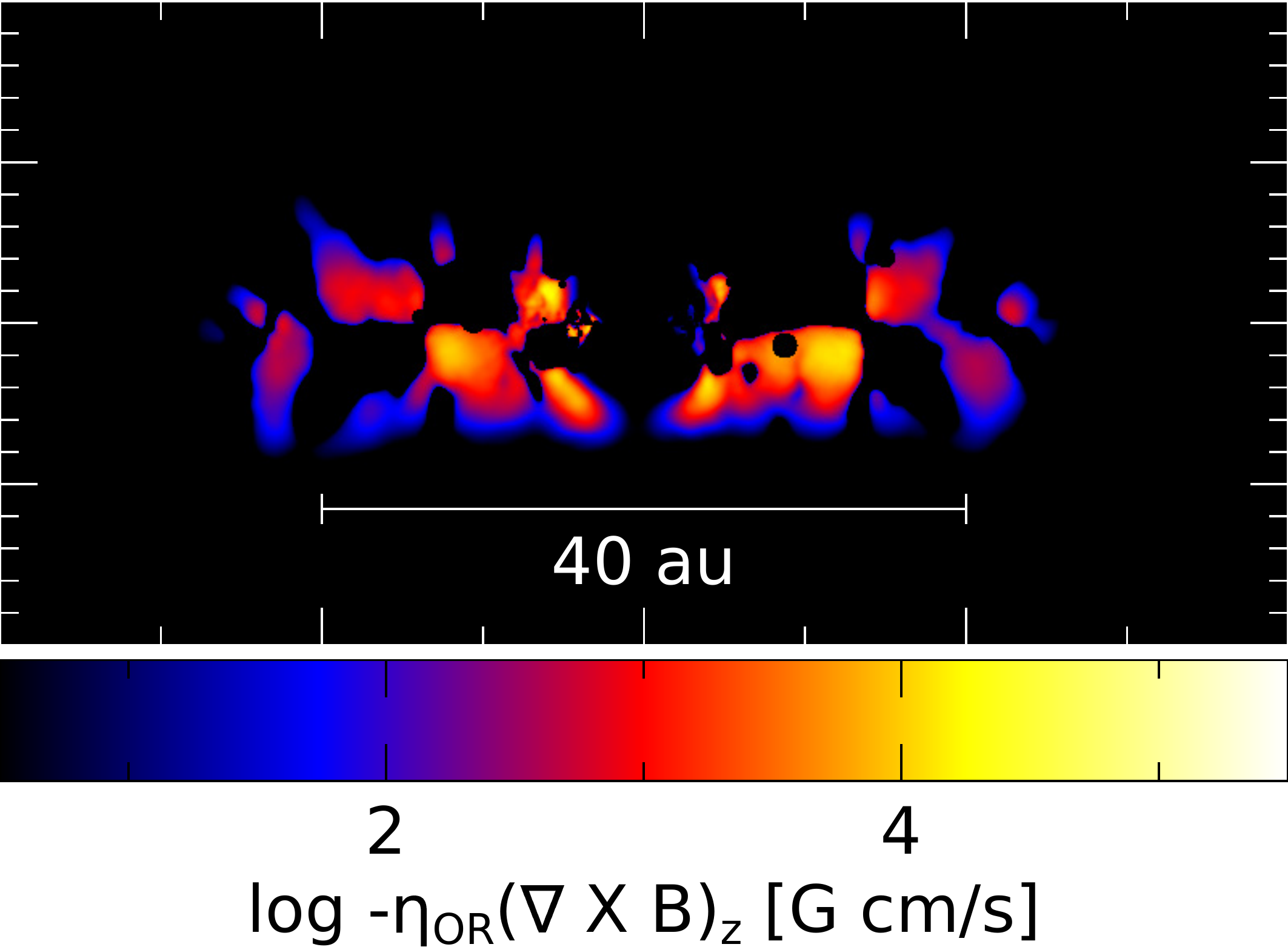}  %Made on DiAL
\includegraphics[width=0.24\textwidth]{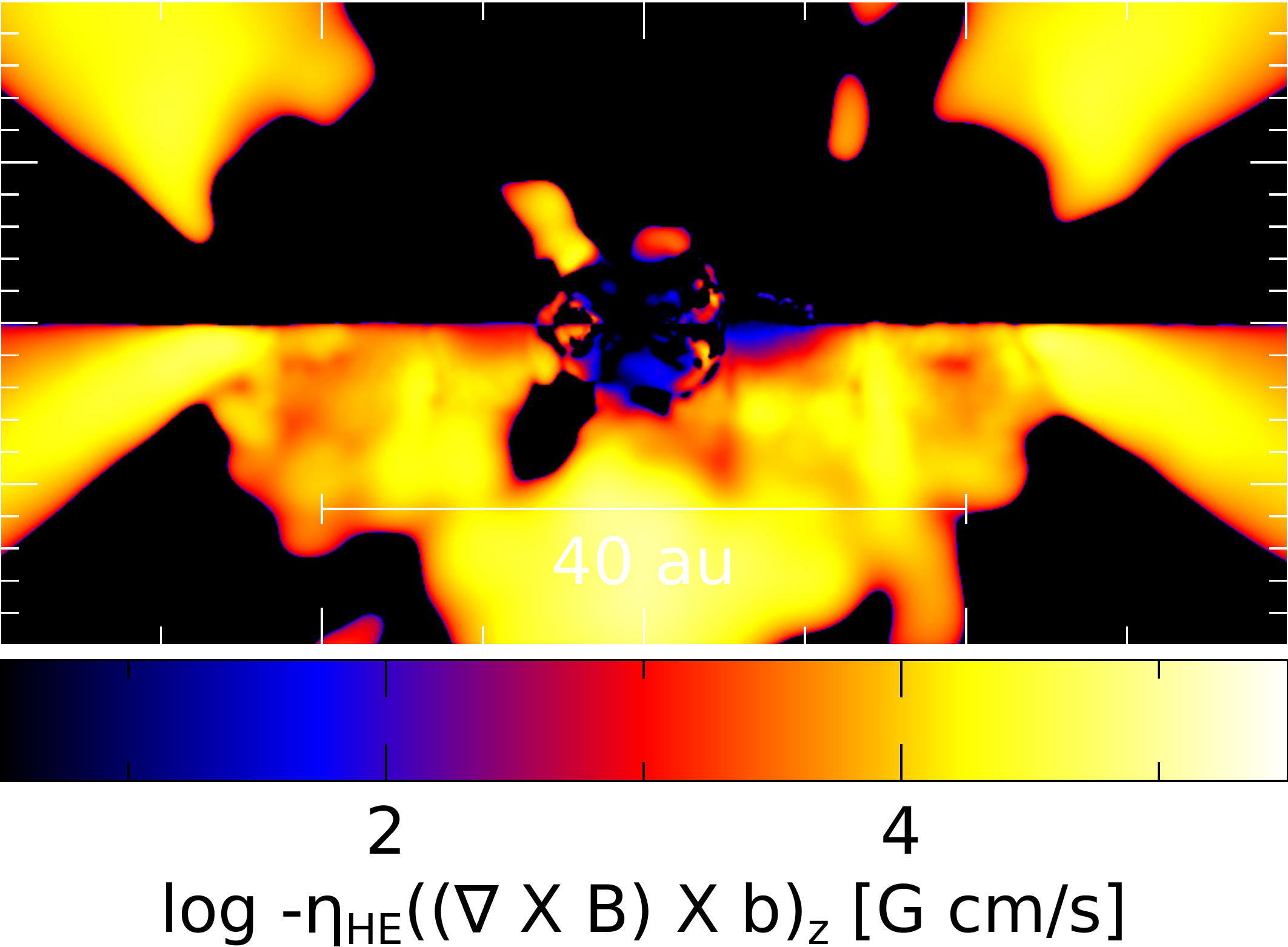}  %Made on DiAL
\includegraphics[width=0.24\textwidth]{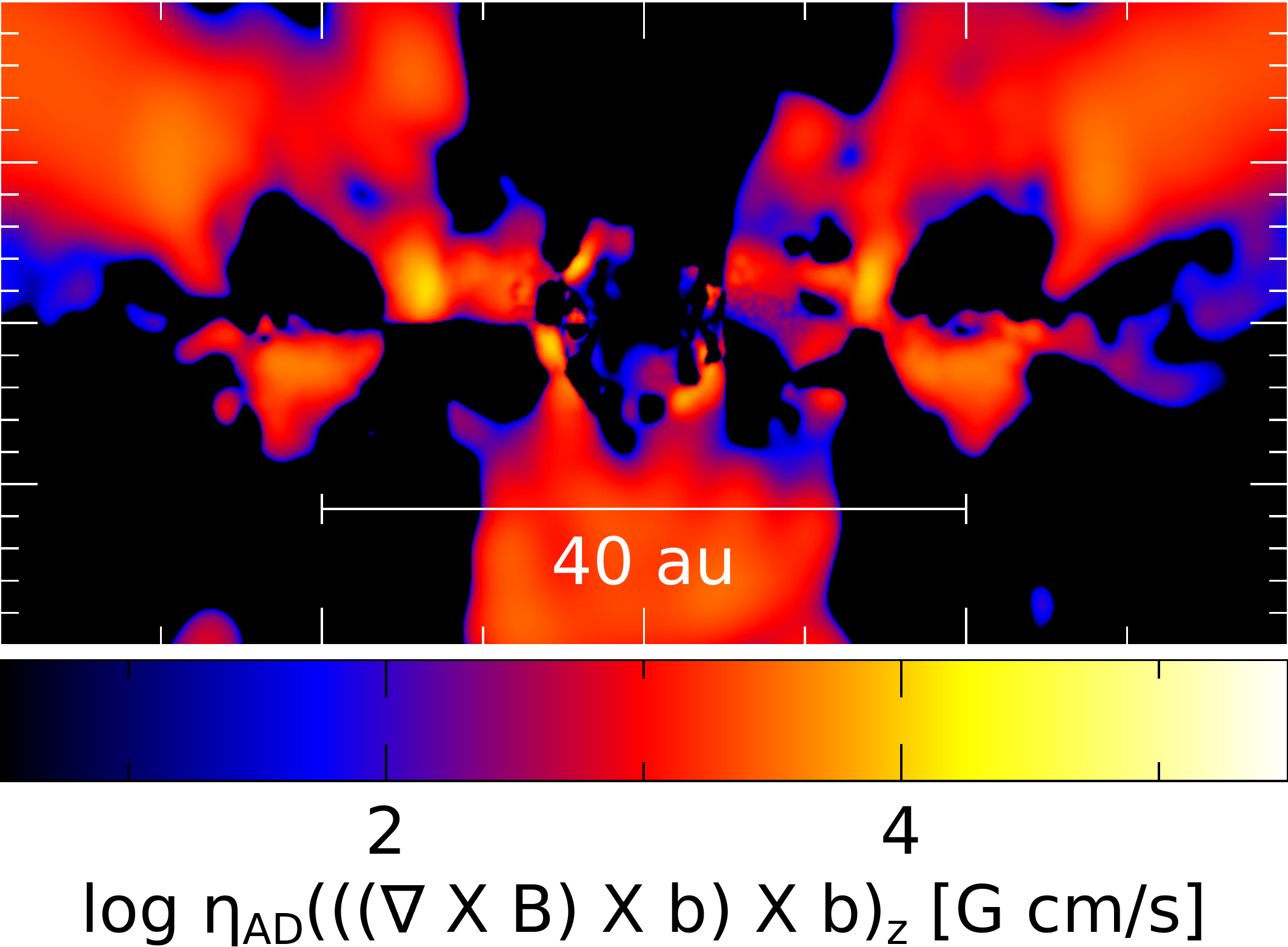}  %Made on DiAL
\caption{The vector contribution to the evolution of the magnetic field for the isolated disc in \citetalias{\wbp2018hd} is the curl of the quantities shown, $\bm{D}$.  $\bm{D}$ is approximately symmetric about 0, with very little of the frames containing $|\bm{D}| < 3.2$~G~cm ~s$^{-1}$, except for the regions surrounding the disc in the second column; therefore most of the black regions refer to $D < 0$ rather than $|\bm{D}| < 3.2$~G~cm ~s$^{-1}$.   At the current time, the ideal component has the largest contribution to the magnetic field followed by the Hall effect.  Each process will contribute to each component of the magnetic field, and the contribution is much more complex than suggested by the ratio plots in \figref{fig:isolated}.}
\label{fig:isolated:comp}
\end{figure*}

This disc formed during the collapse of the first hydrostatic core \citep{Larson1969} and is \sm60~yr old in the images; see \secref{sec:disc} for a discussion on disc ages and their long-term evolution.  %\footnote{Not a typo.}.
The magnetic field in the disc has a strong poloidal component resulting from the initial poloidal field.  However, a toroidal field has been generated in the disc, where its strength is \sm1-10 per cent of the strength of the poloidal component.  The sign of the toroidal field changes vertically across the disc \citepeg{Bai2014,Bai2015,Bai2017}, with $B_\phi > 0$ for $z > 0$ and $B_\phi < 0$ for $z < 0$.

A resolved protostar has formed at the centre of the disc, which heats up and ionises its immediate surroundings, but has not yet created a void.  This leads to a region with a high ionisation fraction and negligible values of the non-ideal coefficients (i.e. $|\eta| < 10^{15}$~\ueta{}).  This region only extends for $r \approx 4$~au, outside of which the disc behaves as a traditional protostellar disc, although it is less flared than typically expected due to both its young age and lack of central cavity.    The disc itself has a low ionisation fraction, which decreases towards the mid-plane.  This occurs organically, without the inclusion of cosmic ray attenuation.  All the coefficients are $|\eta| > 10^{15}$~\ueta{} in the disc ($r > 4$~au), therefore they should all have some effect on its evolution.  Moreover, $\Lambda \ll 1$ and $R_\text{m} \ll 1$ throughout the disc, indicating that the neutral fluid is decoupled from the magnetic field, and that magnetic diffusion is governing the evolution of the disc.  This is a noticeable difference from the idealised discs in \secref{sec:id}, where these dimensionless numbers were typically $\gg 1$ near the disc surface.

In general, Ohmic resistivity is relatively unimportant in this disc, where it is at least an order of magnitude weaker than ambipolar diffusion.   This best agrees with the MRN disc with $B_0 = 1$~G disc from \secref{sec:id}, although the relative importance of Ohmic resistivity is still likely too high in the idealised model, despite reasonable agreement of the magnetic field strengths.  This suggests that, although Ohmic resistivity will affect the evolution of a system, ambipolar diffusion also needs to be considered.  Therefore, not only is ideal MHD an incomplete picture of star formation as we have previously argued, so is resistive MHD where only Ohmic resistivity is considered.

In \citetalias{\wbp2018hd}, we showed that the reasonably sized disc presented here formed when the initial magnetic field and rotation vectors were anti-aligned, while only a small 5~au disc formed with they were aligned.  Thus, the Hall effect is clearly important for disc formation in isolate star formation simulations \citep[see also, e.g.][]{KrasnopolskyLiShang2011,BraidingWardle2012acc,Tsukamoto+2015hall,\wpb2016}.  This is reinforced here where the value of $|\eta_\text{HE}|$ is high, and indeed higher than $\eta_\text{AD}$ in the majority of the disc.  Contrary to expectations, ambipolar diffusion is dominant in a small region near the mid-plane, while Hall is dominant in most of the disc and surrounding environment!  As cautioned in \secref{sec:id}, however, the Hall coefficients are only a few times higher than the ambipolar diffusion coefficient in the disc, reinforcing that both processes are equally important.  

As discussed in \secref{sec:id:comp}, each non-ideal process has a different effect on the different components of the magnetic field.  For a complicated magnetic field geometry such as in a realistic disc, each process will contribute to each component.  In \figref{fig:isolated:comp}, we show the components of $\bm{D}_\text{ideal} \equiv \bm{v} \times \bm{B}$, $\bm{D}_\text{OR} \equiv-\eta_\text{OR}\left(\bm{\nabla}\times\bm{B}\right)$, $\bm{D}_\text{HE} \equiv-\eta_\text{HE} \left(\bm{\nabla}\times\bm{B}\right)\times\bm{\hat{B}}$ and $\bm{D}_\text{AD} \equiv \eta_\text{AD}\left[\left(\bm{\nabla}\times\bm{B}\right)\times\bm{\hat{B}}\right]\times\bm{\hat{B}}$; the curl of these quantities is added to d$\bm{B}/\text{d}t$. \footnote{The curl of these terms is not saved, but the terms presented will imply their component-wise effect.}

At the current time, the ideal component will have the largest contribution to the magnetic field in the displayed region, followed by the Hall effect then ambipolar diffusion then Ohmic resistivity.  This is the same order of contribution concluded from the ratio plots in \figref{fig:isolated}.  However, it is now clear that the ratio of contributions varies spatially and for each component after accounting for the vector.  Throughout the region, the sign of $\bm{D}$ is typically the same for both Ohmic resistivity and ambipolar diffusion, showing that these are complementary processes.  There is no correlation between the sign of $\bm{D}_\text{OR,AD}$ and $\bm{D}_\text{HE}$ or $\bm{D}_\text{ideal}$, indicating that each process has a different effect on the evolution of the magnetic field.

The ratio plots (bottom row of \figref{fig:isolated}) provide convincing evidence that the non-ideal MHD processes (especially the Hall effect and ambipolar diffusion cannot be ignored is disc simulations.  The vector contributions plot (\figref{fig:isolated:comp}) provide further evidence that none of these processes can be ignored since each processes has a different effect on the different components of the magnetic field.  Finally, given that $\Lambda \ll 1$ and $R_\text{m} \ll 1$, non-ideal MHD cannot be ignored in disc formation simulations.

From analysing this disc, we obtain a very different picture than the cartoon (\figref{fig:cartoon}) or the idealised discs (\secref{sec:id}).  This disc shows that Ohmic resistivity is unimportant compared to ambipolar diffusion ($\eta_\text{OR}/\eta_\text{AD} < 0.01$) and that the Hall effect and ambipolar diffusion are similarly important with the Hall effect being slightly stronger.  This reinforces that the typical disc parameterisation in \secref{sec:id} and throughout the literature is a clear oversimplification.  

\subsubsection{Local environment}
\begin{figure}
\centering
\includegraphics[width=0.9\columnwidth]{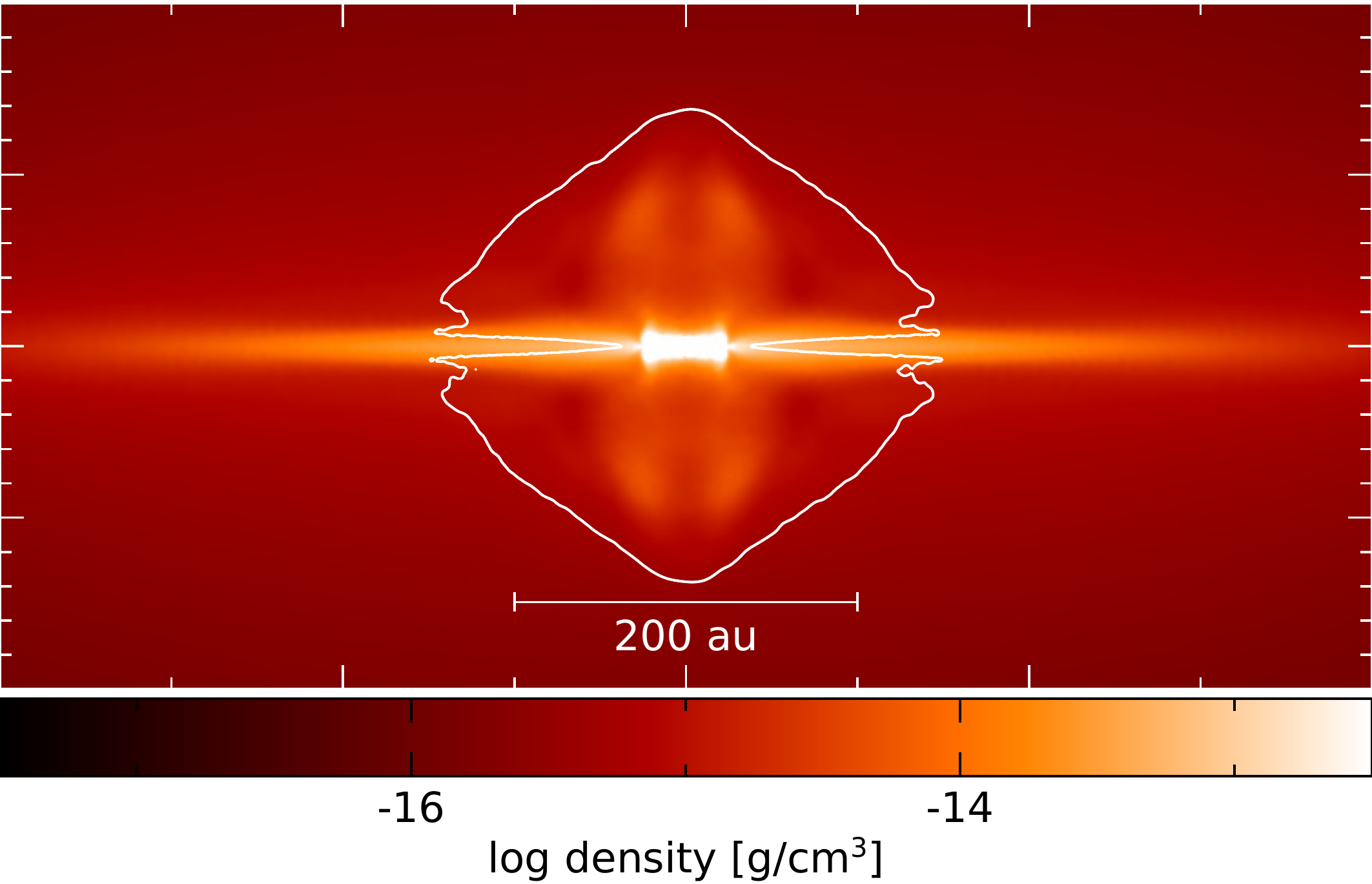}  %Made on DiAL
\includegraphics[width=0.9\columnwidth]{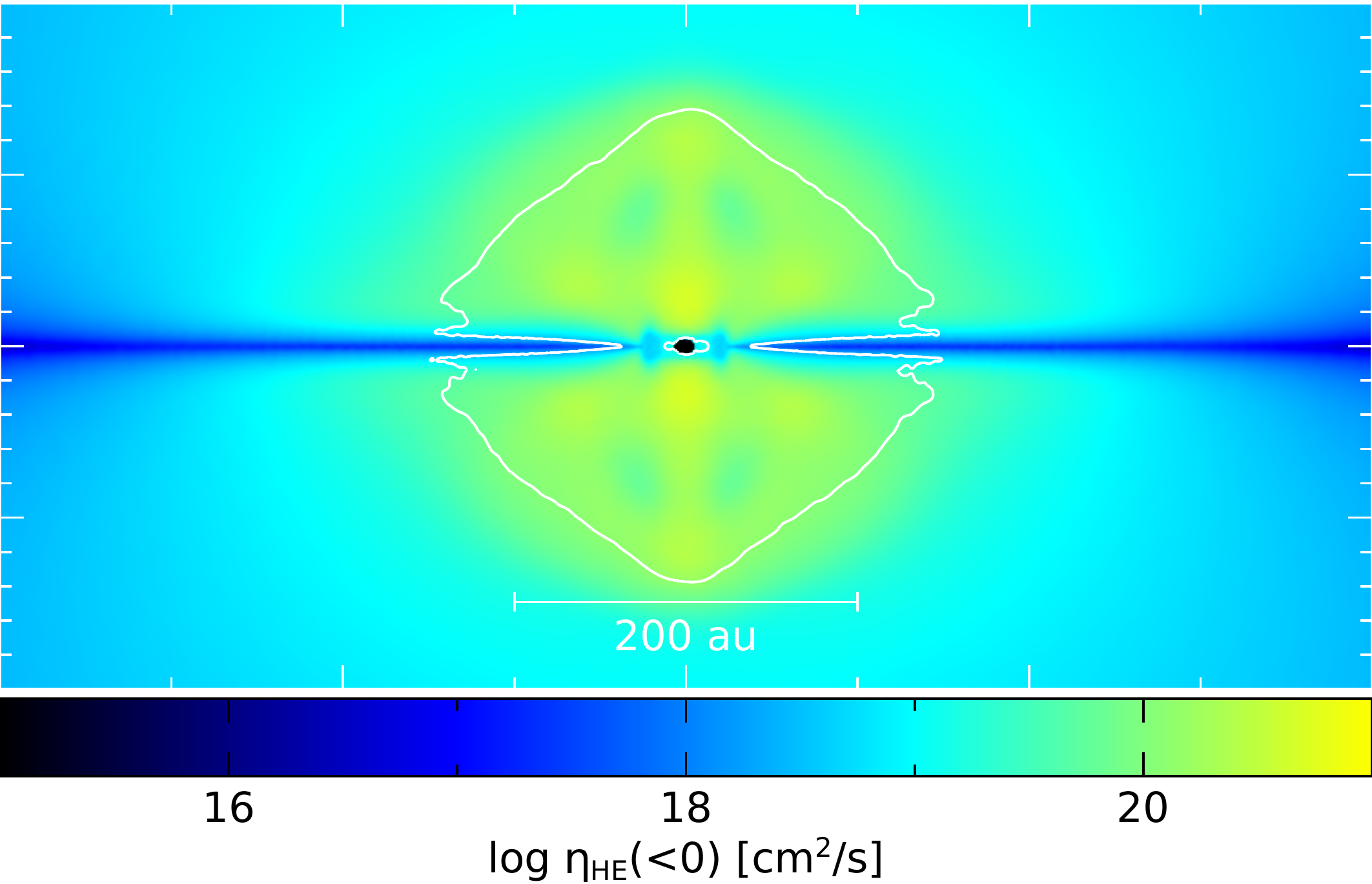}  %Made on DiAL
\includegraphics[width=0.9\columnwidth]{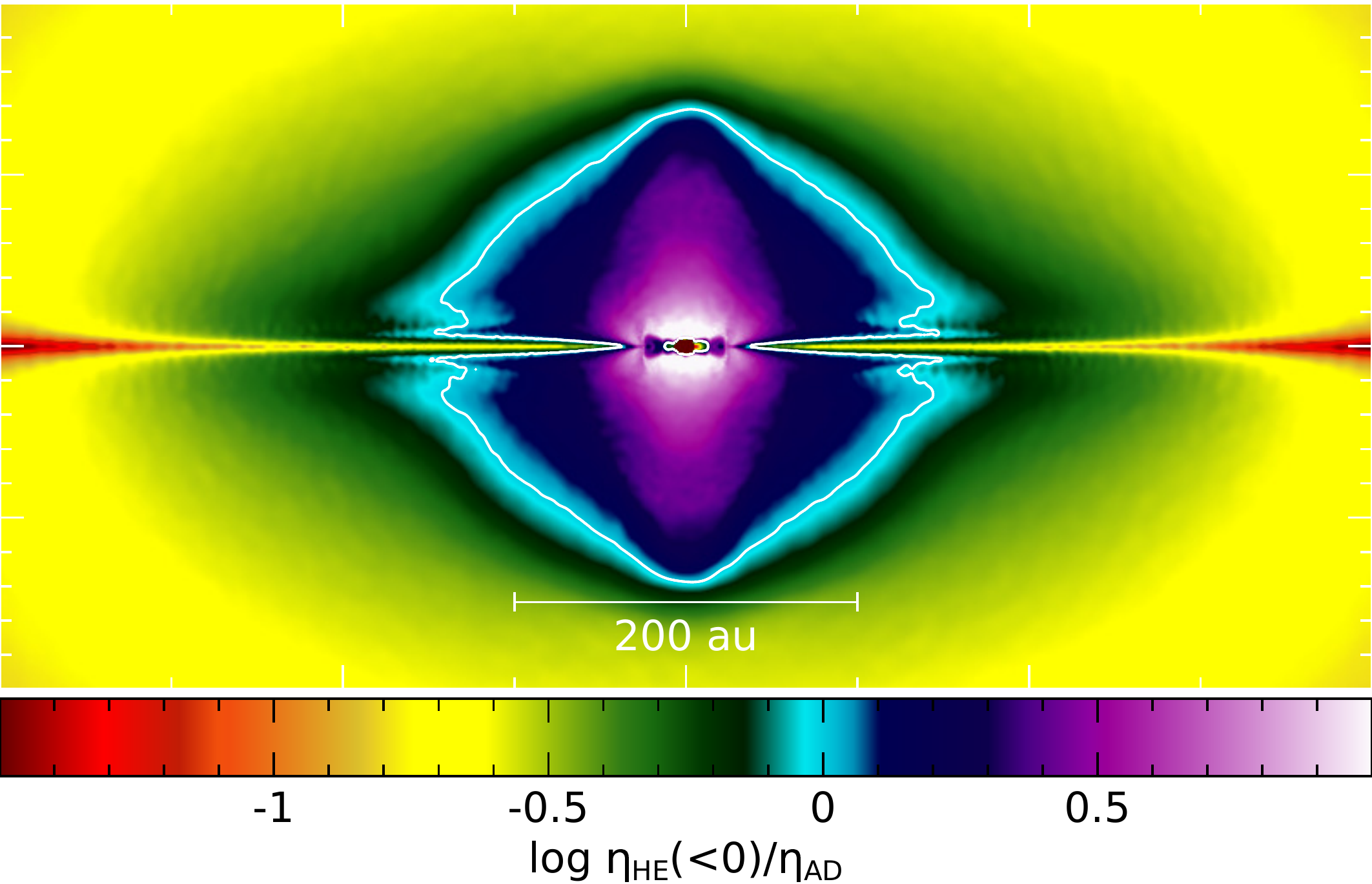}  %Made on DiAL
\caption{The gas density, value of the Hall effect and its relative value compared to ambipolar diffusion for the environment around the isolated disc from \citetalias{\wbp2018hd}.  The images are taken at the same time as those in \figref{fig:isolated}.  Each frame measures 800 x 400 au.  The contour in all three panels is $-\eta_\text{HE}/\eta_\text{AD} = 1$.  The Hall effect is the dominant term in the region surrounding the star and the disc; given its strong values, this shows that the Hall effect is an influential process in the star and disc formation processes.}
\label{fig:isolated:big}
\end{figure}

From the cartoon and idealised discs, it was expected that ambipolar diffusion would at least be the dominant term surrounding the disc, if not in its outer edges.  However, this is not observed in the isolated disc (final panel in \figref{fig:isolated}).  \figref{fig:isolated:big} shows the gas density, the value of $\eta_\text{HE} < 0$ and the ratio of $-\eta_\text{HE}/\eta_\text{AD}$ for the environment surrounding the disc.  At this time, the Hall coefficient in the surrounding environment has a similar value to its disc value and $|\eta_\text{HE}| > \eta_\text{AD}$ within a radius of $r \gtrsim 100$~au.  

It is generally accepted that ambipolar diffusion plays the dominant role in the early stages of star and disc formation \citepeg{Mouschovias1978,Mouschovias1991,MouschoviasCiolek1999,MellonLi2009,LiKrasnopolskyShang2011,Crutcher2012,Bai2017}.  Indeed, in the early stages of this simulation, $\eta_\text{AD} > |\eta_\text{HE}|$, but the ratio is less than 10.  As the collapsing gas enters the first hydrostatic core phase, $|\eta_\text{HE}| > \eta_\text{AD}$ in the pseudo-disc, and this region of $|\eta_\text{HE}| > \eta_\text{AD}$ expands in advance of the first core outflow.  

Thus, it is clear that the Hall effect can modify the star forming environment reasonably early in the star forming process, which accounts for its influence over whether a large or small disc forms.  In agreement with the angular momentum profiles in previous studies \citep{Tsukamoto+2015hall,\wbp2018hd}, this modification of the environment by the Hall effect occurs well before the protostar forms.  

The large values of $|\eta_\text{HE}|$ for such a large region around the forming star and its disc starting at such an important phase in the star formation process reinforces the importance of the Hall effect on star and disc formation.  

%----
\subsection{Structure of discs formed in a cluster environment}
\label{sec:rd:clus}
\begin{figure}
\centering
\includegraphics[width=0.49\textwidth]{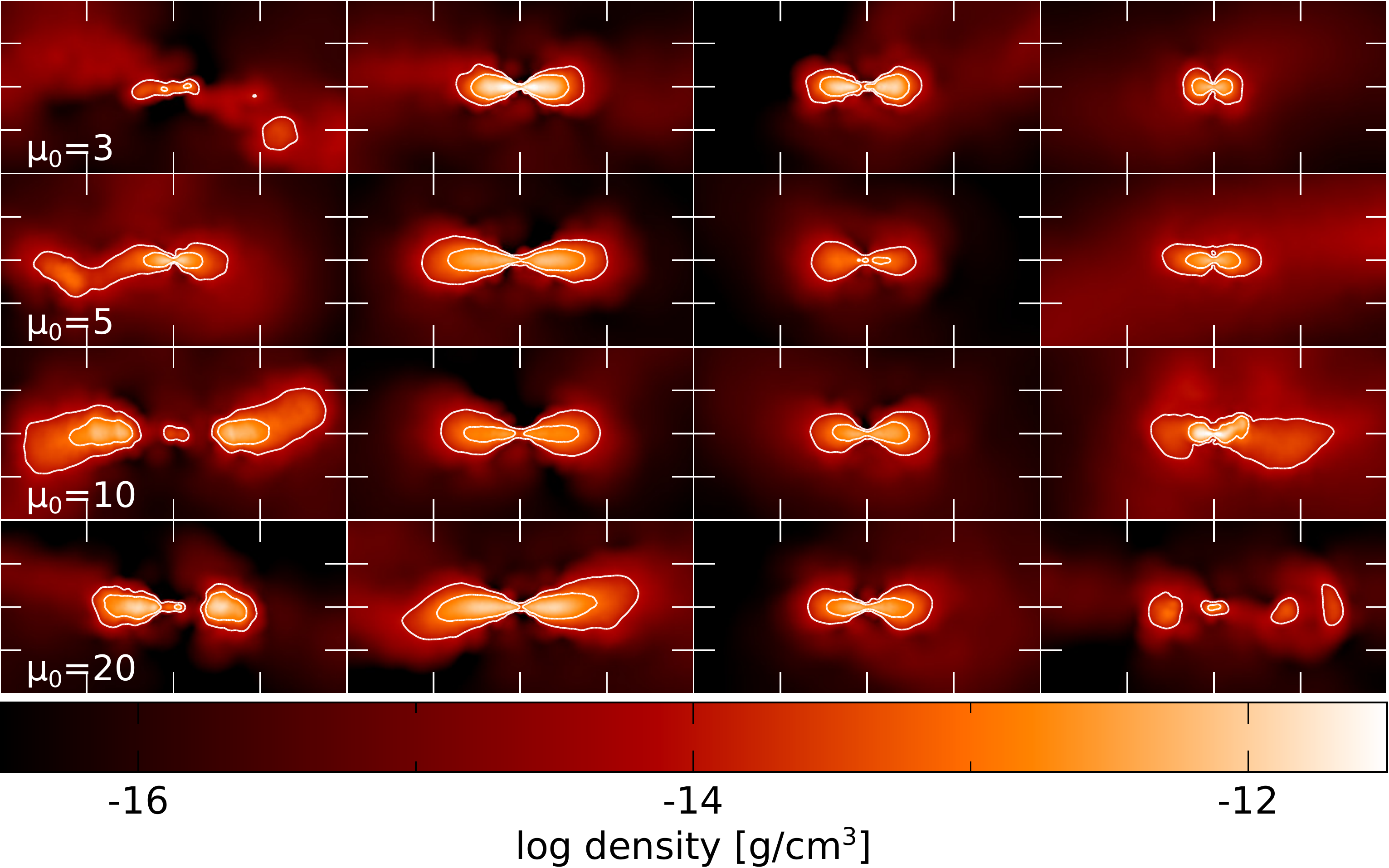}  %Made on DiAL
\includegraphics[width=0.49\textwidth]{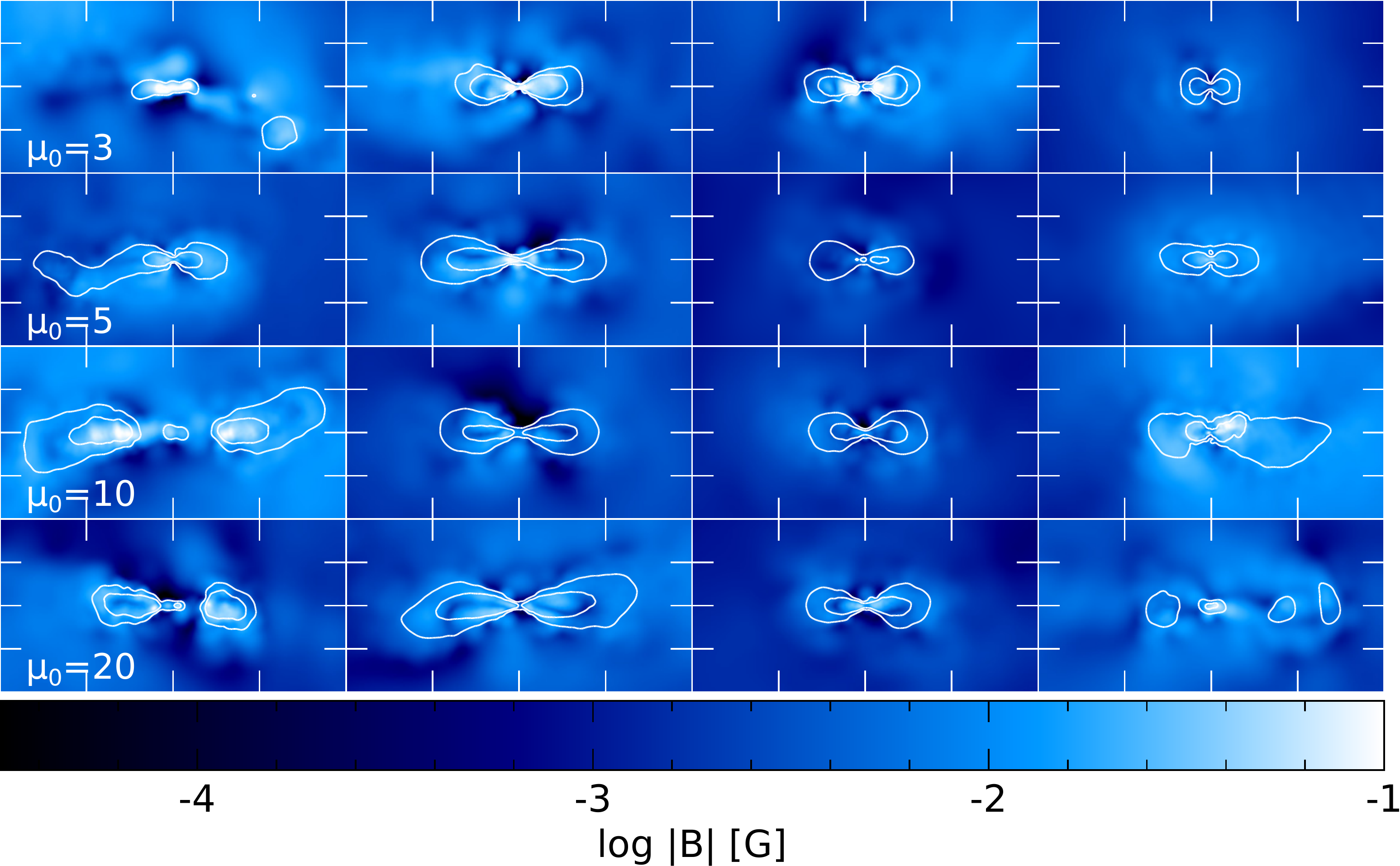}  %Made on DiAL
\includegraphics[width=0.49\textwidth]{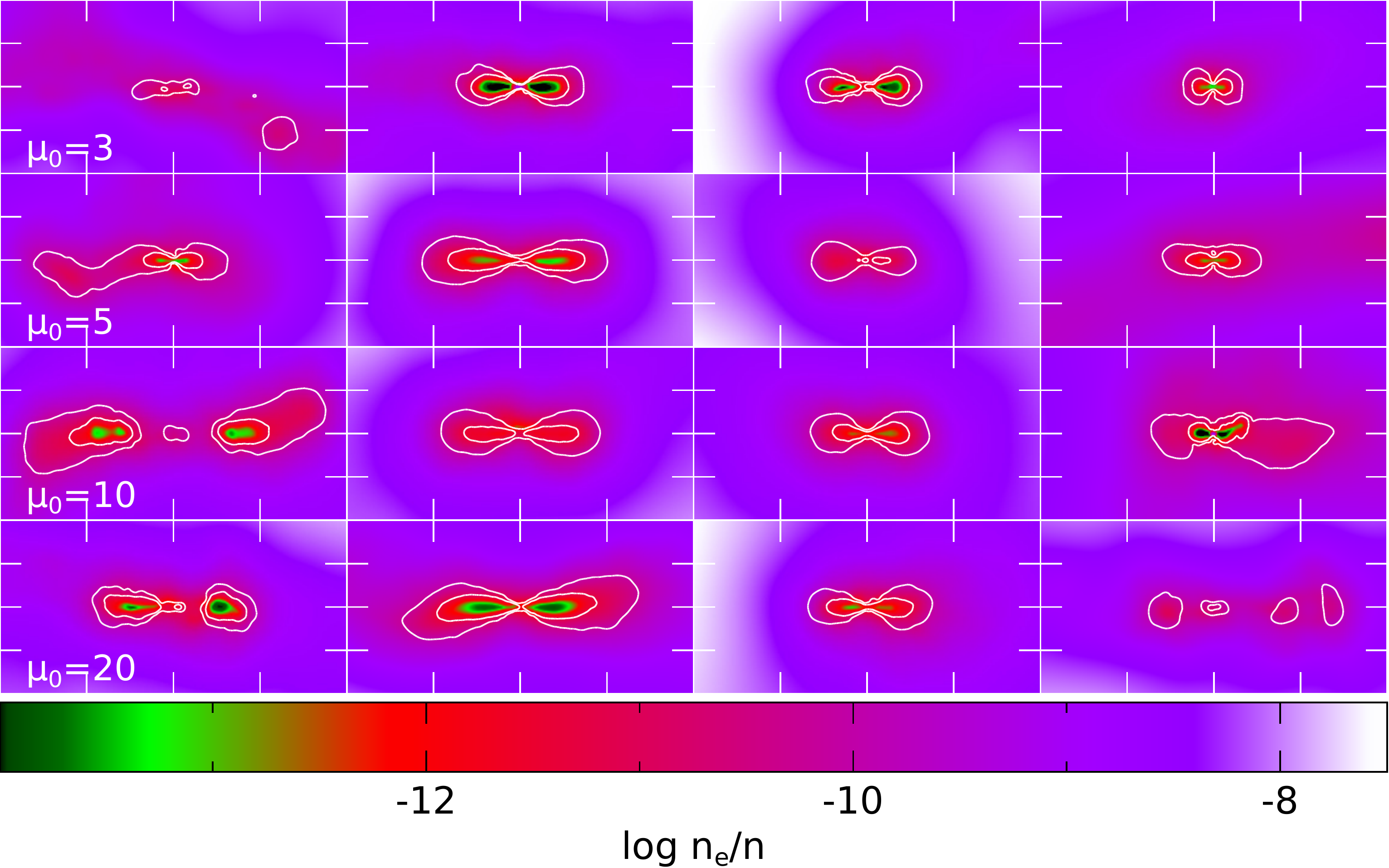}  %Made on DiAL
\caption{Density, magnetic field strength and ionisation fraction of selected discs that formed in the non-ideal MHD simulations from \citet[][herein \citetalias{\wbp2019}]{\wbp2019}.  The images are in a slice through the centre of the star in the $xz$-plane, where the disc has been rotated to lie in the $xy$-plane. Frame sizes are 400 x 200~au.  Each row is from a different simulation, where the initial mass-to-flux ratio in units of the critical value $\mu_0$ is shown in the first column.  In all frames, the contours are $\rho = 10^{-14}$ and $10^{-13}$~\gpercc.  The images are taken at $t = 1.45$~\tff{}.  The discs are larger and less well-defined than in \citetalias{\wbp2018hd} due to the dynamic motion of the gas, interactions from nearby stars, and discs and that several of these are circumsystem discs.}
\label{fig:cluster:rbn}
\end{figure}

The disc in \secref{sec:rd:iso} formed self-consistently, however, it did so in an idealised environment that was initialised with idealised initial conditions.  For less idealised initial conditions of disc forming regions, we next consider the discs that formed in our star cluster formation simulations \citepalias{\wbp2019}.  These disc formed out of a turbulent environment and dynamically interacted with stars and other discs, thus the discs are not as pristine or well-characterised as those we have previously discussed.  The host stars have ages ranging from $10^3$ - $10^5$~yr, but the discs are typically younger than their host star since they are frequently disrupted and replenished due to the high stellar density \citep[see also][]{Bate2018}; thus, these interactions typically prevent the discs from smoothly evolving from Class 0 to Class I.
We selected four discs from each non-ideal MHD simulation, and the gas density, magnetic field strength, ionisation fraction, the non-ideal MHD coefficients and ratios with respect to $\eta_\text{AD}$ are shown in \figsref{fig:cluster:rbn}{fig:cluster:eta}.  %The images are taken at 1.45~t$_\text{ff}$; for reference the gas column density of all the discs is shown in fig.~B1 of \citetalias{\wbp2019}.
\begin{figure*}
\centering
\includegraphics[width=0.49\textwidth]{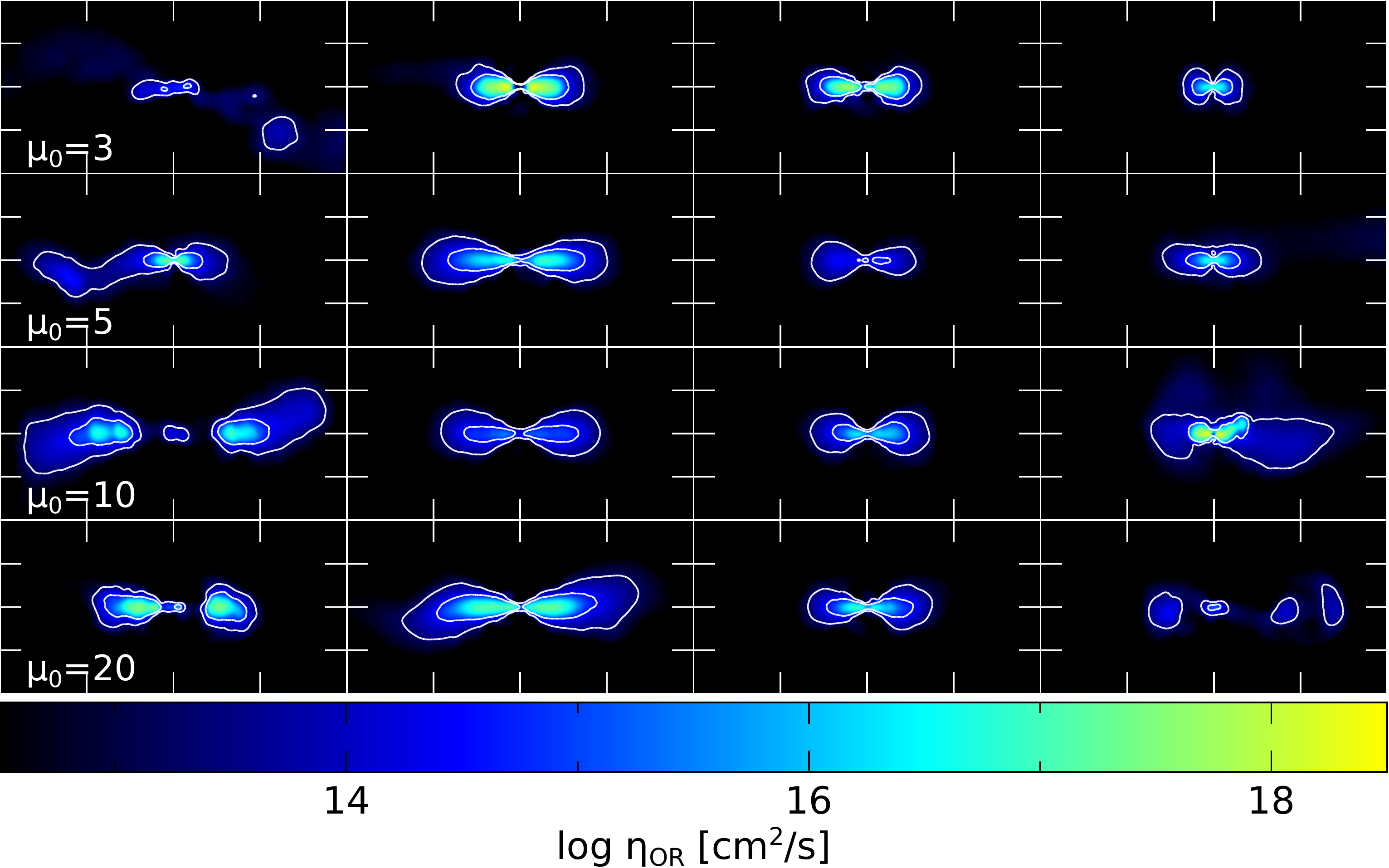}  %Made on DiAL
\includegraphics[width=0.49\textwidth]{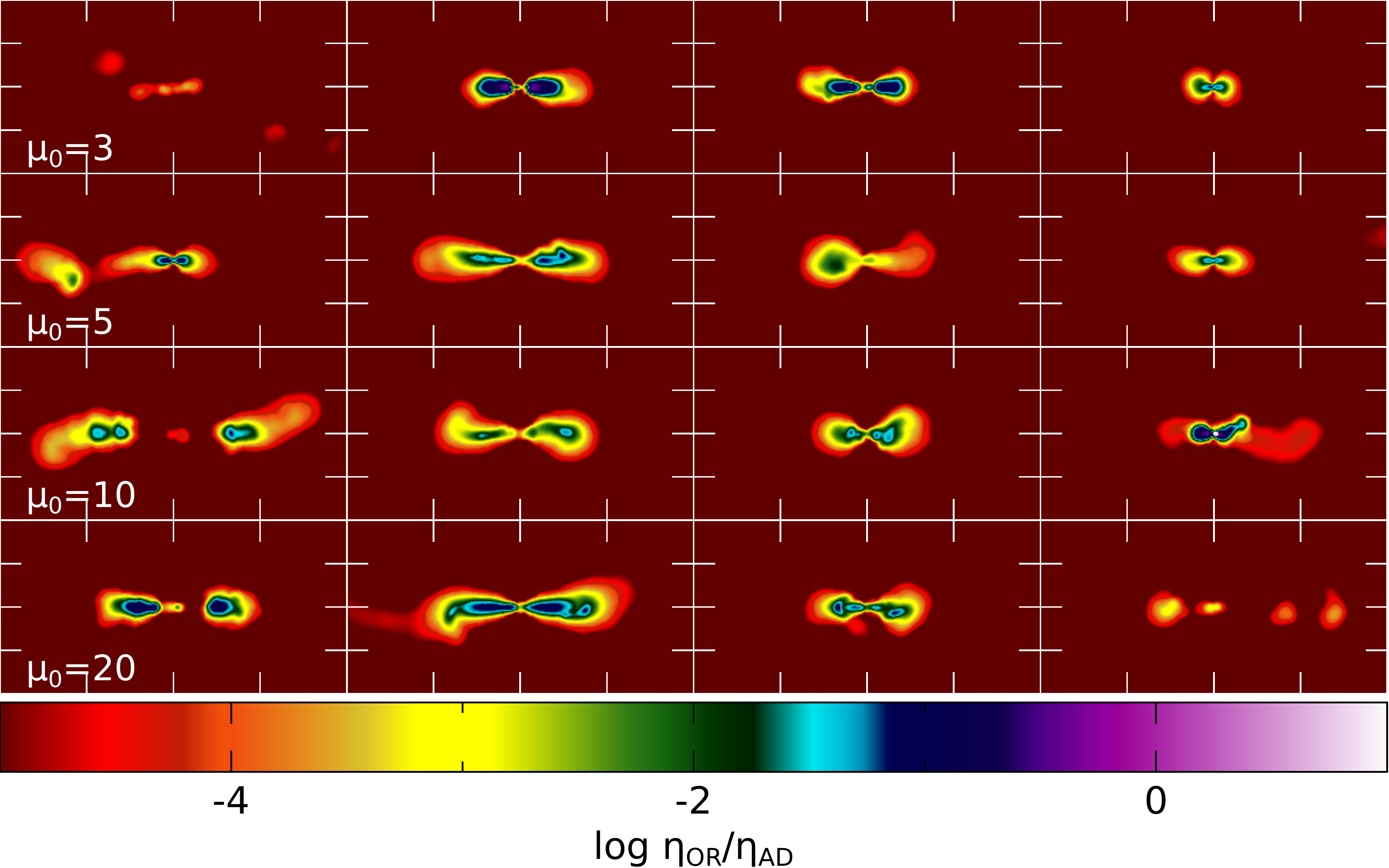}  %Made on DiAL
\includegraphics[width=0.49\textwidth]{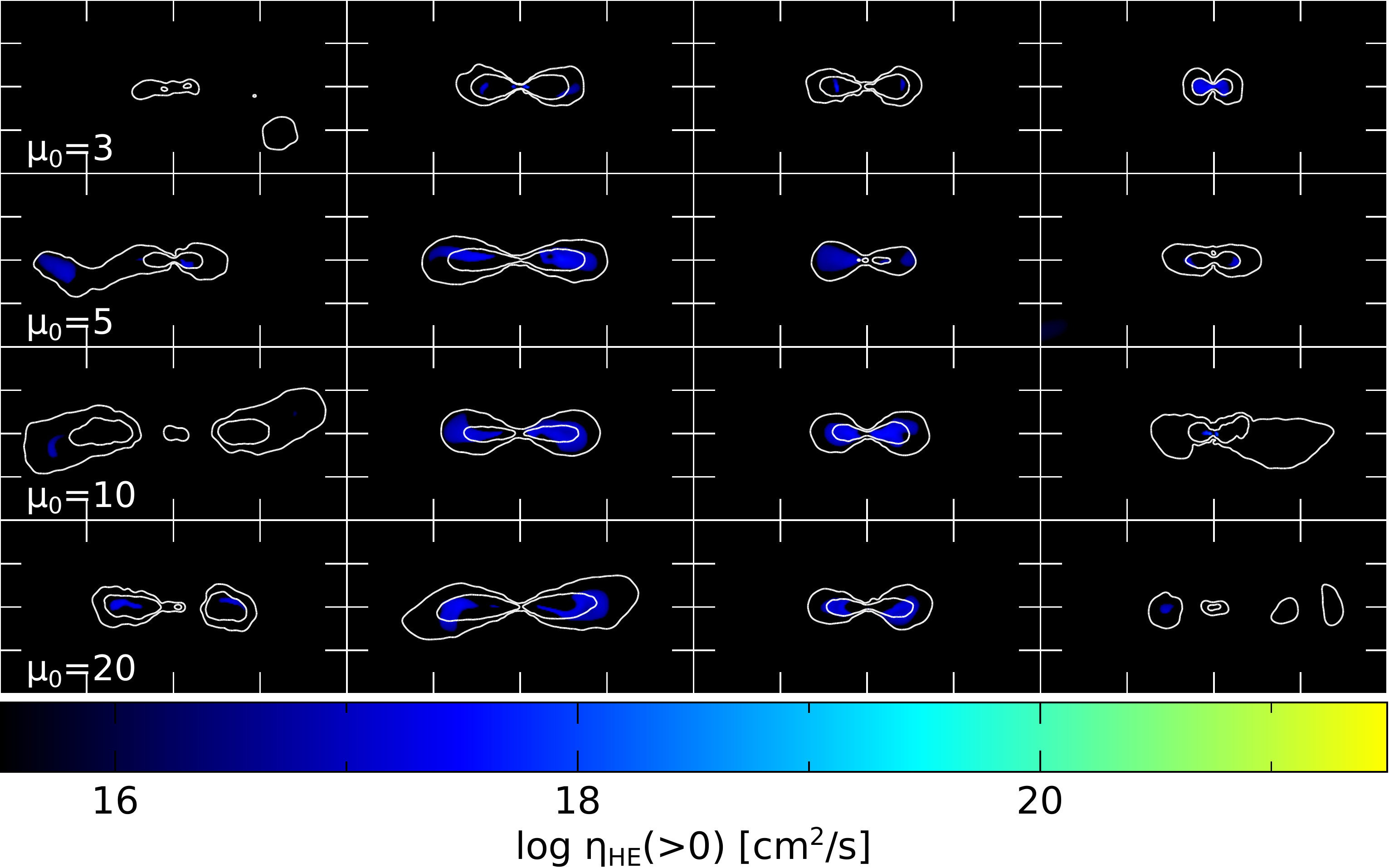}  %Made on DiAL
\includegraphics[width=0.49\textwidth]{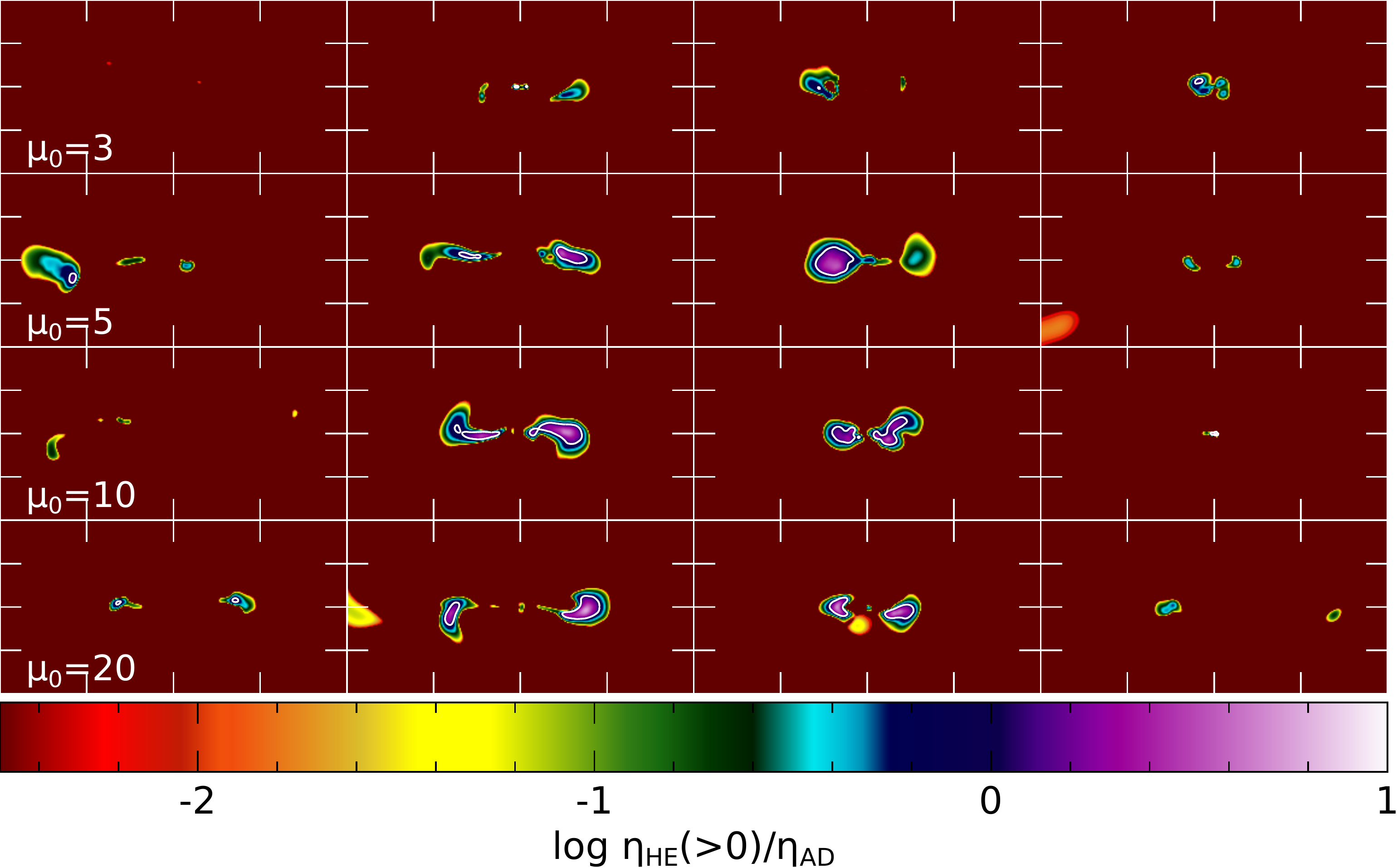}  %Made on DiAL
\includegraphics[width=0.49\textwidth]{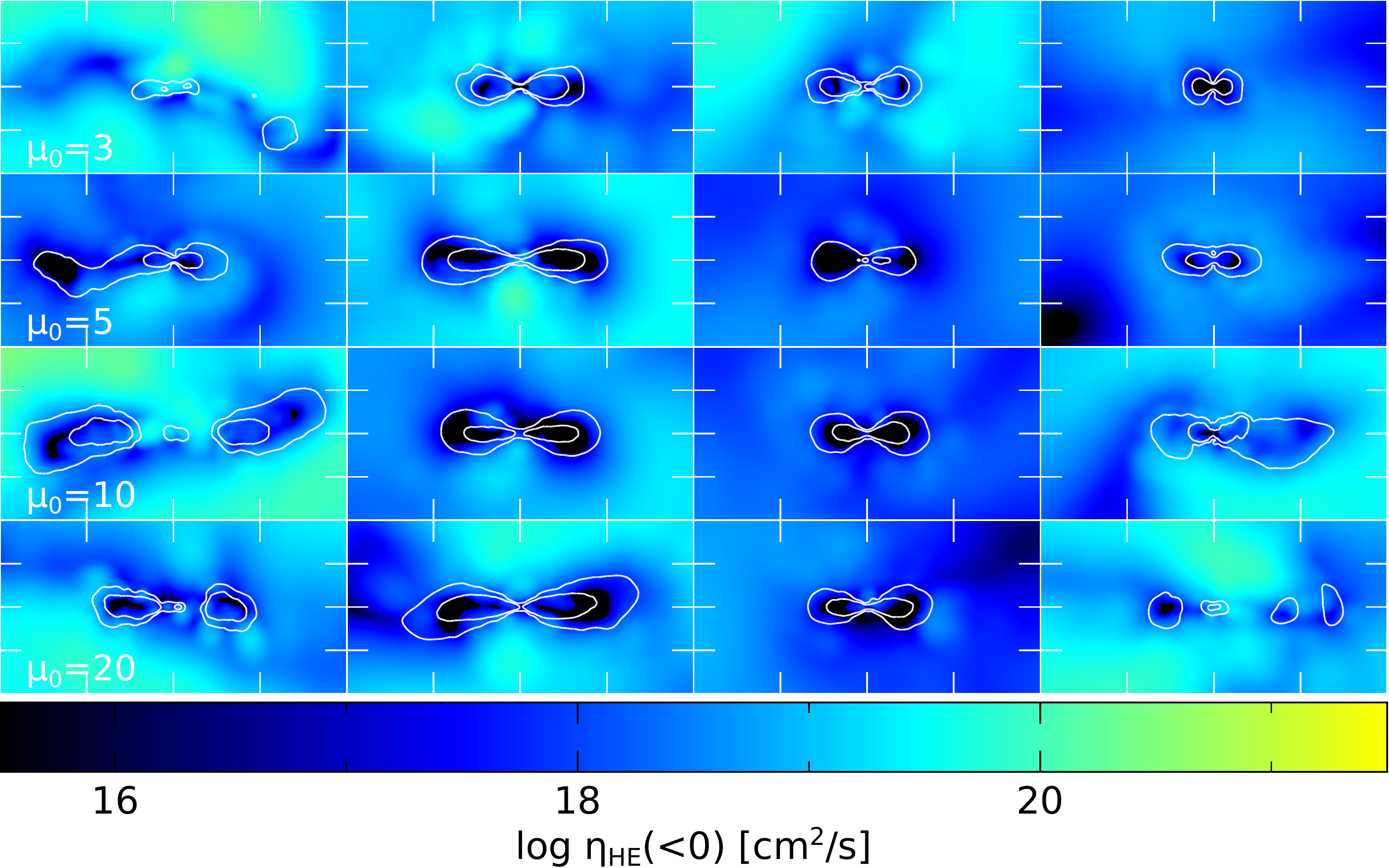}  %Made on DiAL
\includegraphics[width=0.49\textwidth]{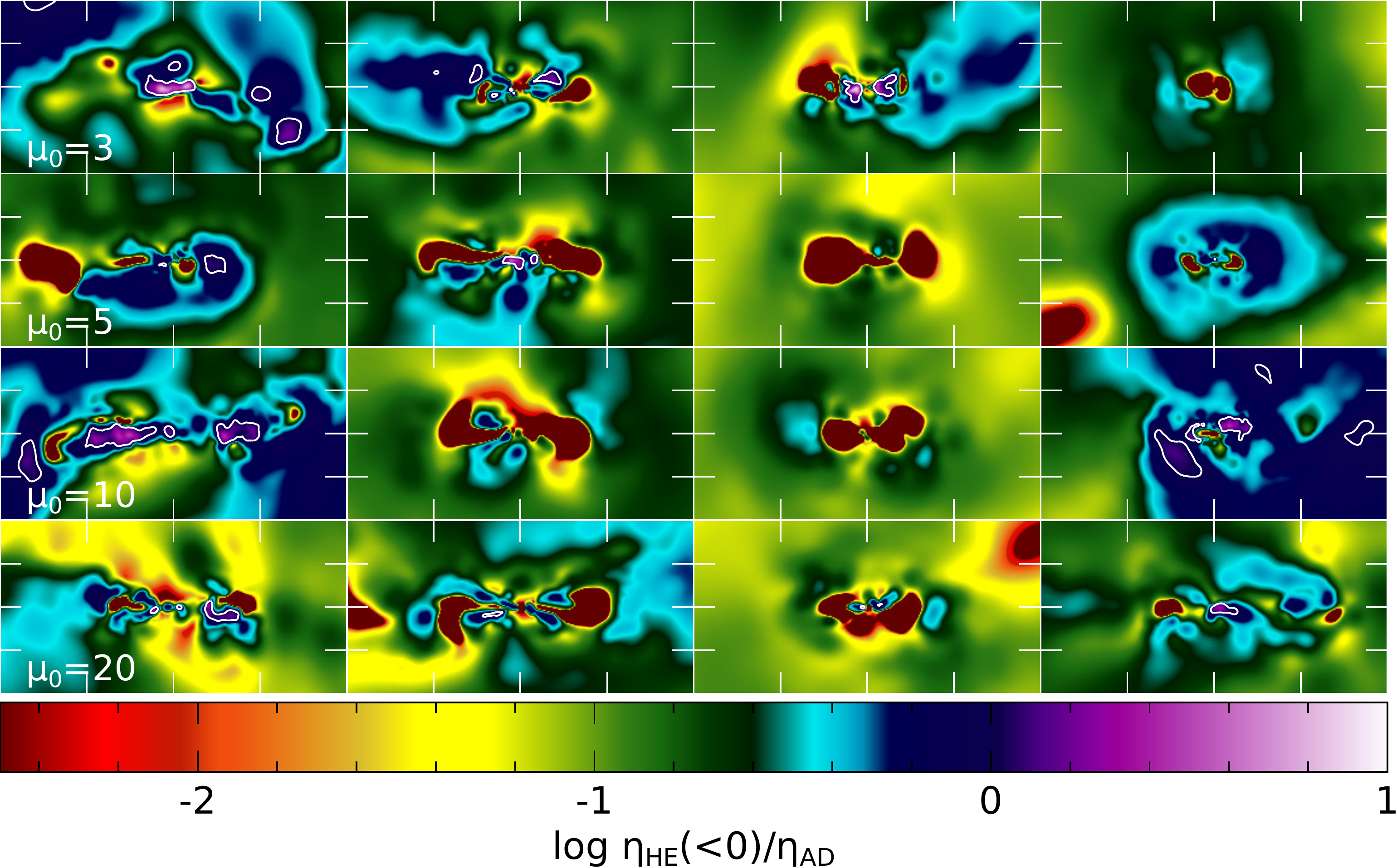}  %Made on DiAL
\includegraphics[width=0.49\textwidth]{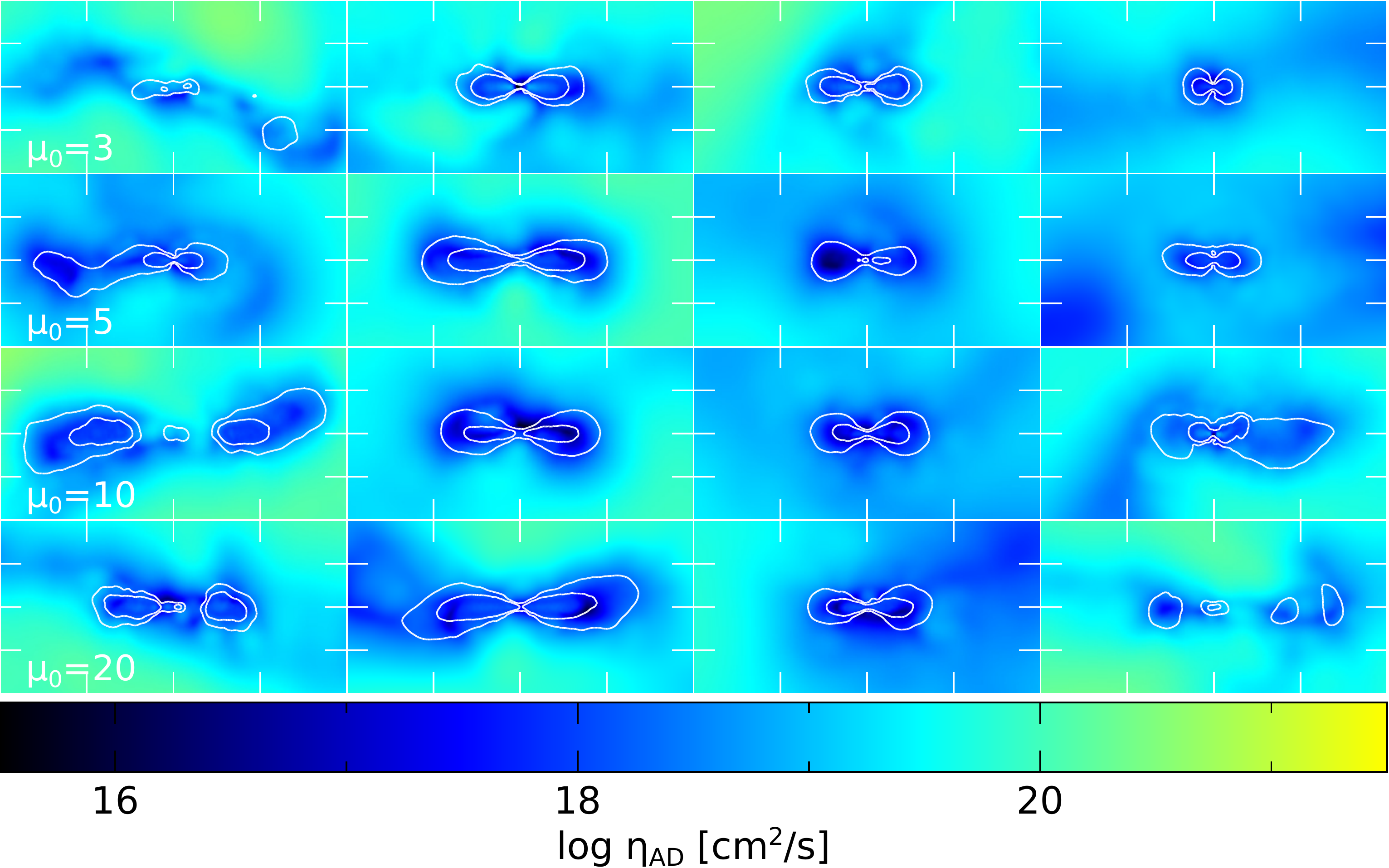}  %Made on DiAL
\caption{The non-ideal MHD coefficients and the ratios with respect to the ambipolar diffusion coefficient for the discs in \figref{fig:cluster:rbn}.  In the ratio plots, the contour is at 1; in the coefficient plots, the contours are $\rho = 10^{-14}$ and $10^{-13}$~\gpercc.  The pale `lines' in the top right ratio plot are log $\eta_\text{OR}/\eta_\text{AD} \approx -1.5$ and not the contour line at  $\eta_\text{OR}/\eta_\text{AD} =1$.  As in \citetalias{\wbp2018hd}, the Hall coefficient is the dominant term in the disc, although in these discs, the Hall coefficient is typically positive, which is a result of the weaker magnetic field in the discs.  Ambipolar diffusion is typically the dominant term in the environment surrounding the disc.} \label{fig:cluster:eta}
\end{figure*}

The discs in \citetalias{\wbp2019} generally have larger radii than the isolated disc; many of these discs surround multiple stars, which naturally creates a large cavity within the disc and extends the radius.  Most of these discs have the traditional flared density profile, although a few have warped outer edges since they are being influenced by an external source (e.g. first column, third row) or are a small protostellar disc surrounded by either a larger circumsystem disc (e.g. fourth column, fourth row) or have tidal tails (e.g. first column, first row).  The magnetic field strengths of these discs are generally weaker than the isolated disc by almost two orders of magnitude.  Thus, we expect the non-ideal effects to be less important in these discs compared to the isolated disc in \secref{sec:rd:iso}.

As expected, the discs are weakly ionised, however, their ionisation fraction is higher than in the isolated disc; moreover, the surrounding gas is also more ionised.  Thus, there are considerable environmental differences between these discs and the isolated disc, resulting primarily from the dynamical environment.  

In the discs, $\eta_\text{OR} \gtrsim 10^{15}$~\ueta{} suggesting that this effect does influence the evolution of the disc.  Similar to the isolated disc, $\eta_\text{OR} < \eta_\text{AD}$, and there is only a small region near the mid-plane where $\eta_\text{OR}/ \eta_\text{AD} > 0.1$. Thus, although Ohmic resistivity will have some effect on the magnetic field evolution, it is a much weaker contribution than ambipolar diffusion, providing further evidence that modelling Ohmic resistivity alone is also an incomplete picture of star formation.

As traditionally expected, ambipolar diffusion is stronger in the surrounding environment and gets weaker towards the disc mid-plane.  In the surrounding environment, ambipolar diffusion is the dominant term, typically by more than an order of magnitude over the other two effects; this differs from the isolated disc which was surrounded by a large region of Hall-dominated gas (recall \figref{fig:isolated:big}).  Unlike the isolate disc, there is not a region near the mid-plane of increased $\eta_\text{AD}$.  Thus, in the cluster simulations, ambipolar diffusion is more influential in the surrounding medium than the disc.

This better matches the idealised discs in \secref{sec:id} where ambipolar diffusion is important only at the edge of the disc; had we also plotted the surrounding environment in those figures, then we would see that ambipolar diffusion is dominant in surrounding environment.  The differences in the surrounding environments between these discs and that in \secref{sec:rd:iso} show how other processes shape the environment and ultimately contribute to the non-ideal processes and determining where each process is dominant.  Since the dominant term in the environment is different between \citetalias{\wbp2018hd} and  \citetalias{\wbp2019}, this further suggests that ambipolar diffusion and the Hall effect are equally important and that neither can be ignored.

The coefficients of the Hall effect differs most from the isolated disc.  In the cluster simulations, the surrounding medium and outer disc have $\eta_\text{HE} < 0$ while the discs themselves have $\eta_\text{HE} > 0$.  This is consistent with the weaker disc magnetic fields and better matches the idealised discs in \secref{sec:id}, although we did not recover $\eta_\text{HE} < 0$ even in the outer disc in the idealised models (except for the single grain model with $a_\text{g} = 0.03$~\mum{}).

In summary, for the well-defined discs, we have the general picture of ambipolar diffusion dominating in the surrounding environment and the outer disc; this is where $\eta_\text{HE} < 0$.  For the remainder of the disc, $\eta_\text{HE} > 0$ is the dominant term.  Ohmic resistivity has reasonably high coefficients in the disc mid-plane but is never the dominant term.  In the disc where $\eta_\text{HE}  > \eta_\text{AD}$, the two values are still within a factor of 10, suggesting that both terms are still important in determining the evolution of a disc.

%----------------------------------------------------------------------------------------------------------------
\section{Discussion}
\label{sec:disc}

This paper has shown the structure of the disc at a very young age (\secref{sec:rd:iso}) and at indeterminate ages (Sections~\ref{sec:id} and \ref{sec:rd:clus}).  Our previous numerical work \citepeg{\wpb2016,\wbp2018hd,WursterBate2019,WursterLewis2020d} has focused on the formation and early evolution of a disc, and has shown the necessity of non-ideal MHD (and specifically the Hall effect) to overcome magnetic braking \citepeg{AllenLiShu2003,Galli+2006} and permit a protostellar disc to form; the sizes of the discs around single stars is in agreement with analytical approximation given by \citet{Hennebelle+2016}.  From these papers \citep[see also][]{Tsukamoto+2015hall,Tsukamoto+2017,Zhao+2018,Vaytet+2018}, it is clear that non-ideal MHD is required to shape the initial conditions of the discs.  Since the ultimate evolution is determined (at least in part) by the initial conditions, then it follows that the evolution of the discs is dependent on non-ideal MHD.  

The Class 0 phase of protostellar discs is $2\times10^4 \lesssim  t/\text{yr} \lesssim 2\times10^5$ \citepeg{Evans+2009,Enoch+2009,Maury+2011,MachidaHosokawa2013,Dunham+2014,KristensenDunham2018}.  
%\citet{KristensenDunham2018} calculates a half-life of 74Myr.
This lifespan is much longer than the \sm60~yr isolated discs \secref{sec:rd:iso}, but similar to the age of the host stars in the cluster environment \secref{sec:rd:clus}; however, the discs in the cluster environment evolve through interactions and typically do not evolve smoothly from Class 0 to Class II discs.    As isolated discs evolve, they undergo many changes that will affect the structure of the disc and the (relative) importance of the non-ideal MHD processes.

As discs evolve, dust grains settle and coagulate \citepeg{WilliamsCieza2011,Krapp+2018,RiolsLesur2018,Riols+2020,RiolsLesurMenard2020}.  This removes the larger grains from the upper disc, which will either increase or decrease the importance of ambipolar diffusion, depending on the remaining disc properties \citepeg{ZhaoCaselliLi2018,Tsukamoto+2020}.  Larger grains in the disc mid-plane will decrease the importance of non-ideal MHD in that region.

While the dust settles, the discs slowly decrease in mass, either through photo-evaporation \citepeg{JohnstoneHollenbachBally1998,AlexanderClarkePringle2006,GortiHollenbach2009,Concharamirez+2019}, magnetocentrifugal winds \citepeg{BlandfordPayne1982,Anderson+2003} or magneto-thermal winds \citepeg{Bai+2016,Bai2017}; although the discs continue to accrete material from the envelope \citep[possibly even driven by the magnetocentrifugal winds; e.g.][]{Bai2014,Bai2015}, the mass loss rate exceeds the accretion rate.  As the inner edge of the disc is photoevaporated, the material from the outer regions replenishes this region; at the same time, the winds remove gas from the surface \citep[for a review, see][]{WilliamsCieza2011}.  This leads to a continual redistribution of the gas, which yields smaller disc heights, leading one to expect a reduced impact of the non-ideal effects, particularly Ohmic resistivity. 

The envelope may be replenished via the disc winds \citepeg{Bai2017} or by accreting gas from larger scales if the discs are not isolated \citepeg{WursterBatePrice2019}, or may be depleted by outflows \citepeg{MachidaHosokawa2013}.  If the envelope is replenished, then the nature of the replenishing material will dictate the importance of the non-ideal processes on the evolution.  If dense gas from the surroundings replenishes the envelope, then the behaviour of the gas in the envelope will be similar to the envelope at earlier ages.  If winds replenish the envelope, then the envelope will be  attenuated; cosmic rays, UV rays and X-rays will be more easily able to ionise it and the upper layers of the disc, leading to a greater importance of ambipolar diffusion \citepeg{Bai2017}.  If the envelope is depleted, then there will be less ambient gas to attenuate the cosmic rays, and the ionisation rate (at least at the surface of the disc) will be higher, resulting in higher ionisation rates and weaker non-ideal MHD processes.  If the disc is thinner due to mass loss, then the cosmic rays may penetrate to the mid-plane and reduce or remove the dead zone.

The disc structure and its surrounding envelope is complex and evolves through time.  As it does, the importance of the non-ideal processes will change, but will remain relevant as long as there is a reasonable amount of gas and a reasonable ionisation fraction.   Further studies are required to determine the precise impact of non-ideal MHD as the disc dissipates.

%----------------------------------------------------------------------------------------------------------------
\section{Summary and Conclusion}
\label{sec:conc}

Non-ideal MHD effects -- namely Ohmic resistivity, ambipolar diffusion and the Hall effect -- are important in strongly magnetised, weakly ionised regions that are populated by dust grains, such as protostellar discs.  It is traditionally assumed that the relative importance of these terms can be understood by the cartoon in \figref{fig:cartoon}, where Ohmic resistivity is dominant near the mid-plane, surrounded by a region that is dominated by the Hall effect, and then the remainder of the disc is dominated by ambipolar diffusion.

In this paper, we created idealised discs to determine where the various non-ideal effects were dominant in an effort to quantitatively recreate \figref{fig:cartoon}.  The disc density and temperature profiles were fixed while we varied the magnetic field strength and the dust grain distribution.  Using our free parameters, we concluded that the traditional view in the cartoon was challenging to reproduce.  We found that in many cases, the Hall effect was the dominant term in the majority of the disc, with only a small region near the edge of the disc being dominated by ambipolar diffusion.  Given our moderate magnetic field strength, the non-ideal effects became weaker as we increased the grain size when modelling a single grain population or when we removed the small grains by using an MRN grain distribution.  As expected, ambipolar diffusion and the Hall effect were stronger for stronger magnetic fields, and all three coefficients  increased in strength towards the mid-plane, especially when we included attenuation of the cosmic ray ionisation.

For most of our idealised discs, the Hall effect with $\eta_\text{HE} > 0$ was the dominant term throughout most of the disc.  However, throughout most of each disc, the coefficients were within a factor of 10 of each other, indicating that all terms were important throughout the disc.  Therefore, none of the three non-ideal processes can be ignored when modelling protostellar discs, and creating such a simple cartoon is a great oversimplification of the non-ideal processes in a disc.  
When accounting for magnetic field geometry, it becomes even more clear that no processes can be ignored.

We also investigated the self-consistently formed protostellar discs from two of our previous studies \citep{\wbp2018hd,\wbp2019}. All of these discs were generally dominated by the Hall effect.  In the simulation that formed a single, high resolution disc \citep{\wbp2018hd}, the Hall coefficient was negative in the disc, and was also the dominant term in the environment surrounding the disc; it became the dominant term in the surrounding environment at the beginning of the first hydrostatic core phase.  In the simulations that formed multiple discs \citep{\wbp2019}, the Hall effect was positive in the disc and negative in the surroundings, but ambipolar diffusion was the dominant term in the surroundings.

In all of our self-consistently formed discs, the values of the Hall effect and ambipolar diffusion were typically within a factor of 10 of one another, suggesting that both are equally important during the formation and evolution of a protostellar disc.  The coefficient for Ohmic resistivity is large enough that it affects the evolution of the system, therefore resistive MHD (i.e. only including Ohmic resistivity) is more realistic than ideal MHD.  However, its strength is generally much lower than ambipolar diffusion, indicating that ambipolar diffusion cannot be ignored in disc formation simulations.  Therefore, in addition to our previous claims that ideal MHD is an incomplete picture of star formation, these results further conclude that resistive MHD is also an incomplete picture.  

In all of our examples (both idealised and self-consistently formed), we have shown that the traditional cartoon is a great oversimplification, since this structure is difficult to reproduce and since the ratio of two non-ideal coefficients is typically less than 10 throughout the disc.  Thus, in agreement with many previous studies in the literature, we conclude that when modelling magnetised discs, non-ideal MHD (including the Hall effect) cannot be ignored.

%----------------------------------------------------------------------------------------------------------------
\section*{Acknowledgements}
We would like to thank the anonymous referee for useful comments that greatly improved the quality of this manuscript, and who made useful suggestions regarding the improvement of \textsc{Nicil}.
We would like to thank Daniel J. Price for hosting \emph{Great Barriers in Planet Formation} (2019), the conference which sparked the idea for this study, and for inviting me to speak at the \emph{3rd Phantom + MCFOST Users Workshop} (2020), which further progressed this study.
We would like to thank Rebecca Nealon for helpful discussions regarding the properties of protostellar discs.  
The data for \citetalias{\wbp2018hd} was first published in \citet{\wbp2018hd} and generated on the DiRAC Complexity machine, jointly funded by STFC and the Large Facilities Capital Fund of BIS (STFC grants ST/K000373/1, ST/K0003259/1, and ST/M006948/1), and the University of Exeter Supercomputer, Isca, a DiRAC Facility jointly funded by STFC, the Large Facilities Capital Fund of BIS, and the University of Exeter.
The data for \citetalias{\wbp2019} was first published in \citet{\wbp2019} and generated on the University of Exeter Supercomputer, Isca, and on the DiRAC Data Intensive service at Leicester, operated by the University of Leicester IT Services, which forms part of the STFC DiRAC HPC Facility (www.dirac.ac.uk). The equipment was funded by BEIS capital funding via STFC capital grants ST/K000373/1 and ST/R002363/1 and STFC DiRAC Operations grant ST/R001014/1. DiRAC is part of the National e-Infrastructure. 
The column density and cross-section figures were made using \textsc{splash} \citep{Price2007}.

%----------------------------------------------------------------------------------------------------------------
\section*{Data availability}
The data underlying \secref{sec:id} can be promptly generated using \textsc{Nicil} v2.1 as summarised in Appendix~\ref{app:nicil:discs}; the code can be downloaded at www.bitbucket.org/jameswurster/nicil, and is subject to the GNU license agreement.
The data underlying \citet[][aka \citetalias{\wbp2018hd}]{\wbp2018hd} are openly available from the University of Exeter's institutional repository at https://doi.org/10.24378/exe.607 \citep{WursterBatePrice2018hddata}.
The data underlying \citet[][aka \citetalias{\wbp2019}]{\wbp2019} is available upon reasonable request.

%--------------------------------------------------------------------------------
\bibliography{DiscsNimhd.bib}

\begin{thebibliography}{}
\makeatletter
\relax
\def\mn@urlcharsother{\let\do\@makeother \do\$\do\&\do\#\do\^\do\_\do\%\do\~}
\def\mn@doi{\begingroup\mn@urlcharsother \@ifnextchar [ {\mn@doi@}
  {\mn@doi@[]}}
\def\mn@doi@[#1]#2{\def\@tempa{#1}\ifx\@tempa\@empty \href
  {http://dx.doi.org/#2} {doi:#2}\else \href {http://dx.doi.org/#2} {#1}\fi
  \endgroup}
\def\mn@eprint#1#2{\mn@eprint@#1:#2::\@nil}
\def\mn@eprint@arXiv#1{\href {http://arxiv.org/abs/#1} {{\tt arXiv:#1}}}
\def\mn@eprint@dblp#1{\href {http://dblp.uni-trier.de/rec/bibtex/#1.xml}
  {dblp:#1}}
\def\mn@eprint@#1:#2:#3:#4\@nil{\def\@tempa {#1}\def\@tempb {#2}\def\@tempc
  {#3}\ifx \@tempc \@empty \let \@tempc \@tempb \let \@tempb \@tempa \fi \ifx
  \@tempb \@empty \def\@tempb {arXiv}\fi \@ifundefined
  {mn@eprint@\@tempb}{\@tempb:\@tempc}{\expandafter \expandafter \csname
  mn@eprint@\@tempb\endcsname \expandafter{\@tempc}}}

\bibitem[\protect\citeauthoryear{{Alexander}, {Clarke}  \&
  {Pringle}}{{Alexander} et~al.}{2006}]{AlexanderClarkePringle2006}
{Alexander} R.~D.,  {Clarke} C.~J.,   {Pringle} J.~E.,  2006, \mn@doi [\mnras]
  {10.1111/j.1365-2966.2006.10293.x}, \href
  {https://ui.adsabs.harvard.edu/abs/2006MNRAS.369..216A} {369, 216}

\bibitem[\protect\citeauthoryear{{Allen}, {Li}  \& {Shu}}{{Allen}
  et~al.}{2003}]{AllenLiShu2003}
{Allen} A.,  {Li} Z.-Y.,   {Shu} F.~H.,  2003, \mn@doi [\apj] {10.1086/379243},
  \href {http://adsabs.harvard.edu/abs/2003ApJ...599..363A} {599, 363}

\bibitem[\protect\citeauthoryear{{Andersen} et~al.,}{{Andersen}
  et~al.}{2019}]{Andersen+2019}
{Andersen} B.~C.,  et~al., 2019, \mn@doi [\apj] {10.3847/1538-4357/ab05c7},
  \href {https://ui.adsabs.harvard.edu/abs/2019ApJ...873...54A} {873, 54}

\bibitem[\protect\citeauthoryear{{Anderson}, {Li}, {Krasnopolsky}  \&
  {Blandford}}{{Anderson} et~al.}{2003}]{Anderson+2003}
{Anderson} J.~M.,  {Li} Z.-Y.,  {Krasnopolsky} R.,   {Blandford} R.~D.,  2003,
  \mn@doi [\apjl] {10.1086/376824}, \href
  {https://ui.adsabs.harvard.edu/abs/2003ApJ...590L.107A} {590, L107}

\bibitem[\protect\citeauthoryear{{Andrews} et~al.,}{{Andrews}
  et~al.}{2018}]{Andrews+2018}
{Andrews} S.~M.,  et~al., 2018, \mn@doi [\apjl] {10.3847/2041-8213/aaf741},
  \href {https://ui.adsabs.harvard.edu/abs/2018ApJ...869L..41A} {869, L41}

\bibitem[\protect\citeauthoryear{{Ansdell} et~al.,}{{Ansdell}
  et~al.}{2016}]{Ansdell+2016}
{Ansdell} M.,  et~al., 2016, \mn@doi [\apj] {10.3847/0004-637X/828/1/46}, \href
  {http://adsabs.harvard.edu/abs/2016ApJ...828...46A} {828, 46}

\bibitem[\protect\citeauthoryear{{Ansdell}, {Williams}, {Manara}, {Miotello},
  {Facchini}, {van der Marel}, {Testi}  \& {van Dishoeck}}{{Ansdell}
  et~al.}{2017}]{Ansdell+2017}
{Ansdell} M.,  {Williams} J.~P.,  {Manara} C.~F.,  {Miotello} A.,  {Facchini}
  S.,  {van der Marel} N.,  {Testi} L.,   {van Dishoeck} E.~F.,  2017, \mn@doi
  [\aj] {10.3847/1538-3881/aa69c0}, \href
  {http://adsabs.harvard.edu/abs/2017AJ....153..240A} {153, 240}

\bibitem[\protect\citeauthoryear{{Ansdell} et~al.,}{{Ansdell}
  et~al.}{2018}]{Ansdell+2018}
{Ansdell} M.,  et~al., 2018, \mn@doi [\apj] {10.3847/1538-4357/aab890}, \href
  {http://adsabs.harvard.edu/abs/2018ApJ...859...21A} {859, 21}

\bibitem[\protect\citeauthoryear{{Bai}}{{Bai}}{2011}]{Bai2011grain}
{Bai} X.-N.,  2011, \mn@doi [\apj] {10.1088/0004-637X/739/1/51}, \href
  {https://ui.adsabs.harvard.edu/abs/2011ApJ...739...51B} {739, 51}

\bibitem[\protect\citeauthoryear{{Bai}}{{Bai}}{2014}]{Bai2014}
{Bai} X.-N.,  2014, \mn@doi [\apj] {10.1088/0004-637X/791/2/137}, \href
  {http://adsabs.harvard.edu/abs/2014ApJ...791..137B} {791, 137}

\bibitem[\protect\citeauthoryear{{Bai}}{{Bai}}{2015}]{Bai2015}
{Bai} X.-N.,  2015, \mn@doi [\apj] {10.1088/0004-637X/798/2/84}, \href
  {http://adsabs.harvard.edu/abs/2015ApJ...798...84B} {798, 84}

\bibitem[\protect\citeauthoryear{{Bai}}{{Bai}}{2017}]{Bai2017}
{Bai} X.-N.,  2017, \mn@doi [\apj] {10.3847/1538-4357/aa7dda}, \href
  {https://ui.adsabs.harvard.edu/abs/2017ApJ...845...75B} {845, 75}

\bibitem[\protect\citeauthoryear{{Bai}, {Ye}, {Goodman}  \& {Yuan}}{{Bai}
  et~al.}{2016}]{Bai+2016}
{Bai} X.-N.,  {Ye} J.,  {Goodman} J.,   {Yuan} F.,  2016, \mn@doi [\apj]
  {10.3847/0004-637X/818/2/152}, \href
  {https://ui.adsabs.harvard.edu/abs/2016ApJ...818..152B} {818, 152}

\bibitem[\protect\citeauthoryear{{Barenfeld}, {Carpenter}, {Ricci}  \&
  {Isella}}{{Barenfeld} et~al.}{2016}]{Barenfeld+2016}
{Barenfeld} S.~A.,  {Carpenter} J.~M.,  {Ricci} L.,   {Isella} A.,  2016,
  \mn@doi [\apj] {10.3847/0004-637X/827/2/142}, \href
  {https://ui.adsabs.harvard.edu/abs/2016ApJ...827..142B} {827, 142}

\bibitem[\protect\citeauthoryear{{Barenfeld}, {Carpenter}, {Sargent}, {Isella}
  \& {Ricci}}{{Barenfeld} et~al.}{2017}]{Barenfeld+2017}
{Barenfeld} S.~A.,  {Carpenter} J.~M.,  {Sargent} A.~I.,  {Isella} A.,
  {Ricci} L.,  2017, \mn@doi [\apj] {10.3847/1538-4357/aa989d}, \href
  {https://ui.adsabs.harvard.edu/abs/2017ApJ...851...85B} {851, 85}

\bibitem[\protect\citeauthoryear{{Bate}}{{Bate}}{2018}]{Bate2018}
{Bate} M.~R.,  2018, \mn@doi [\mnras] {10.1093/mnras/sty169}, \href
  {http://adsabs.harvard.edu/abs/2018MNRAS.475.5618B} {475, 5618}

\bibitem[\protect\citeauthoryear{{Bate} \& {Keto}}{{Bate} \&
  {Keto}}{2015}]{BateKeto2015}
{Bate} M.~R.,  {Keto} E.~R.,  2015, \mn@doi [\mnras] {10.1093/mnras/stv451},
  \href {http://adsabs.harvard.edu/abs/2015MNRAS.449.2643B} {449, 2643}

\bibitem[\protect\citeauthoryear{{Bate}, {Bonnell}  \& {Price}}{{Bate}
  et~al.}{1995}]{BateBonnellPrice1995}
{Bate} M.~R.,  {Bonnell} I.~A.,   {Price} N.~M.,  1995, \mn@doi [\mnras]
  {10.1093/mnras/277.2.362}, \href
  {http://adsabs.harvard.edu/abs/1995MNRAS.277..362B} {277, 362}

\bibitem[\protect\citeauthoryear{{Benz}}{{Benz}}{1990}]{Benz1990}
{Benz} W.,  1990, in {Buchler} J.~R.,  ed., Numerical Modelling of Nonlinear
  Stellar Pulsations Problems and Prospects. Kluwer, Dordrecht, p.~269

\bibitem[\protect\citeauthoryear{{Birnstiel} et~al.,}{{Birnstiel}
  et~al.}{2018}]{Birnstiel+2018}
{Birnstiel} T.,  et~al., 2018, \mn@doi [\apjl] {10.3847/2041-8213/aaf743},
  \href {https://ui.adsabs.harvard.edu/abs/2018ApJ...869L..45B} {869, L45}

\bibitem[\protect\citeauthoryear{{Blandford} \& {Payne}}{{Blandford} \&
  {Payne}}{1982}]{BlandfordPayne1982}
{Blandford} R.~D.,  {Payne} D.~G.,  1982, \mn@doi [\mnras]
  {10.1093/mnras/199.4.883}, \href
  {http://adsabs.harvard.edu/abs/1982MNRAS.199..883B} {199, 883}

\bibitem[\protect\citeauthoryear{{Bohlin}, {Savage}  \& {Drake}}{{Bohlin}
  et~al.}{1978}]{BohlinSavageDrake1978}
{Bohlin} R.~C.,  {Savage} B.~D.,   {Drake} J.~F.,  1978, \mn@doi [\apj]
  {10.1086/156357}, \href {http://adsabs.harvard.edu/abs/1978ApJ...224..132B}
  {224, 132}

\bibitem[\protect\citeauthoryear{{B{\o}rve}, {Omang}  \& {Trulsen}}{{B{\o}rve}
  et~al.}{2001}]{BorveOmangTrulsen2001}
{B{\o}rve} S.,  {Omang} M.,   {Trulsen} J.,  2001, \mn@doi [\apj]
  {10.1086/323228}, \href {http://adsabs.harvard.edu/abs/2001ApJ...561...82B}
  {561, 82}

\bibitem[\protect\citeauthoryear{{Braiding} \& {Wardle}}{{Braiding} \&
  {Wardle}}{2012a}]{BraidingWardle2012sf}
{Braiding} C.~R.,  {Wardle} M.,  2012a, \mn@doi [\mnras]
  {10.1111/j.1365-2966.2012.20601.x}, \href
  {http://adsabs.harvard.edu/abs/2012MNRAS.422..261B} {422, 261}

\bibitem[\protect\citeauthoryear{{Braiding} \& {Wardle}}{{Braiding} \&
  {Wardle}}{2012b}]{BraidingWardle2012acc}
{Braiding} C.~R.,  {Wardle} M.,  2012b, \mn@doi [\mnras]
  {10.1111/j.1365-2966.2012.22001.x}, \href
  {http://adsabs.harvard.edu/abs/2012MNRAS.427.3188B} {427, 3188}

\bibitem[\protect\citeauthoryear{{Concha-Ram{\'\i}rez}, {Wilhelm}, {Portegies
  Zwart}  \& {Haworth}}{{Concha-Ram{\'\i}rez}
  et~al.}{2019}]{Concharamirez+2019}
{Concha-Ram{\'\i}rez} F.,  {Wilhelm} M. J.~C.,  {Portegies Zwart} S.,
  {Haworth} T.~J.,  2019, \mn@doi [\mnras] {10.1093/mnras/stz2973}, \href
  {https://ui.adsabs.harvard.edu/abs/2019MNRAS.490.5678C} {490, 5678}

\bibitem[\protect\citeauthoryear{{Cox} et~al.,}{{Cox} et~al.}{2015}]{Cox+2015}
{Cox} E.~G.,  et~al., 2015, \mn@doi [\apjl] {10.1088/2041-8205/814/2/L28},
  \href {http://adsabs.harvard.edu/abs/2015ApJ...814L..28C} {814, L28}

\bibitem[\protect\citeauthoryear{{Crutcher}}{{Crutcher}}{2012}]{Crutcher2012}
{Crutcher} R.~M.,  2012, \mn@doi [\araa] {10.1146/annurev-astro-081811-125514},
  \href {http://adsabs.harvard.edu/abs/2012ARA%26A..50...29C} {50, 29}

\bibitem[\protect\citeauthoryear{{Desch} \& {Mouschovias}}{{Desch} \&
  {Mouschovias}}{2001}]{DeschMouschovias2001}
{Desch} S.~J.,  {Mouschovias} T.~C.,  2001, \mn@doi [\apj] {10.1086/319703},
  \href {http://adsabs.harvard.edu/abs/2001ApJ...550..314D} {550, 314}

\bibitem[\protect\citeauthoryear{{Dipierro}, {Price}, {Laibe}, {Hirsh},
  {Cerioli}  \& {Lodato}}{{Dipierro} et~al.}{2015}]{Dipierro+2015}
{Dipierro} G.,  {Price} D.,  {Laibe} G.,  {Hirsh} K.,  {Cerioli} A.,   {Lodato}
  G.,  2015, \mn@doi [\mnras] {10.1093/mnrasl/slv105}, \href
  {http://adsabs.harvard.edu/abs/2015MNRAS.453L..73D} {453, L73}

\bibitem[\protect\citeauthoryear{{Draine} \& {Sutin}}{{Draine} \&
  {Sutin}}{1987}]{DraineSutin1987}
{Draine} B.~T.,  {Sutin} B.,  1987, \mn@doi [\apj] {10.1086/165596}, \href
  {https://ui.adsabs.harvard.edu/abs/1987ApJ...320..803D} {320, 803}

\bibitem[\protect\citeauthoryear{{Dullemond} \& {Dominik}}{{Dullemond} \&
  {Dominik}}{2005}]{DullemondDominik2005}
{Dullemond} C.~P.,  {Dominik} C.,  2005, \mn@doi [\aap]
  {10.1051/0004-6361:20042080}, \href
  {https://ui.adsabs.harvard.edu/abs/2005A&A...434..971D} {434, 971}

\bibitem[\protect\citeauthoryear{{Dunham} et~al.,}{{Dunham}
  et~al.}{2014}]{Dunham+2014}
{Dunham} M.~M.,  et~al., 2014, in {Beuther} H.,  {Klessen} R.~S.,  {Dullemond}
  C.~P.,   {Henning} T.,  eds, Protostars and Planets VI. The University of
  Arizona Press, Tucson, AZ, p.~195 (\mn@eprint {arXiv} {1401.1809}),
  \mn@doi{10.2458/azu_uapress_9780816531240-ch009}

\bibitem[\protect\citeauthoryear{{Eisner} et~al.,}{{Eisner}
  et~al.}{2018}]{Eisner+2018}
{Eisner} J.~A.,  et~al., 2018, \mn@doi [\apj] {10.3847/1538-4357/aac3e2}, \href
  {https://ui.adsabs.harvard.edu/abs/2018ApJ...860...77E} {860, 77}

\bibitem[\protect\citeauthoryear{{Enoch}, {Corder}, {Dunham}  \&
  {Duch{\^e}ne}}{{Enoch} et~al.}{2009}]{Enoch+2009}
{Enoch} M.~L.,  {Corder} S.,  {Dunham} M.~M.,   {Duch{\^e}ne} G.,  2009,
  \mn@doi [\apj] {10.1088/0004-637X/707/1/103}, \href
  {https://ui.adsabs.harvard.edu/abs/2009ApJ...707..103E} {707, 103}

\bibitem[\protect\citeauthoryear{{Evans} Neal~J. et~al.,}{{Evans}
  et~al.}{2009}]{Evans+2009}
{Evans} Neal~J. I.,  et~al., 2009, \mn@doi [\apjs]
  {10.1088/0067-0049/181/2/321}, \href
  {https://ui.adsabs.harvard.edu/abs/2009ApJS..181..321E} {181, 321}

\bibitem[\protect\citeauthoryear{{Flock}, {Henning}  \& {Klahr}}{{Flock}
  et~al.}{2012}]{FlockHenningKlahr2012}
{Flock} M.,  {Henning} T.,   {Klahr} H.,  2012, \mn@doi [\apj]
  {10.1088/0004-637X/761/2/95}, \href
  {https://ui.adsabs.harvard.edu/abs/2012ApJ...761...95F} {761, 95}

\bibitem[\protect\citeauthoryear{{Fromang} \& {Papaloizou}}{{Fromang} \&
  {Papaloizou}}{2006}]{FromangPapaloizou2006}
{Fromang} S.,  {Papaloizou} J.,  2006, \mn@doi [\aap]
  {10.1051/0004-6361:20054612}, \href
  {https://ui.adsabs.harvard.edu/abs/2006A&A...452..751F} {452, 751}

\bibitem[\protect\citeauthoryear{{Galli}, {Lizano}, {Shu}  \& {Allen}}{{Galli}
  et~al.}{2006}]{Galli+2006}
{Galli} D.,  {Lizano} S.,  {Shu} F.~H.,   {Allen} A.,  2006, \mn@doi [\apj]
  {10.1086/505257}, \href {http://adsabs.harvard.edu/abs/2006ApJ...647..374G}
  {647, 374}

\bibitem[\protect\citeauthoryear{{Gammie}}{{Gammie}}{1996}]{Gammie1996}
{Gammie} C.~F.,  1996, \mn@doi [\apj] {10.1086/176735}, \href
  {http://adsabs.harvard.edu/abs/1996ApJ...457..355G} {457, 355}

\bibitem[\protect\citeauthoryear{{Garaud} \& {Lin}}{{Garaud} \&
  {Lin}}{2004}]{GaraudLin2004}
{Garaud} P.,  {Lin} D.~N.~C.,  2004, \mn@doi [\apj] {10.1086/420839}, \href
  {https://ui.adsabs.harvard.edu/abs/2004ApJ...608.1050G} {608, 1050}

\bibitem[\protect\citeauthoryear{{Girart}, {Rao}  \& {Marrone}}{{Girart}
  et~al.}{2006}]{GirartRaoMarrone2006}
{Girart} J.~M.,  {Rao} R.,   {Marrone} D.~P.,  2006, \mn@doi [Science]
  {10.1126/science.1129093}, \href
  {https://ui.adsabs.harvard.edu/abs/2006Sci...313..812G} {313, 812}

\bibitem[\protect\citeauthoryear{{Glassgold}, {Lizano}  \& {Galli}}{{Glassgold}
  et~al.}{2017}]{GlassgoldLizanoGalli2017}
{Glassgold} A.~E.,  {Lizano} S.,   {Galli} D.,  2017, \mn@doi [\mnras]
  {10.1093/mnras/stx2145}, \href
  {https://ui.adsabs.harvard.edu/abs/2017MNRAS.472.2447G} {472, 2447}

\bibitem[\protect\citeauthoryear{{Gorti} \& {Hollenbach}}{{Gorti} \&
  {Hollenbach}}{2009}]{GortiHollenbach2009}
{Gorti} U.,  {Hollenbach} D.,  2009, \mn@doi [\apj]
  {10.1088/0004-637X/690/2/1539}, \href
  {https://ui.adsabs.harvard.edu/abs/2009ApJ...690.1539G} {690, 1539}

\bibitem[\protect\citeauthoryear{{Harris} et~al.,}{{Harris}
  et~al.}{2018}]{Harris+2018}
{Harris} R.~J.,  et~al., 2018, \mn@doi [\apj] {10.3847/1538-4357/aac6ec}, \href
  {http://adsabs.harvard.edu/abs/2018ApJ...861...91H} {861, 91}

\bibitem[\protect\citeauthoryear{{Hendler}, {Pascucci}, {Pinilla}, {Tazzari},
  {Carpenter}, {Malhotra}  \& {Testi}}{{Hendler} et~al.}{2020}]{Hendler+2020}
{Hendler} N.,  {Pascucci} I.,  {Pinilla} P.,  {Tazzari} M.,  {Carpenter} J.,
  {Malhotra} R.,   {Testi} L.,  2020, \mn@doi [\apj]
  {10.3847/1538-4357/ab70ba}, \href
  {https://ui.adsabs.harvard.edu/abs/2020ApJ...895..126H} {895, 126}

\bibitem[\protect\citeauthoryear{{Hennebelle}, {Commer{\c c}on}, {Chabrier}  \&
  {Marchand}}{{Hennebelle} et~al.}{2016}]{Hennebelle+2016}
{Hennebelle} P.,  {Commer{\c c}on} B.,  {Chabrier} G.,   {Marchand} P.,  2016,
  \mn@doi [\apjl] {10.3847/2041-8205/830/1/L8}, \href
  {http://adsabs.harvard.edu/abs/2016ApJ...830L...8H} {830, L8}

\bibitem[\protect\citeauthoryear{{Huang} et~al.,}{{Huang}
  et~al.}{2018}]{Huang+2018}
{Huang} J.,  et~al., 2018, \mn@doi [\apj] {10.3847/1538-4357/aaa1e7}, \href
  {https://ui.adsabs.harvard.edu/abs/2018ApJ...852..122H} {852, 122}

\bibitem[\protect\citeauthoryear{{Hull} et~al.,}{{Hull}
  et~al.}{2014}]{Hull+2014}
{Hull} C. L.~H.,  et~al., 2014, \mn@doi [\apjs] {10.1088/0067-0049/213/1/13},
  \href {https://ui.adsabs.harvard.edu/abs/2014ApJS..213...13H} {213, 13}

\bibitem[\protect\citeauthoryear{{Igea} \& {Glassgold}}{{Igea} \&
  {Glassgold}}{1999}]{IgeaGlassgold1999}
{Igea} J.,  {Glassgold} A.~E.,  1999, \mn@doi [\apj] {10.1086/307302}, \href
  {http://adsabs.harvard.edu/abs/1999ApJ...518..848I} {518, 848}

\bibitem[\protect\citeauthoryear{{Ivlev}, {Padovani}, {Galli}  \&
  {Caselli}}{{Ivlev} et~al.}{2015}]{Ivlev+2015}
{Ivlev} A.~V.,  {Padovani} M.,  {Galli} D.,   {Caselli} P.,  2015, \mn@doi
  [\apj] {10.1088/0004-637X/812/2/135}, \href
  {https://ui.adsabs.harvard.edu/abs/2015ApJ...812..135I} {812, 135}

\bibitem[\protect\citeauthoryear{{Johnstone}, {Hollenbach}  \&
  {Bally}}{{Johnstone} et~al.}{1998}]{JohnstoneHollenbachBally1998}
{Johnstone} D.,  {Hollenbach} D.,   {Bally} J.,  1998, \mn@doi [\apj]
  {10.1086/305658}, \href
  {https://ui.adsabs.harvard.edu/abs/1998ApJ...499..758J} {499, 758}

\bibitem[\protect\citeauthoryear{{Kataoka} et~al.,}{{Kataoka}
  et~al.}{2015}]{Kataoka+2015}
{Kataoka} A.,  et~al., 2015, \mn@doi [\apj] {10.1088/0004-637X/809/1/78}, \href
  {http://adsabs.harvard.edu/abs/2015ApJ...809...78K} {809, 78}

\bibitem[\protect\citeauthoryear{{Kataoka}, {Tsukagoshi}, {Pohl}, {Muto},
  {Nagai}, {Stephens}, {Tomisaka}  \& {Momose}}{{Kataoka}
  et~al.}{2017}]{Kataoka+2017}
{Kataoka} A.,  {Tsukagoshi} T.,  {Pohl} A.,  {Muto} T.,  {Nagai} H.,
  {Stephens} I.~W.,  {Tomisaka} K.,   {Momose} M.,  2017, \mn@doi [\apjl]
  {10.3847/2041-8213/aa7e33}, \href
  {http://adsabs.harvard.edu/abs/2017ApJ...844L...5K} {844, L5}

\bibitem[\protect\citeauthoryear{{Krapp}, {Gressel}, {Ben{\'\i}tez-Llambay},
  {Downes}, {Mohandas}  \& {Pessah}}{{Krapp} et~al.}{2018}]{Krapp+2018}
{Krapp} L.,  {Gressel} O.,  {Ben{\'\i}tez-Llambay} P.,  {Downes} T.~P.,
  {Mohandas} G.,   {Pessah} M.~E.,  2018, \mn@doi [\apj]
  {10.3847/1538-4357/aadcf0}, \href
  {https://ui.adsabs.harvard.edu/abs/2018ApJ...865..105K} {865, 105}

\bibitem[\protect\citeauthoryear{{Krasnopolsky}, {Li}  \&
  {Shang}}{{Krasnopolsky} et~al.}{2011}]{KrasnopolskyLiShang2011}
{Krasnopolsky} R.,  {Li} Z.-Y.,   {Shang} H.,  2011, \mn@doi [\apj]
  {10.1088/0004-637X/733/1/54}, \href
  {http://adsabs.harvard.edu/abs/2011ApJ...733...54K} {733, 54}

\bibitem[\protect\citeauthoryear{{Kristensen} \& {Dunham}}{{Kristensen} \&
  {Dunham}}{2018}]{KristensenDunham2018}
{Kristensen} L.~E.,  {Dunham} M.~M.,  2018, \mn@doi [\aap]
  {10.1051/0004-6361/201731584}, \href
  {https://ui.adsabs.harvard.edu/abs/2018A&A...618A.158K} {618, A158}

\bibitem[\protect\citeauthoryear{{Kunz} \& {Mouschovias}}{{Kunz} \&
  {Mouschovias}}{2009}]{KunzMouschovias2009}
{Kunz} M.~W.,  {Mouschovias} T.~C.,  2009, \mn@doi [\apj]
  {10.1088/0004-637X/693/2/1895}, \href
  {http://adsabs.harvard.edu/abs/2009ApJ...693.1895K} {693, 1895}

\bibitem[\protect\citeauthoryear{{Kwon}, {Looney}  \& {Mundy}}{{Kwon}
  et~al.}{2011}]{KwonLooneyMundy2011}
{Kwon} W.,  {Looney} L.~W.,   {Mundy} L.~G.,  2011, \mn@doi [\apj]
  {10.1088/0004-637X/741/1/3}, \href
  {https://ui.adsabs.harvard.edu/abs/2011ApJ...741....3K} {741, 3}

\bibitem[\protect\citeauthoryear{{Kwon}, {Looney}, {Mundy}  \& {Welch}}{{Kwon}
  et~al.}{2015}]{Kwon+2015}
{Kwon} W.,  {Looney} L.~W.,  {Mundy} L.~G.,   {Welch} W.~J.,  2015, \mn@doi
  [\apj] {10.1088/0004-637X/808/1/102}, \href
  {https://ui.adsabs.harvard.edu/abs/2015ApJ...808..102K} {808, 102}

\bibitem[\protect\citeauthoryear{{Kwon}, {Stephens}, {Tobin}, {Looney}, {Li},
  {van der Tak}  \& {Crutcher}}{{Kwon} et~al.}{2019}]{Kwon+2019}
{Kwon} W.,  {Stephens} I.~W.,  {Tobin} J.~J.,  {Looney} L.~W.,  {Li} Z.-Y.,
  {van der Tak} F. F.~S.,   {Crutcher} R.~M.,  2019, \mn@doi [\apj]
  {10.3847/1538-4357/ab24c8}, \href
  {https://ui.adsabs.harvard.edu/abs/2019ApJ...879...25K} {879, 25}

\bibitem[\protect\citeauthoryear{{Larson}}{{Larson}}{1969}]{Larson1969}
{Larson} R.~B.,  1969, \mnras, \href
  {http://adsabs.harvard.edu/abs/1969MNRAS.145..271L} {145, 271}

\bibitem[\protect\citeauthoryear{{Lazarian}}{{Lazarian}}{2007}]{Lazarian2007}
{Lazarian} A.,  2007, \mn@doi [\jqsrt] {10.1016/j.jqsrt.2007.01.038}, \href
  {http://adsabs.harvard.edu/abs/2007JQSRT.106..225L} {106, 225}

\bibitem[\protect\citeauthoryear{{Lee}, {Li}, {Ching}, {Lai}  \& {Yang}}{{Lee}
  et~al.}{2018a}]{Lee+2018}
{Lee} C.-F.,  {Li} Z.-Y.,  {Ching} T.-C.,  {Lai} S.-P.,   {Yang} H.,  2018a,
  \mn@doi [\apj] {10.3847/1538-4357/aaa769}, \href
  {http://adsabs.harvard.edu/abs/2018ApJ...854...56L} {854, 56}

\bibitem[\protect\citeauthoryear{{Lee}, {Li}, {Hirano}, {Shang}, {Ho}  \&
  {Zhang}}{{Lee} et~al.}{2018b}]{Lee+2018young}
{Lee} C.-F.,  {Li} Z.-Y.,  {Hirano} N.,  {Shang} H.,  {Ho} P. T.~P.,   {Zhang}
  Q.,  2018b, \mn@doi [\apj] {10.3847/1538-4357/aad2da}, \href
  {https://ui.adsabs.harvard.edu/abs/2018ApJ...863...94L} {863, 94}

\bibitem[\protect\citeauthoryear{{Lenzuni}, {Gail}  \& {Henning}}{{Lenzuni}
  et~al.}{1995}]{LenzuniGailHenning1995}
{Lenzuni} P.,  {Gail} H.-P.,   {Henning} T.,  1995, \mn@doi [\apj]
  {10.1086/175922}, \href {http://adsabs.harvard.edu/abs/1995ApJ...447..848L}
  {447, 848}

\bibitem[\protect\citeauthoryear{{Lesur}, {Kunz}  \& {Fromang}}{{Lesur}
  et~al.}{2014}]{LesurKunzFromang2014}
{Lesur} G.,  {Kunz} M.~W.,   {Fromang} S.,  2014, \mn@doi [\aap]
  {10.1051/0004-6361/201423660}, \href
  {https://ui.adsabs.harvard.edu/abs/2014A&A...566A..56L} {566, A56}

\bibitem[\protect\citeauthoryear{{Li}, {Krasnopolsky}  \& {Shang}}{{Li}
  et~al.}{2011}]{LiKrasnopolskyShang2011}
{Li} Z.-Y.,  {Krasnopolsky} R.,   {Shang} H.,  2011, \mn@doi [\apj]
  {10.1088/0004-637X/738/2/180}, \href
  {http://adsabs.harvard.edu/abs/2011ApJ...738..180L} {738, 180}

\bibitem[\protect\citeauthoryear{{Lin}}{{Lin}}{2014}]{Lin2014}
{Lin} M.-K.,  2014, \mn@doi [\apj] {10.1088/0004-637X/790/1/13}, \href
  {http://adsabs.harvard.edu/abs/2014ApJ...790...13L} {790, 13}

\bibitem[\protect\citeauthoryear{{Loomis} et~al.,}{{Loomis}
  et~al.}{2020}]{Loomis+2020}
{Loomis} R.~A.,  et~al., 2020, \mn@doi [\apj] {10.3847/1538-4357/ab7cc8}, \href
  {https://ui.adsabs.harvard.edu/abs/2020ApJ...893..101L} {893, 101}

\bibitem[\protect\citeauthoryear{{Mac Low} \& {Klessen}}{{Mac Low} \&
  {Klessen}}{2004}]{MaclowKlessen2004}
{Mac Low} M.-M.,  {Klessen} R.~S.,  2004, \mn@doi [Reviews of Modern Physics]
  {10.1103/RevModPhys.76.125}, \href
  {https://ui.adsabs.harvard.edu/abs/2004RvMP...76..125M} {76, 125}

\bibitem[\protect\citeauthoryear{{Machida} \& {Hosokawa}}{{Machida} \&
  {Hosokawa}}{2013}]{MachidaHosokawa2013}
{Machida} M.~N.,  {Hosokawa} T.,  2013, \mn@doi [\mnras]
  {10.1093/mnras/stt291}, \href
  {http://adsabs.harvard.edu/abs/2013MNRAS.431.1719M} {431, 1719}

\bibitem[\protect\citeauthoryear{{Machida}, {Inutsuka}  \&
  {Matsumoto}}{{Machida} et~al.}{2006}]{MachidaInutsukaMatsumoto2006}
{Machida} M.~N.,  {Inutsuka} S.-i.,   {Matsumoto} T.,  2006, \mn@doi [\apjl]
  {10.1086/507179}, \href {http://adsabs.harvard.edu/abs/2006ApJ...647L.151M}
  {647, L151}

\bibitem[\protect\citeauthoryear{{Marchand}, {Masson}, {Chabrier},
  {Hennebelle}, {Commer{\c c}on}  \& {Vaytet}}{{Marchand}
  et~al.}{2016}]{Marchand+2016}
{Marchand} P.,  {Masson} J.,  {Chabrier} G.,  {Hennebelle} P.,  {Commer{\c
  c}on} B.,   {Vaytet} N.,  2016, \mn@doi [\aap] {10.1051/0004-6361/201526780},
  \href {http://adsabs.harvard.edu/abs/2016A%26A...592A..18M} {592, A18}

\bibitem[\protect\citeauthoryear{{Marchand}, {Tomida}, {Commer{\c{c}}on}  \&
  {Chabrier}}{{Marchand} et~al.}{2019}]{Marchand+2019}
{Marchand} P.,  {Tomida} K.,  {Commer{\c{c}}on} B.,   {Chabrier} G.,  2019,
  \mn@doi [\aap] {10.1051/0004-6361/201936215}, \href
  {https://ui.adsabs.harvard.edu/abs/2019A&A...631A..66M} {631, A66}

\bibitem[\protect\citeauthoryear{{Mathis}, {Rumpl}  \& {Nordsieck}}{{Mathis}
  et~al.}{1977}]{MathisRumplNordsieck1977}
{Mathis} J.~S.,  {Rumpl} W.,   {Nordsieck} K.~H.,  1977, \mn@doi [\apj]
  {10.1086/155591}, \href {http://adsabs.harvard.edu/abs/1977ApJ...217..425M}
  {217, 425}

\bibitem[\protect\citeauthoryear{{Maury}, {Andr{\'e}}, {Men'shchikov},
  {K{\"o}nyves}  \& {Bontemps}}{{Maury} et~al.}{2011}]{Maury+2011}
{Maury} A.~J.,  {Andr{\'e}} P.,  {Men'shchikov} A.,  {K{\"o}nyves} V.,
  {Bontemps} S.,  2011, \mn@doi [\aap] {10.1051/0004-6361/201117132}, \href
  {https://ui.adsabs.harvard.edu/abs/2011A&A...535A..77M} {535, A77}

\bibitem[\protect\citeauthoryear{{Maury} et~al.,}{{Maury}
  et~al.}{2018}]{Maury+2018}
{Maury} A.~J.,  et~al., 2018, \mn@doi [\mnras] {10.1093/mnras/sty574}, \href
  {http://adsabs.harvard.edu/abs/2018MNRAS.477.2760M} {477, 2760}

\bibitem[\protect\citeauthoryear{{McElroy}, {Walsh}, {Markwick}, {Cordiner},
  {Smith}  \& {Millar}}{{McElroy} et~al.}{2013}]{Mcelroy+2013}
{McElroy} D.,  {Walsh} C.,  {Markwick} A.~J.,  {Cordiner} M.~A.,  {Smith} K.,
  {Millar} T.~J.,  2013, \mn@doi [\aap] {10.1051/0004-6361/201220465}, \href
  {http://adsabs.harvard.edu/abs/2013A%26A...550A..36M} {550, A36}

\bibitem[\protect\citeauthoryear{{Mellon} \& {Li}}{{Mellon} \&
  {Li}}{2009}]{MellonLi2009}
{Mellon} R.~R.,  {Li} Z.-Y.,  2009, \mn@doi [\apj]
  {10.1088/0004-637X/698/1/922}, \href
  {http://adsabs.harvard.edu/abs/2009ApJ...698..922M} {698, 922}

\bibitem[\protect\citeauthoryear{{Mestel}}{{Mestel}}{1999}]{Mestel1999}
{Mestel} L.,  1999, {Stellar magnetism}.
Clarendon, Oxford

\bibitem[\protect\citeauthoryear{{Mouschovias}}{{Mouschovias}}{1978}]{Mouschovias1978}
{Mouschovias} T.~C.,  1978, in Stars and Star Systems. p.~A34

\bibitem[\protect\citeauthoryear{{Mouschovias}}{{Mouschovias}}{1991}]{Mouschovias1991}
{Mouschovias} T.~C.,  1991, \mn@doi [\apj] {10.1086/170035}, \href
  {http://adsabs.harvard.edu/abs/1991ApJ...373..169M} {373, 169}

\bibitem[\protect\citeauthoryear{{Mouschovias} \& {Ciolek}}{{Mouschovias} \&
  {Ciolek}}{1999}]{MouschoviasCiolek1999}
{Mouschovias} T.~C.,  {Ciolek} G.~E.,  1999, in {Lada} C.~J.,  {Kylafis} N.~D.,
   eds,  NATO Advanced Science Institutes (ASI) Series C Vol. 540, NATO
  Advanced Science Institutes (ASI) Series C. p.~305

\bibitem[\protect\citeauthoryear{{Myers} \& {Goodman}}{{Myers} \&
  {Goodman}}{1988}]{MyersGoodman1988}
{Myers} P.~C.,  {Goodman} A.~A.,  1988, \mn@doi [\apjl] {10.1086/185116}, \href
  {http://adsabs.harvard.edu/abs/1988ApJ...326L..27M} {326, L27}

\bibitem[\protect\citeauthoryear{{Nakano}, {Nishi}  \& {Umebayashi}}{{Nakano}
  et~al.}{2002}]{NakanoNishiUmebayashi2002}
{Nakano} T.,  {Nishi} R.,   {Umebayashi} T.,  2002, \mn@doi [\apj]
  {10.1086/340587}, \href {http://adsabs.harvard.edu/abs/2002ApJ...573..199N}
  {573, 199}

\bibitem[\protect\citeauthoryear{{Pascucci} et~al.,}{{Pascucci}
  et~al.}{2016}]{Pascucci+2016}
{Pascucci} I.,  et~al., 2016, \mn@doi [\apj] {10.3847/0004-637X/831/2/125},
  \href {https://ui.adsabs.harvard.edu/abs/2016ApJ...831..125P} {831, 125}

\bibitem[\protect\citeauthoryear{{Pinte} et~al.,}{{Pinte}
  et~al.}{2008}]{Pinte+2008}
{Pinte} C.,  et~al., 2008, \mn@doi [\aap] {10.1051/0004-6361:200810121}, \href
  {https://ui.adsabs.harvard.edu/abs/2008A&A...489..633P} {489, 633}

\bibitem[\protect\citeauthoryear{{Pinte}, {Dent}, {M{\'e}nard}, {Hales},
  {Hill}, {Cortes}  \& {de Gregorio-Monsalvo}}{{Pinte}
  et~al.}{2016}]{Pinte+2016}
{Pinte} C.,  {Dent} W.~R.~F.,  {M{\'e}nard} F.,  {Hales} A.,  {Hill} T.,
  {Cortes} P.,   {de Gregorio-Monsalvo} I.,  2016, \mn@doi [\apj]
  {10.3847/0004-637X/816/1/25}, \href
  {https://ui.adsabs.harvard.edu/abs/2016ApJ...816...25P} {816, 25}

\bibitem[\protect\citeauthoryear{{Price}}{{Price}}{2007}]{Price2007}
{Price} D.~J.,  2007, \mn@doi [\pasa] {10.1071/AS07022}, \href
  {http://adsabs.harvard.edu/abs/2007PASA...24..159P} {24, 159}

\bibitem[\protect\citeauthoryear{{Price}}{{Price}}{2012}]{Price2012}
{Price} D.~J.,  2012, \mn@doi [Journal of Computational Physics]
  {10.1016/j.jcp.2010.12.011}, \href
  {http://adsabs.harvard.edu/abs/2012JCoPh.231..759P} {231, 759}

\bibitem[\protect\citeauthoryear{{Price} \& {Laibe}}{{Price} \&
  {Laibe}}{2015}]{PriceLaibe2015}
{Price} D.~J.,  {Laibe} G.,  2015, \mn@doi [\mnras] {10.1093/mnras/stv996},
  \href {http://adsabs.harvard.edu/abs/2015MNRAS.451..813P} {451, 813}

\bibitem[\protect\citeauthoryear{{Price} \& {Monaghan}}{{Price} \&
  {Monaghan}}{2007}]{PriceMonaghan2007}
{Price} D.~J.,  {Monaghan} J.~J.,  2007, \mn@doi [\mnras]
  {10.1111/j.1365-2966.2006.11241.x}, \href
  {http://adsabs.harvard.edu/abs/2007MNRAS.374.1347P} {374, 1347}

\bibitem[\protect\citeauthoryear{{Price} et~al.,}{{Price}
  et~al.}{2018}]{Phantom2018}
{Price} D.~J.,  et~al., 2018, \mn@doi [\pasa] {10.1017/pasa.2018.25}, \href
  {http://adsabs.harvard.edu/abs/2018PASA...35...31P} {35, e031}

\bibitem[\protect\citeauthoryear{{Priestley}, {Wurster}  \& {Viti}}{{Priestley}
  et~al.}{2019}]{PriestleyWursterViti2019}
{Priestley} F.~D.,  {Wurster} J.,   {Viti} S.,  2019, \mn@doi [\mnras]
  {10.1093/mnras/stz1869}, \href
  {https://ui.adsabs.harvard.edu/abs/2019MNRAS.488.2357P} {488, 2357}

\bibitem[\protect\citeauthoryear{{Pringle}}{{Pringle}}{1981}]{Pringle1981}
{Pringle} J.~E.,  1981, \mn@doi [\araa] {10.1146/annurev.aa.19.090181.001033},
  \href {https://ui.adsabs.harvard.edu/abs/1981ARA&A..19..137P} {19, 137}

\bibitem[\protect\citeauthoryear{{Rao}, {Girart}, {Lai}  \& {Marrone}}{{Rao}
  et~al.}{2014}]{Rao+2014}
{Rao} R.,  {Girart} J.~M.,  {Lai} S.-P.,   {Marrone} D.~P.,  2014, \mn@doi
  [\apjl] {10.1088/2041-8205/780/1/L6}, \href
  {http://adsabs.harvard.edu/abs/2014ApJ...780L...6R} {780, L6}

\bibitem[\protect\citeauthoryear{{Riols} \& {Lesur}}{{Riols} \&
  {Lesur}}{2018}]{RiolsLesur2018}
{Riols} A.,  {Lesur} G.,  2018, \mn@doi [\aap] {10.1051/0004-6361/201833212},
  \href {https://ui.adsabs.harvard.edu/abs/2018A&A...617A.117R} {617, A117}

\bibitem[\protect\citeauthoryear{{Riols}, {Roux}, {Latter}  \& {Lesur}}{{Riols}
  et~al.}{2020a}]{Riols+2020}
{Riols} A.,  {Roux} B.,  {Latter} H.,   {Lesur} G.,  2020a, \mn@doi [\mnras]
  {10.1093/mnras/staa567}, \href
  {https://ui.adsabs.harvard.edu/abs/2020MNRAS.493.4631R} {493, 4631}

\bibitem[\protect\citeauthoryear{{Riols}, {Lesur}  \& {Menard}}{{Riols}
  et~al.}{2020b}]{RiolsLesurMenard2020}
{Riols} A.,  {Lesur} G.,   {Menard} F.,  2020b, \mn@doi [\aap]
  {10.1051/0004-6361/201937418}, \href
  {https://ui.adsabs.harvard.edu/abs/2020A&A...639A..95R} {639, A95}

\bibitem[\protect\citeauthoryear{{Sadavoy} et~al.,}{{Sadavoy}
  et~al.}{2018}]{Sadavoy+2018}
{Sadavoy} S.~I.,  et~al., 2018, \mn@doi [\apj] {10.3847/1538-4357/aac21a},
  \href {http://adsabs.harvard.edu/abs/2018ApJ...859..165S} {859, 165}

\bibitem[\protect\citeauthoryear{{Sano} \& {Stone}}{{Sano} \&
  {Stone}}{2002a}]{SanoStone2002a}
{Sano} T.,  {Stone} J.~M.,  2002a, \mn@doi [\apj] {10.1086/339504}, \href
  {http://adsabs.harvard.edu/abs/2002ApJ...570..314S} {570, 314}

\bibitem[\protect\citeauthoryear{{Sano} \& {Stone}}{{Sano} \&
  {Stone}}{2002b}]{SanoStone2002b}
{Sano} T.,  {Stone} J.~M.,  2002b, \mn@doi [\apj] {10.1086/342172}, \href
  {http://adsabs.harvard.edu/abs/2002ApJ...577..534S} {577, 534}

\bibitem[\protect\citeauthoryear{{Segura-Cox}, {Looney}, {Stephens},
  {Fern{\'a}ndez-L{\'o}pez}, {Kwon}, {Tobin}, {Li}  \& {Crutcher}}{{Segura-Cox}
  et~al.}{2015}]{Seguracox+2015}
{Segura-Cox} D.~M.,  {Looney} L.~W.,  {Stephens} I.~W.,
  {Fern{\'a}ndez-L{\'o}pez} M.,  {Kwon} W.,  {Tobin} J.~J.,  {Li} Z.-Y.,
  {Crutcher} R.,  2015, \mn@doi [\apjl] {10.1088/2041-8205/798/1/L2}, \href
  {http://adsabs.harvard.edu/abs/2015ApJ...798L...2S} {798, L2}

\bibitem[\protect\citeauthoryear{{Simon}, {Bai}, {Flaherty}  \&
  {Hughes}}{{Simon} et~al.}{2018}]{Simon+2018}
{Simon} J.~B.,  {Bai} X.-N.,  {Flaherty} K.~M.,   {Hughes} A.~M.,  2018,
  \mn@doi [\apj] {10.3847/1538-4357/aad86d}, \href
  {https://ui.adsabs.harvard.edu/abs/2018ApJ...865...10S} {865, 10}

\bibitem[\protect\citeauthoryear{{Spitzer} \& {Tomasko}}{{Spitzer} \&
  {Tomasko}}{1968}]{SpitzerTomasko1968}
{Spitzer} Jr. L.,  {Tomasko} M.~G.,  1968, \mn@doi [\apj] {10.1086/149610},
  \href {http://adsabs.harvard.edu/abs/1968ApJ...152..971S} {152, 971}

\bibitem[\protect\citeauthoryear{{Stephens} et~al.,}{{Stephens}
  et~al.}{2013}]{Stephens+2013}
{Stephens} I.~W.,  et~al., 2013, \mn@doi [\apjl] {10.1088/2041-8205/769/1/L15},
  \href {https://ui.adsabs.harvard.edu/abs/2013ApJ...769L..15S} {769, L15}

\bibitem[\protect\citeauthoryear{{Stephens} et~al.,}{{Stephens}
  et~al.}{2014}]{Stephens+2014}
{Stephens} I.~W.,  et~al., 2014, \mn@doi [\nat] {10.1038/nature13850}, \href
  {http://adsabs.harvard.edu/abs/2014Natur.514..597S} {514, 597}

\bibitem[\protect\citeauthoryear{{Stephens} et~al.,}{{Stephens}
  et~al.}{2017}]{Stephens+2017}
{Stephens} I.~W.,  et~al., 2017, \mn@doi [\apj] {10.3847/1538-4357/aa998b},
  \href {http://adsabs.harvard.edu/abs/2017ApJ...851...55S} {851, 55}

\bibitem[\protect\citeauthoryear{{Tazaki}, {Lazarian}  \& {Nomura}}{{Tazaki}
  et~al.}{2017}]{TazakiLazarianNomura2017}
{Tazaki} R.,  {Lazarian} A.,   {Nomura} H.,  2017, \mn@doi [\apj]
  {10.3847/1538-4357/839/1/56}, \href
  {https://ui.adsabs.harvard.edu/abs/2017ApJ...839...56T} {839, 56}

\bibitem[\protect\citeauthoryear{{Tazzari} et~al.,}{{Tazzari}
  et~al.}{2017}]{Tazzari+2017}
{Tazzari} M.,  et~al., 2017, \mn@doi [\aap] {10.1051/0004-6361/201730890},
  \href {https://ui.adsabs.harvard.edu/abs/2017A&A...606A..88T} {606, A88}

\bibitem[\protect\citeauthoryear{{Tobin} et~al.,}{{Tobin}
  et~al.}{2015}]{Tobin+2015}
{Tobin} J.~J.,  et~al., 2015, \mn@doi [\apj] {10.1088/0004-637X/805/2/125},
  \href {http://adsabs.harvard.edu/abs/2015ApJ...805..125T} {805, 125}

\bibitem[\protect\citeauthoryear{{Tobin} et~al.,}{{Tobin}
  et~al.}{2020}]{Tobin+2020}
{Tobin} J.~J.,  et~al., 2020, \mn@doi [\apj] {10.3847/1538-4357/ab6f64}, \href
  {https://ui.adsabs.harvard.edu/abs/2020ApJ...890..130T} {890, 130}

\bibitem[\protect\citeauthoryear{{Tomida}, {Tomisaka}, {Matsumoto}, {Hori},
  {Okuzumi}, {Machida}  \& {Saigo}}{{Tomida} et~al.}{2013}]{Tomida+2013}
{Tomida} K.,  {Tomisaka} K.,  {Matsumoto} T.,  {Hori} Y.,  {Okuzumi} S.,
  {Machida} M.~N.,   {Saigo} K.,  2013, \mn@doi [\apj]
  {10.1088/0004-637X/763/1/6}, \href
  {http://adsabs.harvard.edu/abs/2013ApJ...763....6T} {763, 6}

\bibitem[\protect\citeauthoryear{{Tricco} \& {Price}}{{Tricco} \&
  {Price}}{2012}]{TriccoPrice2012}
{Tricco} T.~S.,  {Price} D.~J.,  2012, \mn@doi [Journal of Computational
  Physics] {10.1016/j.jcp.2012.06.039}, \href
  {http://adsabs.harvard.edu/abs/2012JCoPh.231.7214T} {231, 7214}

\bibitem[\protect\citeauthoryear{{Tricco}, {Price}  \& {Bate}}{{Tricco}
  et~al.}{2016}]{TriccoPriceBate2016}
{Tricco} T.~S.,  {Price} D.~J.,   {Bate} M.~R.,  2016, \mn@doi [Journal of
  Computational Physics] {10.1016/j.jcp.2016.06.053}, \href
  {http://adsabs.harvard.edu/abs/2016JCoPh.322..326T} {322, 326}

\bibitem[\protect\citeauthoryear{{Tsukamoto}, {Iwasaki}, {Okuzumi}, {Machida}
  \& {Inutsuka}}{{Tsukamoto} et~al.}{2015a}]{Tsukamoto+2015oa}
{Tsukamoto} Y.,  {Iwasaki} K.,  {Okuzumi} S.,  {Machida} M.~N.,   {Inutsuka}
  S.,  2015a, \mn@doi [\mnras] {10.1093/mnras/stv1290}, \href
  {http://adsabs.harvard.edu/abs/2015MNRAS.452..278T} {452, 278}

\bibitem[\protect\citeauthoryear{{Tsukamoto}, {Iwasaki}, {Okuzumi}, {Machida}
  \& {Inutsuka}}{{Tsukamoto} et~al.}{2015b}]{Tsukamoto+2015hall}
{Tsukamoto} Y.,  {Iwasaki} K.,  {Okuzumi} S.,  {Machida} M.~N.,   {Inutsuka}
  S.,  2015b, \mn@doi [\apjl] {10.1088/2041-8205/810/2/L26}, \href
  {http://adsabs.harvard.edu/abs/2015ApJ...810L..26T} {810, L26}

\bibitem[\protect\citeauthoryear{{Tsukamoto}, {Okuzumi}, {Iwasaki}, {Machida}
  \& {Inutsuka}}{{Tsukamoto} et~al.}{2017}]{Tsukamoto+2017}
{Tsukamoto} Y.,  {Okuzumi} S.,  {Iwasaki} K.,  {Machida} M.~N.,   {Inutsuka}
  S.-i.,  2017, \mn@doi [\pasj] {10.1093/pasj/psx113}, \href
  {http://adsabs.harvard.edu/abs/2017PASJ...69...95T} {69, 95}

\bibitem[\protect\citeauthoryear{{Tsukamoto}, {Okuzumi}, {Iwasaki}, {Machida}
  \& {Inutsuka}}{{Tsukamoto} et~al.}{2018}]{Tsukamoto+2018}
{Tsukamoto} Y.,  {Okuzumi} S.,  {Iwasaki} K.,  {Machida} M.~N.,   {Inutsuka}
  S.,  2018, \mn@doi [\apj] {10.3847/1538-4357/aae4dc}, \href
  {http://adsabs.harvard.edu/abs/2018ApJ...868...22T} {868, 22}

\bibitem[\protect\citeauthoryear{{Tsukamoto}, {Machida}, {Susa}, {Nomura}  \&
  {Inutsuka}}{{Tsukamoto} et~al.}{2020}]{Tsukamoto+2020}
{Tsukamoto} Y.,  {Machida} M.~N.,  {Susa} H.,  {Nomura} H.,   {Inutsuka} S.,
  2020, \mn@doi [\apj] {10.3847/1538-4357/ab93d0}, \href
  {https://ui.adsabs.harvard.edu/abs/2020ApJ...896..158T} {896, 158}

\bibitem[\protect\citeauthoryear{{Turner} \& {Sano}}{{Turner} \&
  {Sano}}{2008}]{TurnerSano2008}
{Turner} N.~J.,  {Sano} T.,  2008, \mn@doi [\apjl] {10.1086/589540}, \href
  {http://adsabs.harvard.edu/abs/2008ApJ...679L.131T} {679, L131}

\bibitem[\protect\citeauthoryear{{Tychoniec} et~al.,}{{Tychoniec}
  et~al.}{2018}]{Tychoniec+2018}
{Tychoniec} {\L}.,  et~al., 2018, \mn@doi [\apjs] {10.3847/1538-4365/aaceae},
  \href {http://adsabs.harvard.edu/abs/2018ApJS..238...19T} {238, 19}

\bibitem[\protect\citeauthoryear{{Umebayashi} \& {Nakano}}{{Umebayashi} \&
  {Nakano}}{1981}]{UmebayashiNakano1981}
{Umebayashi} T.,  {Nakano} T.,  1981, \pasj, \href
  {http://adsabs.harvard.edu/abs/1981PASJ...33..617U} {33, 617}

\bibitem[\protect\citeauthoryear{{Umebayashi} \& {Nakano}}{{Umebayashi} \&
  {Nakano}}{2009}]{UmebayashiNakano2009}
{Umebayashi} T.,  {Nakano} T.,  2009, \mn@doi [\apj]
  {10.1088/0004-637X/690/1/69}, \href
  {http://adsabs.harvard.edu/abs/2009ApJ...690...69U} {690, 69}

\bibitem[\protect\citeauthoryear{{Vaytet}, {Commer{\c c}on}, {Masson},
  {Gonz{\'a}lez}  \& {Chabrier}}{{Vaytet} et~al.}{2018}]{Vaytet+2018}
{Vaytet} N.,  {Commer{\c c}on} B.,  {Masson} J.,  {Gonz{\'a}lez} M.,
  {Chabrier} G.,  2018, \mn@doi [\aap] {10.1051/0004-6361/201732075}, \href
  {http://adsabs.harvard.edu/abs/2018A%26A...615A...5V} {615, A5}

\bibitem[\protect\citeauthoryear{{Villenave} et~al.,}{{Villenave}
  et~al.}{2020}]{Villenave+2020}
{Villenave} M.,  et~al., 2020, \mn@doi [\aap] {10.1051/0004-6361/202038087},
  \href {https://ui.adsabs.harvard.edu/abs/2020A&A...642A.164V} {642, A164}

\bibitem[\protect\citeauthoryear{{Wang}, {Bai}  \& {Goodman}}{{Wang}
  et~al.}{2019}]{WangBaiGoodman2019}
{Wang} L.,  {Bai} X.-N.,   {Goodman} J.,  2019, \mn@doi [\apj]
  {10.3847/1538-4357/ab06fd}, \href
  {https://ui.adsabs.harvard.edu/abs/2019ApJ...874...90W} {874, 90}

\bibitem[\protect\citeauthoryear{{Wardle}}{{Wardle}}{1997}]{Wardle1997}
{Wardle} M.,  1997, in {Wickramasinghe} D.~T.,  {Bicknell} G.~V.,   {Ferrario}
  L.,  eds,  Astronomical Society of the Pacific Conference Series Vol. 121,
  IAU Colloq. 163: Accretion Phenomena and Related Outflows. p.~561 (\mn@eprint
  {arXiv} {astro-ph/9707228})

\bibitem[\protect\citeauthoryear{{Wardle}}{{Wardle}}{2004}]{Wardle2004}
{Wardle} M.,  2004, \mn@doi [\apss] {10.1023/B:ASTR.0000045033.80068.1f}, \href
  {https://ui.adsabs.harvard.edu/abs/2004Ap&SS.292..317W} {292, 317}

\bibitem[\protect\citeauthoryear{{Wardle}}{{Wardle}}{2007}]{Wardle2007}
{Wardle} M.,  2007, \mn@doi [\apss] {10.1007/s10509-007-9575-8}, \href
  {http://adsabs.harvard.edu/abs/2007Ap%26SS.311...35W} {311, 35}

\bibitem[\protect\citeauthoryear{{Wardle} \& {Ng}}{{Wardle} \&
  {Ng}}{1999}]{WardleNg1999}
{Wardle} M.,  {Ng} C.,  1999, \mn@doi [\mnras]
  {10.1046/j.1365-8711.1999.02211.x}, \href
  {http://adsabs.harvard.edu/abs/1999MNRAS.303..239W} {303, 239}

\bibitem[\protect\citeauthoryear{{Wardle} \& {Salmeron}}{{Wardle} \&
  {Salmeron}}{2012}]{WardleSalmeron2012}
{Wardle} M.,  {Salmeron} R.,  2012, \mn@doi [\mnras]
  {10.1111/j.1365-2966.2011.20022.x}, \href
  {https://ui.adsabs.harvard.edu/abs/2012MNRAS.422.2737W} {422, 2737}

\bibitem[\protect\citeauthoryear{{Whitehouse} \& {Bate}}{{Whitehouse} \&
  {Bate}}{2006}]{WhitehouseBate2006}
{Whitehouse} S.~C.,  {Bate} M.~R.,  2006, \mn@doi [\mnras]
  {10.1111/j.1365-2966.2005.09950.x}, \href
  {http://cdsads.u-strasbg.fr/abs/2006MNRAS.367...32W} {367, 32}

\bibitem[\protect\citeauthoryear{{Whitehouse}, {Bate}  \&
  {Monaghan}}{{Whitehouse} et~al.}{2005}]{WhitehouseBateMonaghan2005}
{Whitehouse} S.~C.,  {Bate} M.~R.,   {Monaghan} J.~J.,  2005, \mn@doi [\mnras]
  {10.1111/j.1365-2966.2005.09683.x}, \href
  {http://adsabs.harvard.edu/abs/2005MNRAS.364.1367W} {364, 1367}

\bibitem[\protect\citeauthoryear{{Williams} \& {Cieza}}{{Williams} \&
  {Cieza}}{2011}]{WilliamsCieza2011}
{Williams} J.~P.,  {Cieza} L.~A.,  2011, \mn@doi [\araa]
  {10.1146/annurev-astro-081710-102548}, \href
  {https://ui.adsabs.harvard.edu/abs/2011ARA&A..49...67W} {49, 67}

\bibitem[\protect\citeauthoryear{{Wurster}}{{Wurster}}{2016}]{Wurster2016}
{Wurster} J.,  2016, \mn@doi [\pasa] {10.1017/pasa.2016.34}, \href
  {http://adsabs.harvard.edu/abs/2016PASA...33...41W} {33, e041}

\bibitem[\protect\citeauthoryear{{Wurster} \& {Bate}}{{Wurster} \&
  {Bate}}{2019}]{WursterBate2019}
{Wurster} J.,  {Bate} M.~R.,  2019, \mn@doi [\mnras] {10.1093/mnras/stz1023},
  \href {http://adsabs.harvard.edu/abs/2019MNRAS.486.2587W} {486, 2587}

\bibitem[\protect\citeauthoryear{{Wurster} \& {Lewis}}{{Wurster} \&
  {Lewis}}{2020a}]{WursterLewis2020d}
{Wurster} J.,  {Lewis} B.~T.,  2020a, \mn@doi [\mnras]
  {10.1093/mnras/staa1339}, \href
  {https://ui.adsabs.harvard.edu/abs/2020MNRAS.495.3795W} {495, 3795}

\bibitem[\protect\citeauthoryear{{Wurster} \& {Lewis}}{{Wurster} \&
  {Lewis}}{2020b}]{WursterLewis2020sc}
{Wurster} J.,  {Lewis} B.~T.,  2020b, \mn@doi [\mnras]
  {10.1093/mnras/staa1340}, \href
  {https://ui.adsabs.harvard.edu/abs/2020MNRAS.495.3807W} {495, 3807}

\bibitem[\protect\citeauthoryear{{Wurster}, {Price}  \& {Ayliffe}}{{Wurster}
  et~al.}{2014}]{WursterPriceAyliffe2014}
{Wurster} J.,  {Price} D.~J.,   {Ayliffe} B.,  2014, \mn@doi [\mnras]
  {10.1093/mnras/stu1524}, \href
  {http://adsabs.harvard.edu/abs/2014MNRAS.444.1104W} {444, 1104}

\bibitem[\protect\citeauthoryear{{Wurster}, {Price}  \& {Bate}}{{Wurster}
  et~al.}{2016}]{WursterPriceBate2016}
{Wurster} J.,  {Price} D.~J.,   {Bate} M.~R.,  2016, \mn@doi [\mnras]
  {10.1093/mnras/stw013}, \href
  {http://adsabs.harvard.edu/abs/2016MNRAS.457.1037W} {457, 1037}

\bibitem[\protect\citeauthoryear{{Wurster}, {Bate}, {Price}  \&
  {Tricco}}{{Wurster} et~al.}{2017a}]{Wurster+2017}
{Wurster} J.,  {Bate} M.~R.,  {Price} D.~J.,   {Tricco} T.~S.,  2017a, in
  {Crespo} A.~J.~C.,  {Gesteira} M.~G.,   {Altomare} C.,  eds, Proc. SPHERIC
  2017: 12th International SPHERIC Workshop. Universidade de Vigo, Spain
  (\mn@eprint {arXiv} {1706.07721})

\bibitem[\protect\citeauthoryear{{Wurster}, {Price}  \& {Bate}}{{Wurster}
  et~al.}{2017b}]{WursterPriceBate2017}
{Wurster} J.,  {Price} D.~J.,   {Bate} M.~R.,  2017b, \mn@doi [\mnras]
  {10.1093/mnras/stw3181}, \href
  {http://adsabs.harvard.edu/abs/2017MNRAS.466.1788W} {466, 1788}

\bibitem[\protect\citeauthoryear{{Wurster}, {Bate}  \& {Price}}{{Wurster}
  et~al.}{2018a}]{WursterBatePrice2018sd}
{Wurster} J.,  {Bate} M.~R.,   {Price} D.~J.,  2018a, \mn@doi [\mnras]
  {10.1093/mnras/stx3339}, \href
  {http://adsabs.harvard.edu/abs/2018MNRAS.475.1859W} {475, 1859}

\bibitem[\protect\citeauthoryear{{Wurster}, {Bate}  \& {Price}}{{Wurster}
  et~al.}{2018b}]{WursterBatePrice2018ion}
{Wurster} J.,  {Bate} M.~R.,   {Price} D.~J.,  2018b, \mn@doi [\mnras]
  {10.1093/mnras/sty392}, \href
  {http://adsabs.harvard.edu/abs/2018MNRAS.476.2063W} {476, 2063}

\bibitem[\protect\citeauthoryear{{Wurster}, {Bate}  \& {Price}}{{Wurster}
  et~al.}{2018c}]{WursterBatePrice2018hd}
{Wurster} J.,  {Bate} M.~R.,   {Price} D.~J.,  2018c, \mn@doi [\mnras]
  {10.1093/mnras/sty2212}, \href
  {http://adsabs.harvard.edu/abs/2018MNRAS.480.4434W} {480, 4434}

\bibitem[\protect\citeauthoryear{{Wurster}, {Bate}  \& {Price}}{{Wurster}
  et~al.}{2018d}]{WursterBatePrice2018ff}
{Wurster} J.,  {Bate} M.~R.,   {Price} D.~J.,  2018d, \mn@doi [\mnras]
  {10.1093/mnras/sty2438}, \href
  {http://adsabs.harvard.edu/abs/2018MNRAS.481.2450W} {481, 2450}

\bibitem[\protect\citeauthoryear{{Wurster}, {Bate}  \& {Price}}{{Wurster}
  et~al.}{2019}]{WursterBatePrice2019}
{Wurster} J.,  {Bate} M.~R.,   {Price} D.~J.,  2019, \mn@doi [\mnras]
  {10.1093/mnras/stz2215}, \href
  {https://ui.adsabs.harvard.edu/abs/2019MNRAS.489.1719W} {489, 1719}

\bibitem[\protect\citeauthoryear{{Xu} \& {Bai}}{{Xu} \&
  {Bai}}{2016}]{XuBai2016}
{Xu} R.,  {Bai} X.-N.,  2016, \mn@doi [\apj] {10.3847/0004-637X/819/1/68},
  \href {https://ui.adsabs.harvard.edu/abs/2016ApJ...819...68X} {819, 68}

\bibitem[\protect\citeauthoryear{{Xu}, {Bai}, {{\"O}berg}  \& {Zhang}}{{Xu}
  et~al.}{2019}]{Xu+2019}
{Xu} R.,  {Bai} X.-N.,  {{\"O}berg} K.,   {Zhang} H.,  2019, \mn@doi [\apj]
  {10.3847/1538-4357/aafdfe}, \href
  {https://ui.adsabs.harvard.edu/abs/2019ApJ...872..107X} {872, 107}

\bibitem[\protect\citeauthoryear{{Zhao}, {Caselli}, {Li}, {Krasnopolsky},
  {Shang}  \& {Nakamura}}{{Zhao} et~al.}{2016}]{Zhao+2016}
{Zhao} B.,  {Caselli} P.,  {Li} Z.-Y.,  {Krasnopolsky} R.,  {Shang} H.,
  {Nakamura} F.,  2016, \mn@doi [\mnras] {10.1093/mnras/stw1124}, \href
  {http://adsabs.harvard.edu/abs/2016MNRAS.460.2050Z} {460, 2050}

\bibitem[\protect\citeauthoryear{{Zhao}, {Caselli}, {Li}  \&
  {Krasnopolsky}}{{Zhao} et~al.}{2018a}]{Zhao+2018}
{Zhao} B.,  {Caselli} P.,  {Li} Z.-Y.,   {Krasnopolsky} R.,  2018a, \mn@doi
  [\mnras] {10.1093/mnras/stx2617}, \href
  {http://adsabs.harvard.edu/abs/2018MNRAS.473.4868Z} {473, 4868}

\bibitem[\protect\citeauthoryear{{Zhao}, {Caselli}  \& {Li}}{{Zhao}
  et~al.}{2018b}]{ZhaoCaselliLi2018}
{Zhao} B.,  {Caselli} P.,   {Li} Z.-Y.,  2018b, \mn@doi [\mnras]
  {10.1093/mnras/sty1165}, \href
  {http://adsabs.harvard.edu/abs/2018MNRAS.478.2723Z} {478, 2723}

\bibitem[\protect\citeauthoryear{{Zhao}, {Caselli}, {Li}, {Krasnopolsky},
  {Shang}  \& {Lam}}{{Zhao} et~al.}{2020}]{Zhao+2020}
{Zhao} B.,  {Caselli} P.,  {Li} Z.-Y.,  {Krasnopolsky} R.,  {Shang} H.,   {Lam}
  K.~H.,  2020, \mn@doi [\mnras] {10.1093/mnras/staa041}, \href
  {https://ui.adsabs.harvard.edu/abs/2020MNRAS.492.3375Z} {492, 3375}

\bibitem[\protect\citeauthoryear{{\emph{[dataset]*} Wurster}, {Bate}  \&
  {Price}}{{\emph{[dataset]*} Wurster}
  et~al.}{2018}]{WursterBatePrice2018hddata}
{\emph{[dataset]*} Wurster} J.,  {Bate} M.~R.,   {Price} D.~J.,  2018, {Hall
  effect-driven formation of gravitationally unstable discs in magnetized
  molecular cloud cores (dataset)}, University of Exeter's institutional
  repository, \mn@doi{https://doi.org/10.24378/exe.607}

\makeatother
\end{thebibliography}
%--------------------------------------------------------------------------------
\appendix

\section{Nicil v2.1}
\label{app:nicil}

\textsc{Nicil}: Non-Ideal magnetohydrodynamics Coefficients and Ionisation Library, was first presented in \citet{Wurster2016} and used in many of our subsequent studies \citep{\wpb2017,\wbp2018sd,\wbp2018ion,\wbp2018hd,\wbp2018ff,PriestleyWursterViti2019,WursterBate2019,\wbp2019,WursterLewis2020d,WursterLewis2020sc}.   Over the past few years, it has undergone many modifications with a near total overhaul recently being performed.  Here, we will summarise the changes.

The complete library can be downloaded at www.bitbucket.org/jameswurster/nicil.  The current version\footnote{Version 2.0 was defined prior to writing this manuscript, and subsequent changes (including recommendations from the referee) warranted the increase in version number.} is v2.1 (commit 201dc39).  This is a free library under the GNU license agreement: free to use, modify, and share, does not come with a warranty, and this paper and \citet{Wurster2016} must be cited if \textsc{Nicil} or any modified version thereof is used in a study.

\subsection{Implementation into a parent code}
This code is designed to be embedded in a parent code and executed at runtime using the local values to calculate the non-ideal coefficients and other required properties.  Many of the subroutines have changed since \citet{Wurster2016}, thus the implementation instructions in section 5 of that paper are outdated.  Please refer to \texttt{IMPLEMENTATION.txt} in the \textsc{Nicil} repository for up-to-date installation instructions.  

The \textsc{Nicil} library also comes with several test programmes that can be independently run for a quick understanding of \textsc{Nicil} and how the non-ideal MHD coefficients behave in certain environments.  These programmes are listed and summarised in the \texttt{README} file in the repository; two of these examples are briefly mentioned below.

\subsection{Overview}
Rather than ionising two proxy chemicals, v2.1 includes 6 cosmic ray reactions and 30 chemical reactions that involve neutral gas species (H, H$_2$, He, C, O, O$_2$, Mg, Si, S, CO, HCO), positive ions (H$^+$, H$_3^+$, He$^+$, C$^+$, O$^+$, O$_2^+$, Mg$^+$, Si$^+$, S$^+$, HCO$^+$) and electrons.  The reaction rates are taken from the UMIST database \citep{Mcelroy+2013}.  The hydrogen, helium, oxygen, carbon and magnesium compounds are typically used when modelling the gravitational collapse of a cloud to form a star and disc \citepeg{Tsukamoto+2015oa,Marchand+2016,Tsukamoto+2018,ZhaoCaselliLi2018} since they are relatively abundant in molecular clouds; silicon and sulfur compounds are important charge carriers in the upper regions of the disc \citepeg{Xu+2019,WangBaiGoodman2019}.  Given that \textsc{Nicil} is designed to be run at runtime, a careful choice of species and reactions was made to reasonably represent the chemical network while allowing the calculations to be performed efficiently.

The cosmic ray ionisation rate remains a free parameter, but the default value has been increased slightly to $\zeta_{0,\text{H}_2} = 1.2\times10^{-17}$~s$^{-1}$ for the dominant reaction involving the ionisation of H$_2$; 
%($2H_2 + cr  \rightarrow H_3^+ + H + e$); 
all other cosmic ray ionisation reactions are scaled to this value.  We have included an optional density- and temperature-dependent cosmic ray ionisation rate, 
\begin{equation}
\zeta(\rho,T) = \zeta_{0,\text{H}_2} e^{-\Sigma/\Sigma_\text{CR}} + \zeta_\text{min},
\end{equation}
which is designed to mimic cosmic ray attenuation as it passes through the gas.  The minimum ionisation rate, $\zeta_\text{min}$, is set by radionuclide decay, with the default value of $\zeta_\text{min} = 1.1\times10^{-22}$~s$^{-1}$ to match the decay of $^{40}$K \citep{ZhaoCaselliLi2018}.  The attenuation depth is $\Sigma_\text{CR} =  96$ g cm$^{-2}$, and
\begin{equation}
\Sigma = \sqrt{\frac{kT\rho}{\pi G m_\text{n}}},
\end{equation}
where $T$ and $\rho$ are the local temperature and gas density respectively, $m_\text{n}$ the mass of an average neutral gas particle, $k$ is the Boltzmann constant, and $G$ is the gravitational constant  \citep{UmebayashiNakano2009}.  This approximation of the column density is included for numerical efficiency, however, a future version of \textsc{Nicil} will likely allow the user to instead pass in the exact column density, $\Sigma$, through which the cosmic ray has passed.  Note that this equation is a factor of 2 lower than the expression from \citet{NakanoNishiUmebayashi2002} and \citet{Zhao+2016}.

As before, we include dust grains which can either be a single species or multiple species with a distribution of sizes.  Each dust species includes three populations --  a positively, negatively and neutrally charged population; modelling only a single charge is reasonable for dense cores \citepeg{DraineSutin1987,Ivlev+2015,Marchand+2016}.  For each grain species, we include grain-electron, grain-ion and grain-grain reactions, where the grains can interact with their own and with other grain species;  the reaction rates are given in \citet{KunzMouschovias2009}.

The total number density of the grain populations is calculated either from a constant and global dust-to-gas ratio $f_\text{dg}$, or from the local $f_\text{dg}$ value passed in to \textsc{Nicil} if the parent code is evolving both gas and dust.  If there is more than one grain size and a constant $f_\text{dg}$, then the default grain distribution follows the MRN \citep*{MathisRumplNordsieck1977} distribution, which is scaled such that the total mass of grains is equal to $f_\text{dg}$ times the density pass it.

We assume that each grain is composed of 88.3 per cent carbon, 11.2 per cent Silicates and 0.5 per cent Aluminium Oxide, and that these grains evaporate as the gas temperature increases from 725 to 1700~K; no dust remains for $T > 1700$~K.  The grain composition and evaporation fraction are interpreted from \citet{LenzuniGailHenning1995}.  This evaporation only occurs for constant $f_\text{dg}$ since any evaporation should be accounted for in the parent code if it is evolving dust. 

To determine the number densities, \textsc{Nicil} first calculates the number densities of the grains, then from the remaining input mass density, it calculates the neutral number densities of the chemical species.  Next, it uses the Saha equation to determine the ion populations of H$^+$, H$_2^+$, He$^+$, C$^+$, O$^+$, Mg$^+$, Si$^+$, S$^+$, K$^+$ and Na$^+$; note that this list varies slightly from the list at the beginning of this section.  This is a relatively efficient calculation since we have derived a single equation with 10 terms where the electron number density is the only unknown value.  Using the remaining neutral gas of these species, the neutral gas of the other neutral species listed above, and the dust grains, \textsc{Nicil} calculates the number densities of the charged species assuming equilibrium chemistry where cosmic rays are the external ionisation source.  This iterative method requires calculating and inverting the Jacobian to determine each of the number densities.  The Jacobian is a $(10+2n_\text{a})^2$ matrix, where $n_\text{a}$ is the number of grain species; thus, this matrix rapidly gets larger (and hence slower to calculate) as grain populations are added.  Using the number densities from both thermal and ionisation chemistry, the non-ideal MHD coefficients are calculated.

Both thermal and ionisation chemistry requires an iterative process to calculate the number densities, thus the parent code is required to store the number densities of the positive ions and the charged grains for efficiency.  For large simulations, this can be memory intensive, however, it is more efficient than requiring the iterations to begin from an initial guess for every calculation and it permits a wide range of input properties (e.g. the dust-to-gas ratios for multiple species).

\subsection{Timestepping}
\label{app:nicil:dt}
The non-ideal timestep is given by
\begin{equation}
\text{d}t_\text{nimhd} = \min\left( C_\text{diff}\frac{h^2}{\max\left(\eta_\text{OR}, \eta_\text{AD}\right)}, C_\text{hall}\frac{h^2}{\left|\eta_\text{HE}\right|}\right).
\end{equation}
where $h$ is the smoothing length in SPH or the cell width in a grid code.  Our previous default coefficients were $C_\text{diff} = C_\text{hall} = 1/2\pi \approx 0.159$, however, subsequent tests showed that under some extreme circumstances\footnote{Our extreme tests included the C-shock test with the Wendland $\mathcal{C}^4$ kernel and global time-stepping for ambipolar diffusion, and a collapse to stellar densities calculation that included only the Hall effect.}, this value is too high and the simulation becomes unstable.  Further tests suggested separate coefficients for the diffusive and dissipative terms are required, with $C_\text{diff} = 0.12$ and $C_\text{hall} = 1/4\pi \approx 0.0796$.

Both SPH codes we have used, \textsc{Phantom} \citep{Phantom2018} and \textsc{sphNG} \citep{Benz1990}, use individual time-stepping, which decreases the timestep of an individual particle by up to a factor of two depending on the particle's individual timestep and pre-selected timesteps at which particles are actually evolved.  The results we have previously presented in the literature are from stable simulations where extreme circumstances were never encountered.

\subsection{Test results}
\subsubsection{Non-ideal coefficients vs density}
Our primary test of the \textsc{Nicil} library includes calculating the number densities and non-ideal MHD coefficients over a range of densities.  For this test, we assume a barotropic equation of state \citep{MachidaInutsukaMatsumoto2006}:
\begin{equation}
\label{eq:nicil:T}
T =  T_0\sqrt{1+\left(\frac{n}{n_1}\right)^{2\Gamma_1}}\left(1+\frac{n}{n_2}\right)^{\Gamma_2}\left(1+\frac{n}{n_3}\right)^{\Gamma_3}
\end{equation}
where $T_0 = 10$ K, $n$ is the total number density, $n_1 = 10^{11}$, $n_2 = 10^{16}$ and $n_3 = 10^{21}$ cm$^{-3}$, $\Gamma_1 = 0.4$, $\Gamma_2 = -0.3$ and $\Gamma_3 = 0.56667$, and that the magnetic field varies as \citep{LiKrasnopolskyShang2011}
\begin{equation}
\label{eq:nicil:B}
\left(\frac{B}{\text{G}}\right) = 1.34 \times 10^{-7} \sqrt{n}.
\end{equation}

\begin{figure}
\centering
\includegraphics[width=0.95\columnwidth]{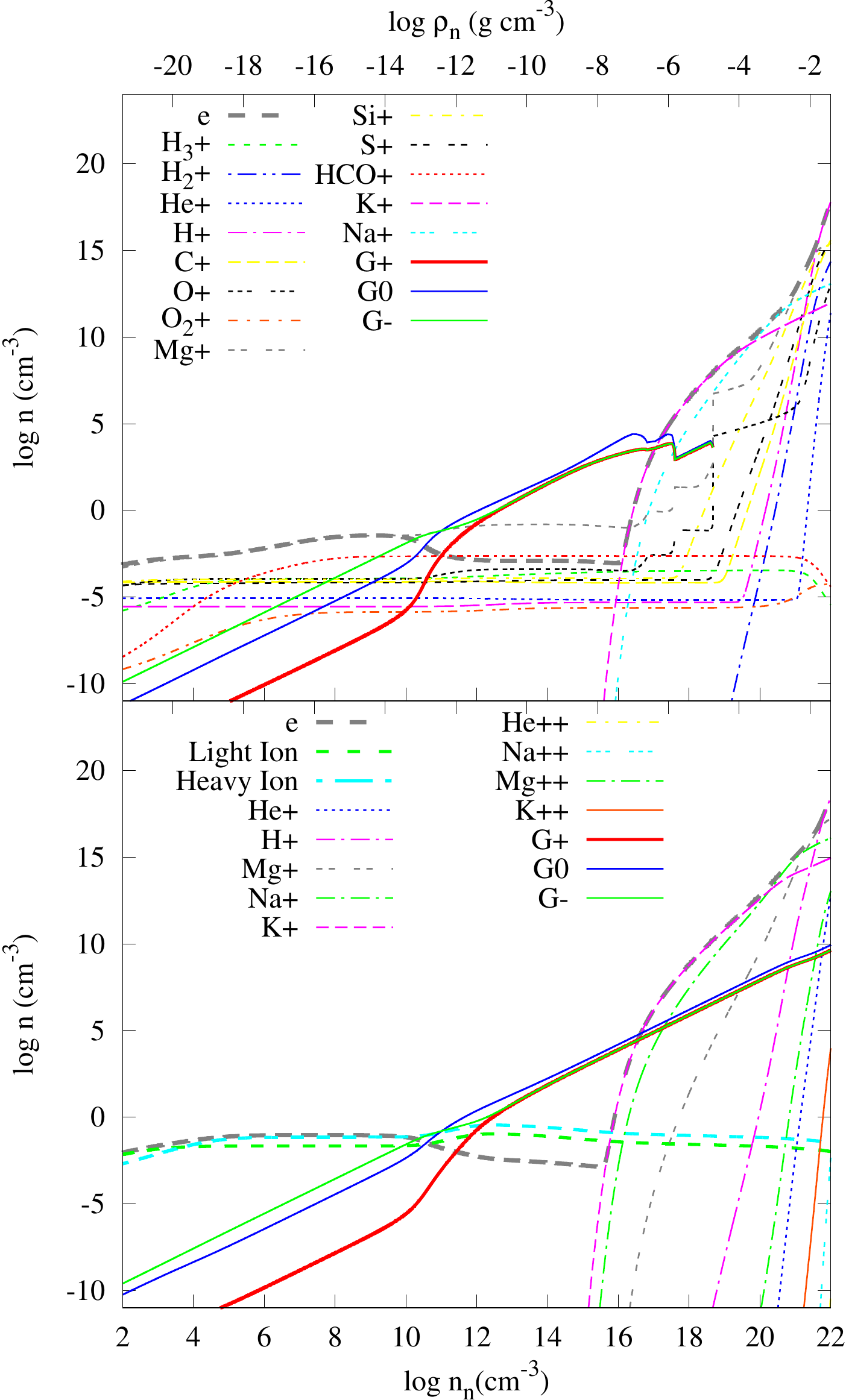}  %Made on Mythos
\caption{Number densities of charged species and neutral grains for v2.1 (top panel) and v1.2.6 (bottom panel) assuming a barotropic equation of state and $B \propto \rho^{1/2}$.  At low temperatures in v1.2.6, species are represented by a light and heavy ion, while in v2.1, all species and reactions are explicitly calculated.  V2.1 only allows for single ionisation of the elements.}
\label{fig:nicil:n}
\end{figure}

\figsref{fig:nicil:n}{fig:nicil:eta} show the number densities and non-ideal coefficients, respectively, for v2.1 (git commit: 201dc39) and v1.2.6 (git commit: 679b501).
\begin{figure}
\centering
\includegraphics[width=0.95\columnwidth]{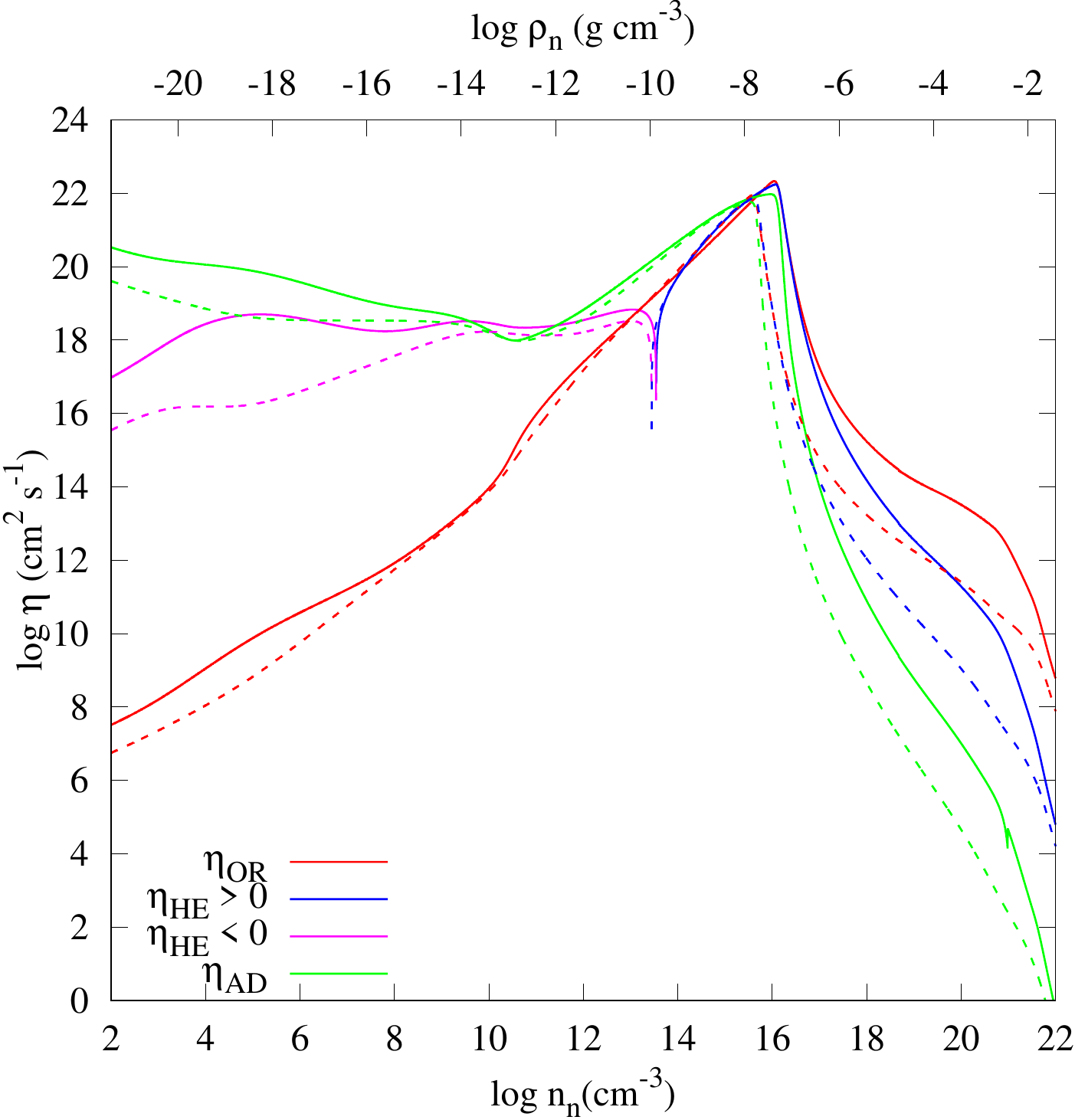}  %Made on Mythos
\caption{The non-ideal MHD coefficients for v2.1 (solid lines) and v1.2.6 (dashed lines) assuming a barotropic equation of state and $B \propto \rho^{1/2}$.  V2.1 is slightly more resistive at low (\rhols{-14}) and high (\rhogs{-8}) densities.  Protostellar discs have densities between these two values, where the coefficients are similar for both versions.}
\label{fig:nicil:eta}
\end{figure}
The additional species in v2.1 yield non-ideal coefficients that are generally more resistive.  However, this increased resistivity is typically outside of density regime of protostellar discs.  This suggests that this regime is more dependent on the grain populations than the gas species.  

\subsubsection{Collapse to stellar densities}
\begin{figure}
\centering
\includegraphics[width=0.45\columnwidth]{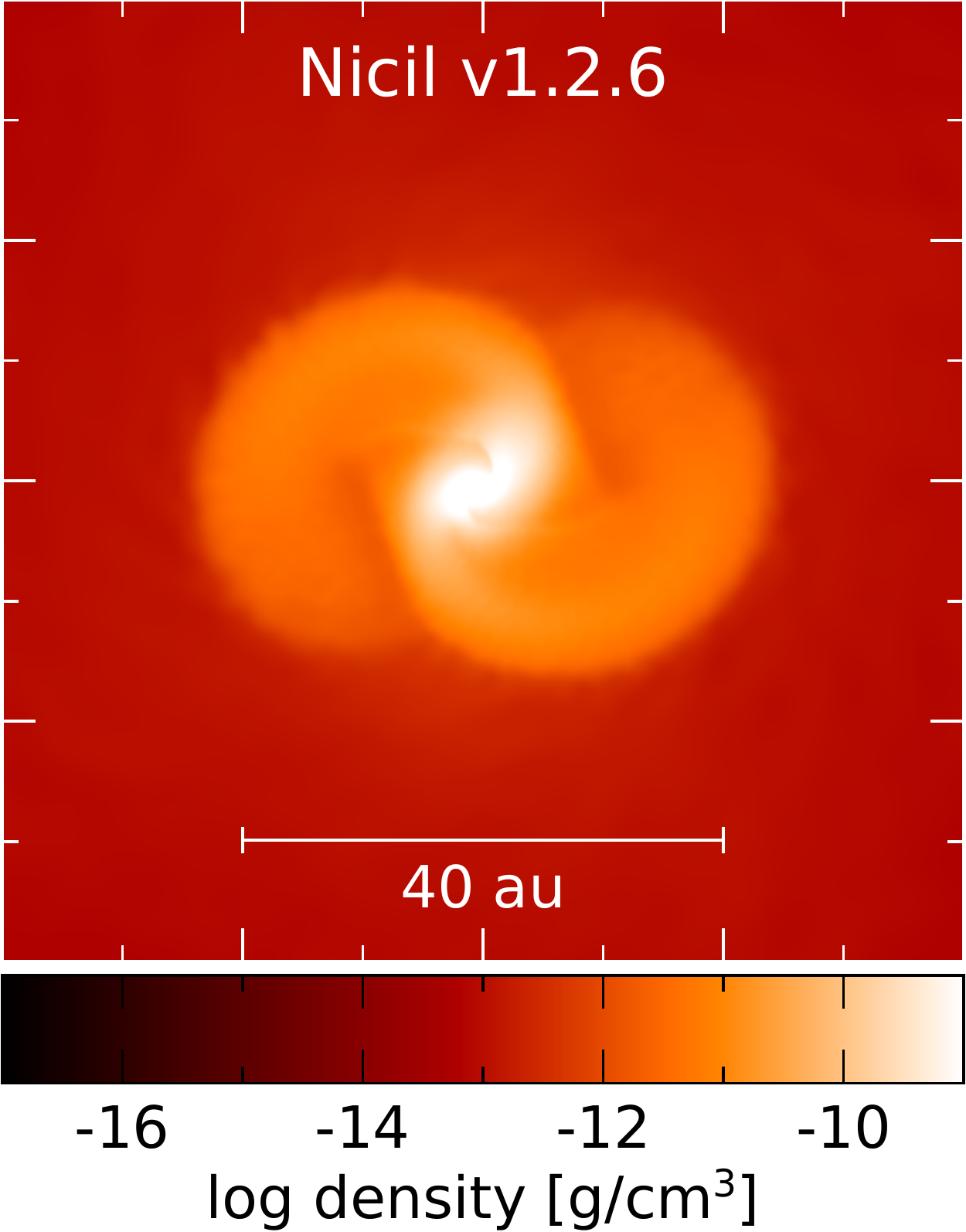}  %Made on Dial after importing data from Kennedy
\includegraphics[width=0.45\columnwidth]{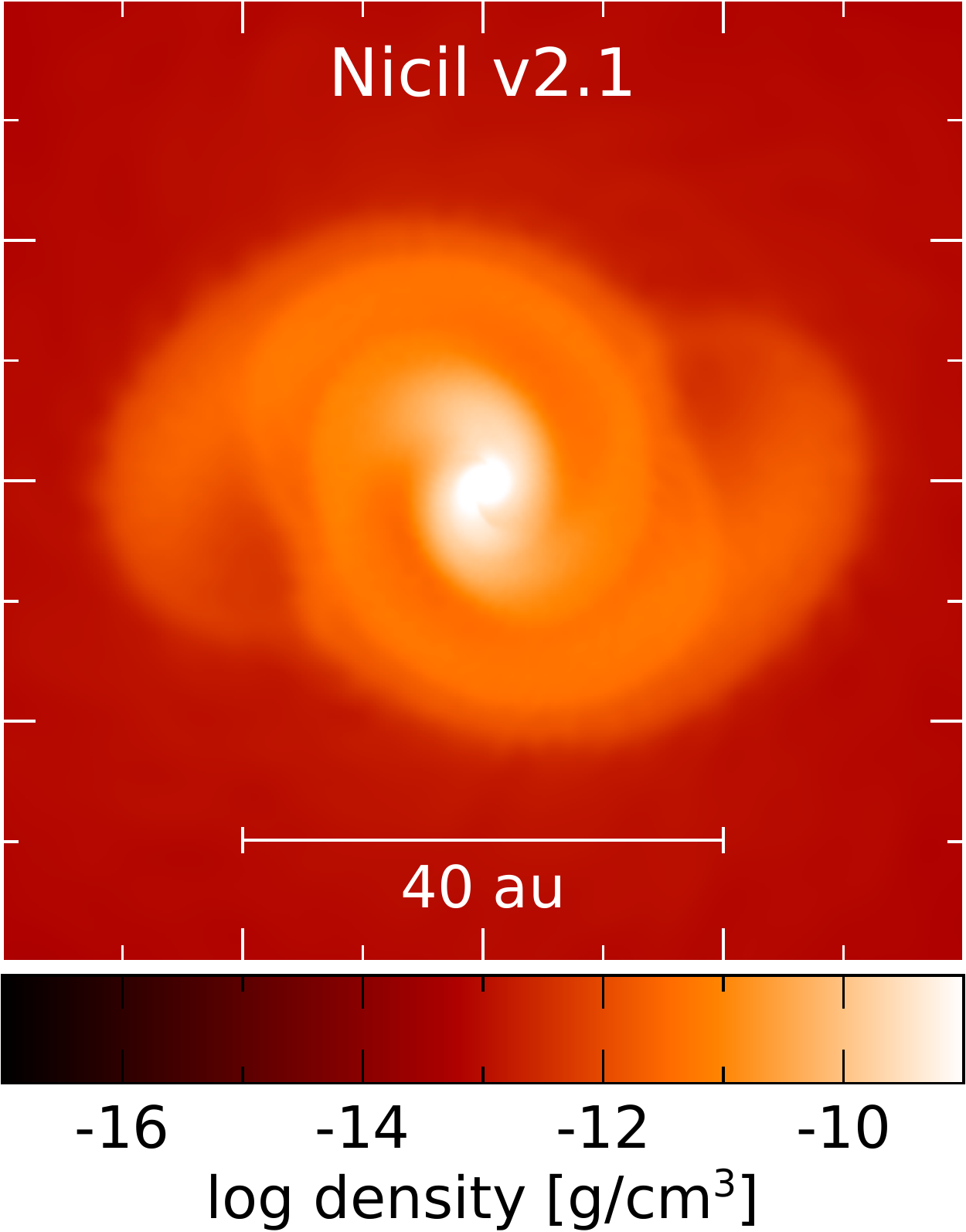}  %Made on Dial after importing data from Kennedy
\includegraphics[width=0.45\columnwidth]{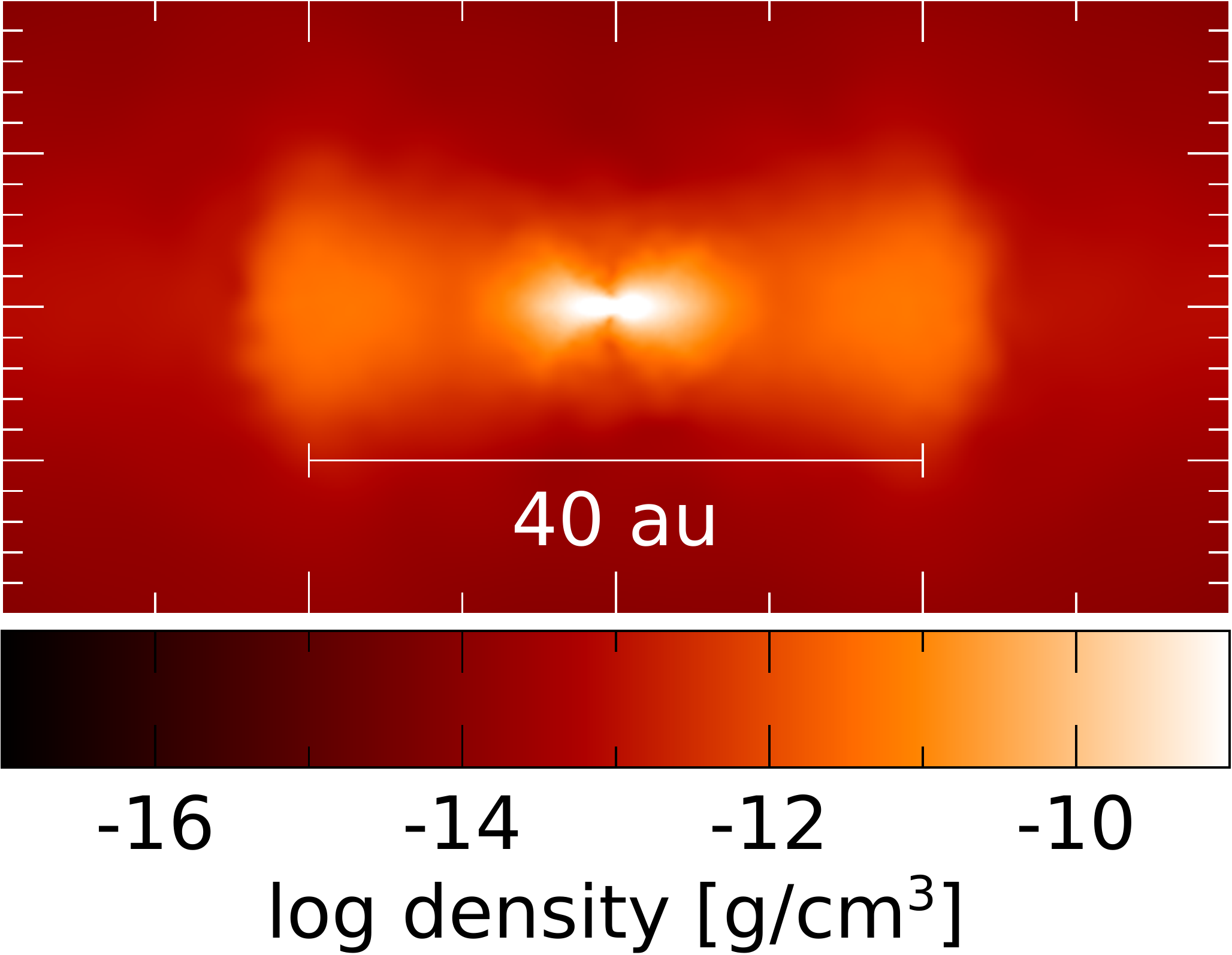}  %Made on Dial after importing data from Kennedy
\includegraphics[width=0.45\columnwidth]{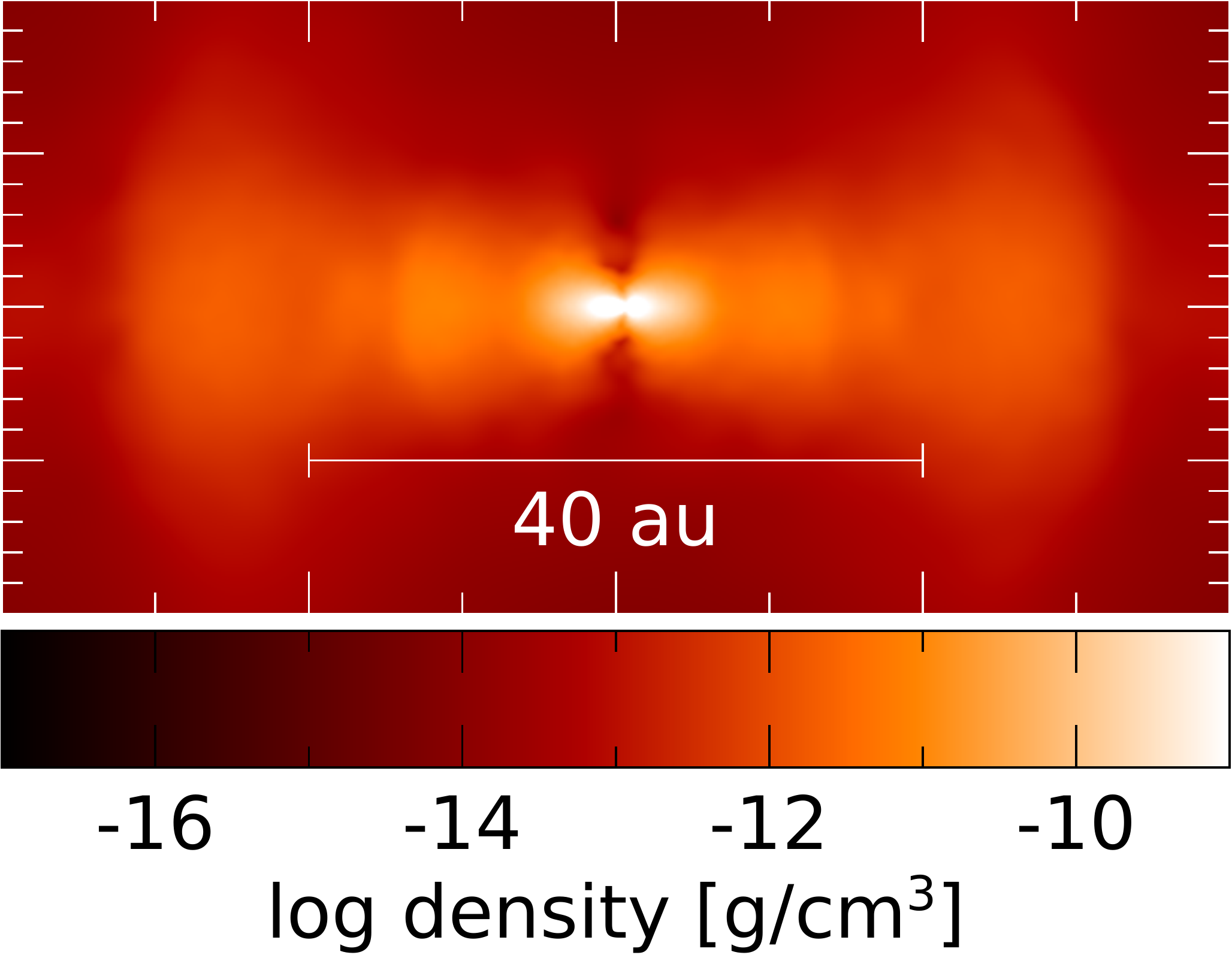}  %Made on Dial after importing data from Kennedy
\includegraphics[width=0.45\columnwidth]{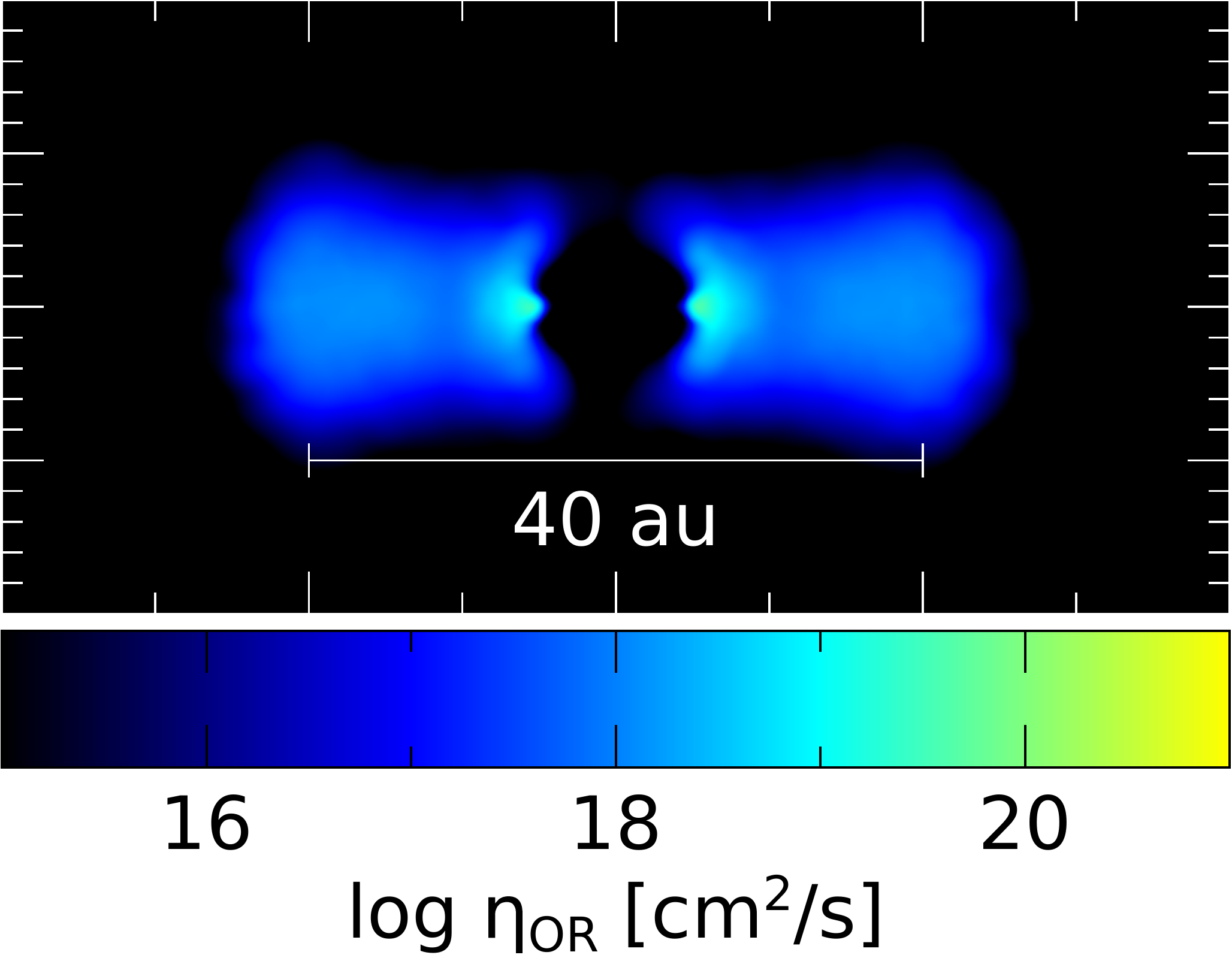}  %Made on Dial after importing data from Kennedy
\includegraphics[width=0.45\columnwidth]{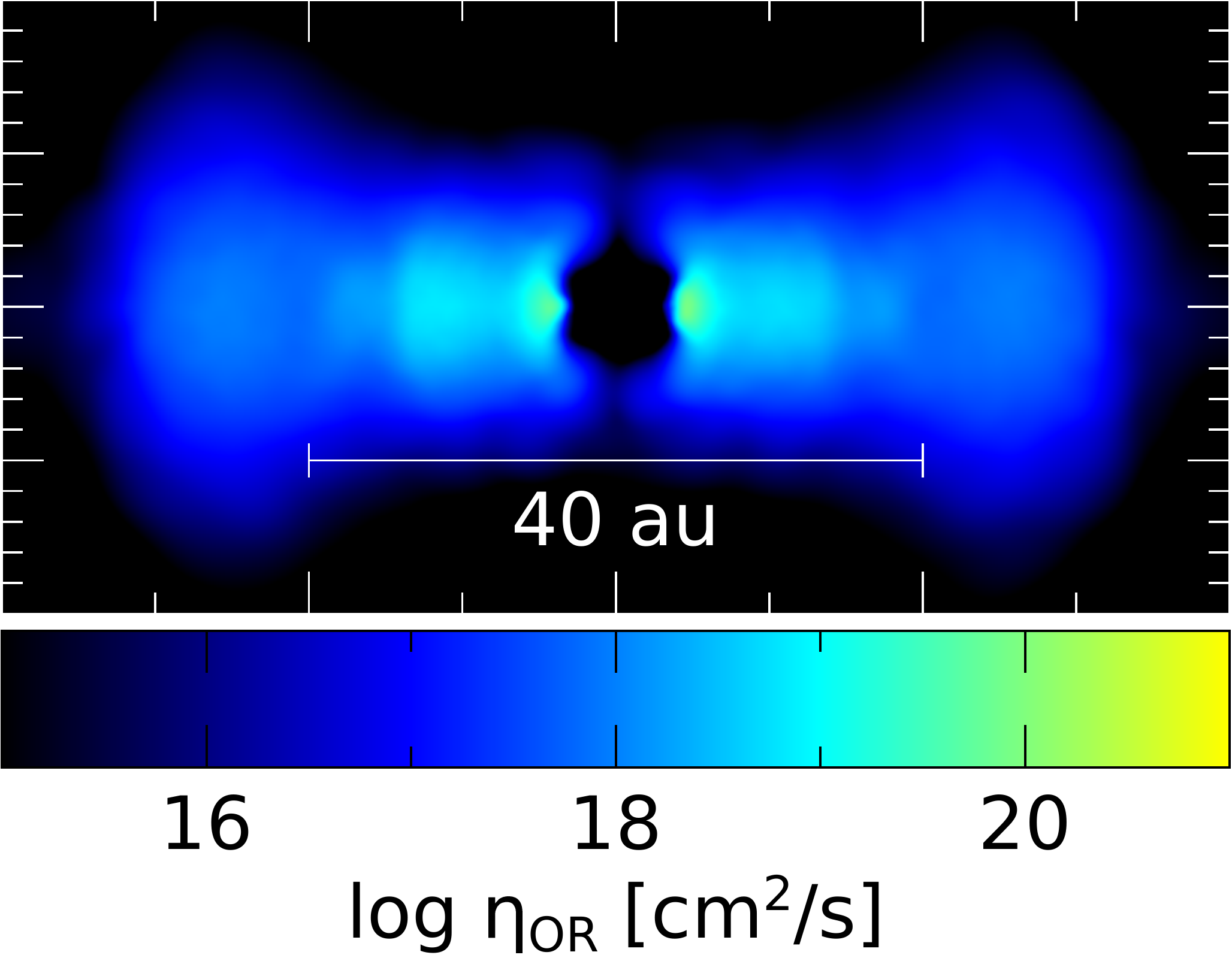}  %Made on Dial after importing data from Kennedy
\includegraphics[width=0.45\columnwidth]{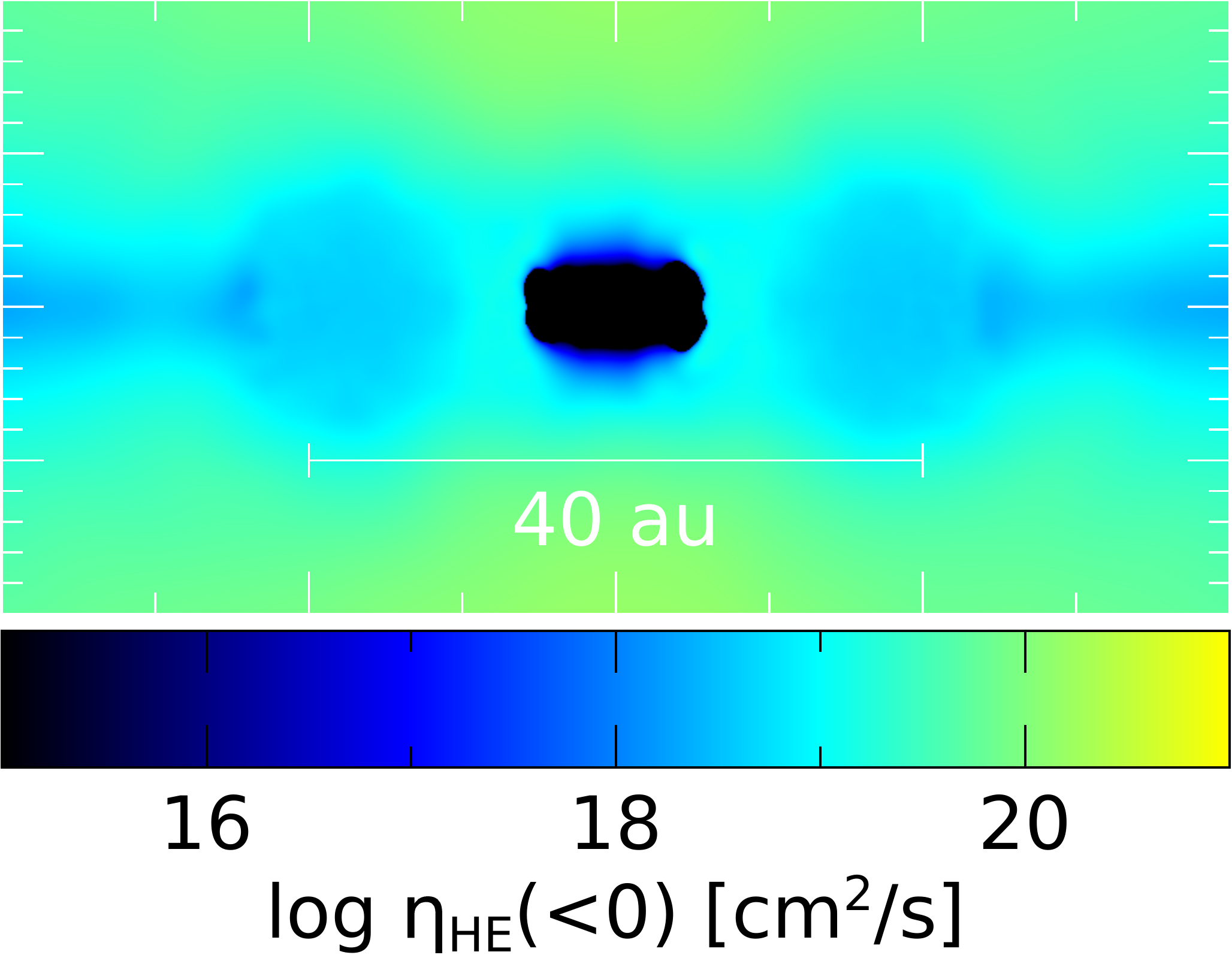}  %Made on Dial after importing data from Kennedy
\includegraphics[width=0.45\columnwidth]{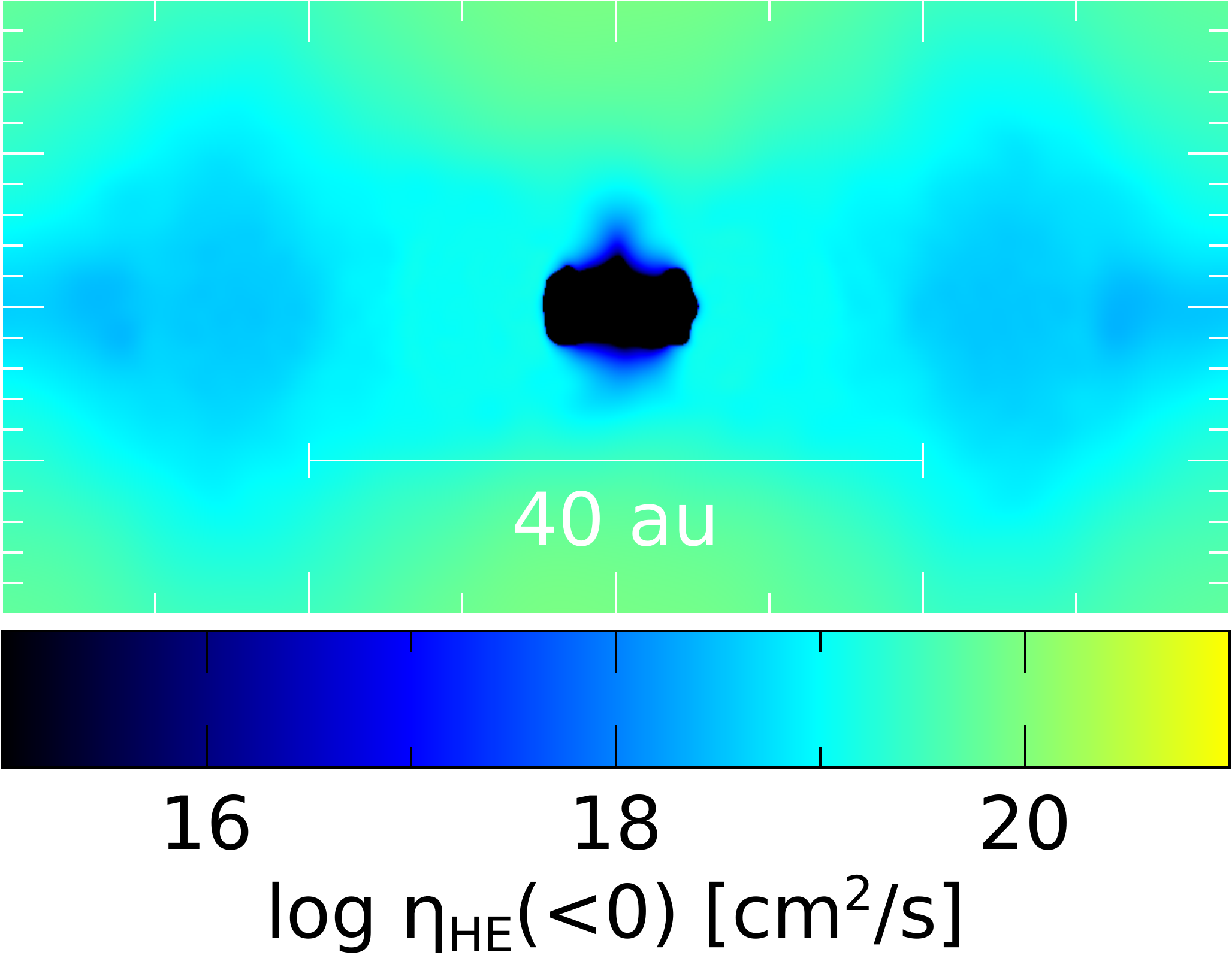}  %Made on Dial after importing data from Kennedy
\includegraphics[width=0.45\columnwidth]{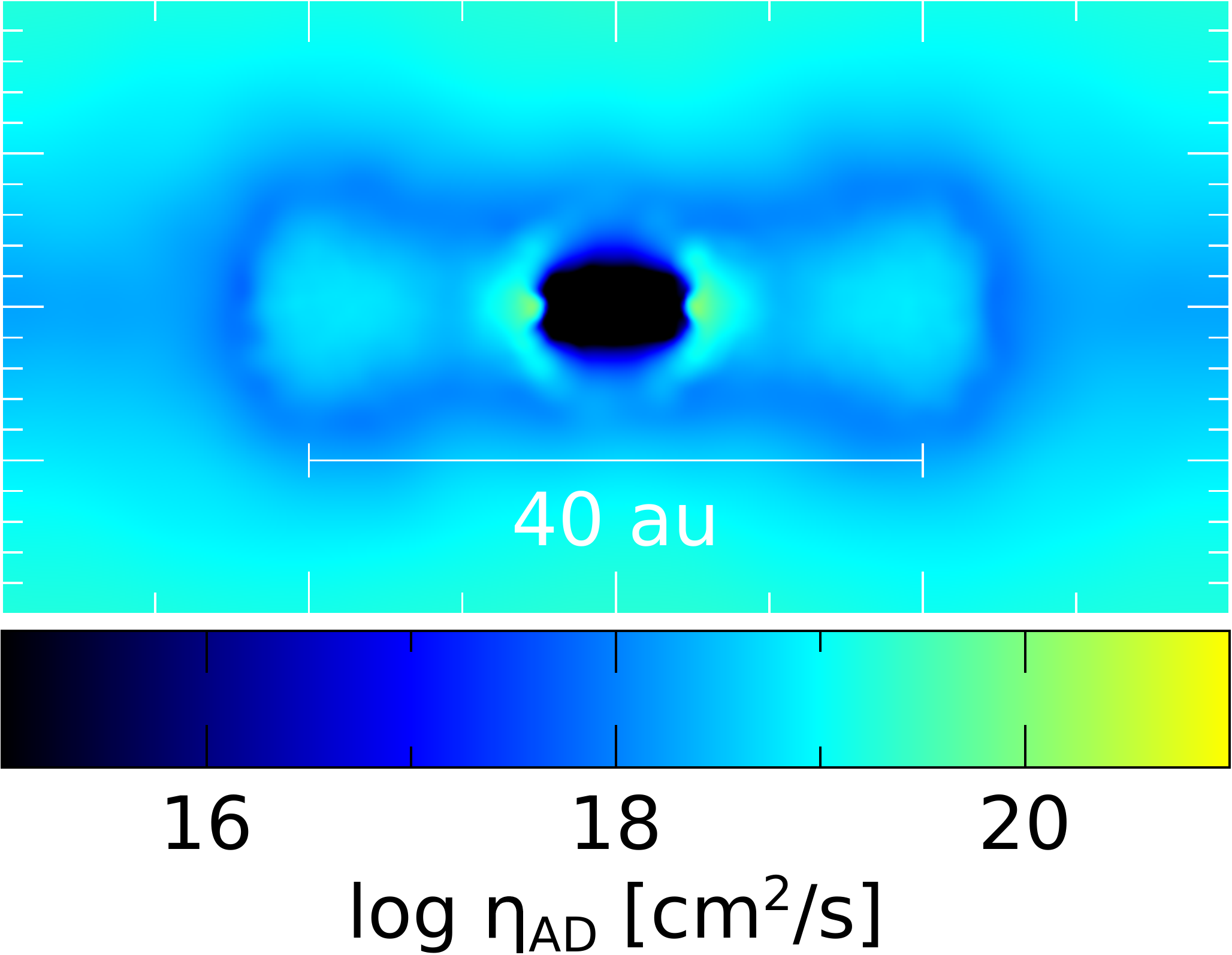}  %Made on Dial after importing data from Kennedy
\includegraphics[width=0.45\columnwidth]{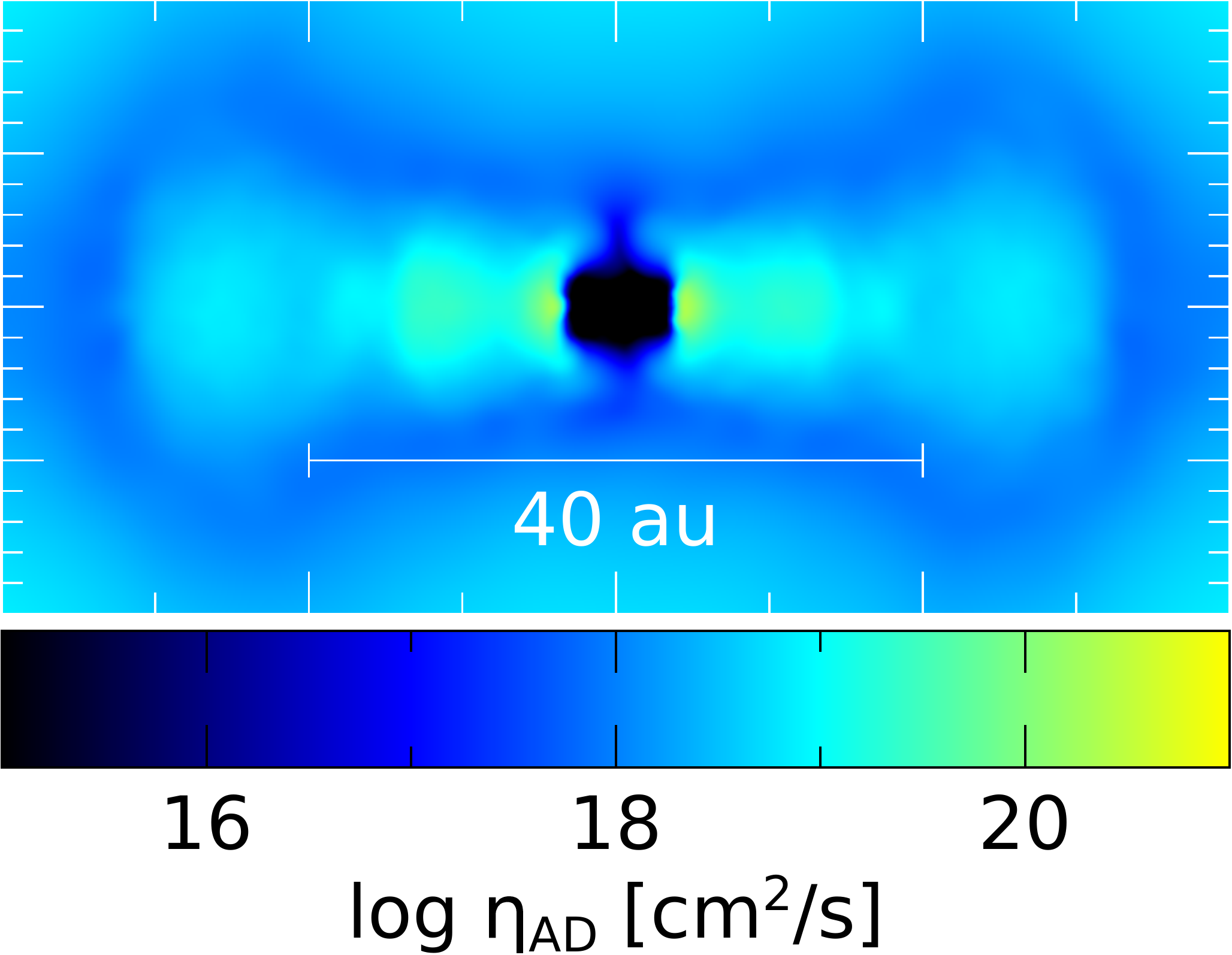}  %Made on Dial after importing data from Kennedy
\caption{Comparing the protostellar disc formed during the gravitational collapse of a gas cloud.  Both simulations were carried out using the same version of \textsc{sphNG} with $3\times10^5$ particles in the gas cloud, where only the version of \textsc{Nicil} has been changed.  All panels are slices taken through the centre of the core and have \rhoxeq{-4}.  The disc is slightly larger in v2.1, and the non-ideal coefficients are slightly stronger.  Although \textsc{Nicil} v2.1 yields a quantitatively different description of the disc, the qualitative conclusions reached here and in our previous papers remains unchanged.}
\label{fig:nicildisc}
\end{figure}

To test the effect of v2.1 on star and disc formation, we repeat the simulation discussed in \secref{sec:rd:iso}, but using only $3\times10^5$ particles in the initial cloud; see \citet{WursterBatePrice2018hd} for the setup and initial conditions.  As shown in \figref{fig:nicil:eta}, v2.1 is more resistive than v1.2.6 at lower densities, and the evolution reflects this greater resistivity in the initial cloud.  The first and second collapse is slightly delayed in v2.1 compared to v1.2.6.  Both models yield a disc with radii of \sm$25$~au, although the disc in v2.1 is sightly more massive and the $m=2$ instability is slightly less well-defined (top row of \figref{fig:nicildisc}).  The non-ideal MHD coefficients are larger over a larger vertical range in v2.1, however, their relative importance is similar as in v1.2.6 and presented in this paper (bottom three row of \figref{fig:nicildisc}). 

As when any numerical code or algorithm is updated, upgrading to \textsc{Nicil} v2.1 will yield quantitatively different results than v1.2.6.  However, the qualitative description of star and disc formation remains unchanged, and the conclusions of this paper and our previous studies remain unaffected.

\subsection{Disc plots}
\label{app:nicil:discs}
To calculate the properties of the discs from \secref{sec:id}, compile \textsc{Nicil} and run \\
\texttt{./nicil\_ex\_eta  Wurster2021\_discs/disc\_param\_*.in}  \\
where \texttt{*} is the name of the input file the user wishes to run.  Parameter files exist for all nine discs discussed in \secref{sec:id}.
Alternatively, to investigate other disc parameters, the user can create and use their own input file by running the \textsc{Python} script \texttt{generate\_disc.py} and following the prompts.

The disc properties can be plotted using the included graphing script, \texttt{plot\_results.py}, which writes and executes a \textsc{GNUplot} script.
%----------------------------------------------------------------------------------------------------------------
\section{The sign of $\eta_\text{HE}$}
\label{app:signeta}

\begin{figure}
\centering
\includegraphics[width=0.95\columnwidth]{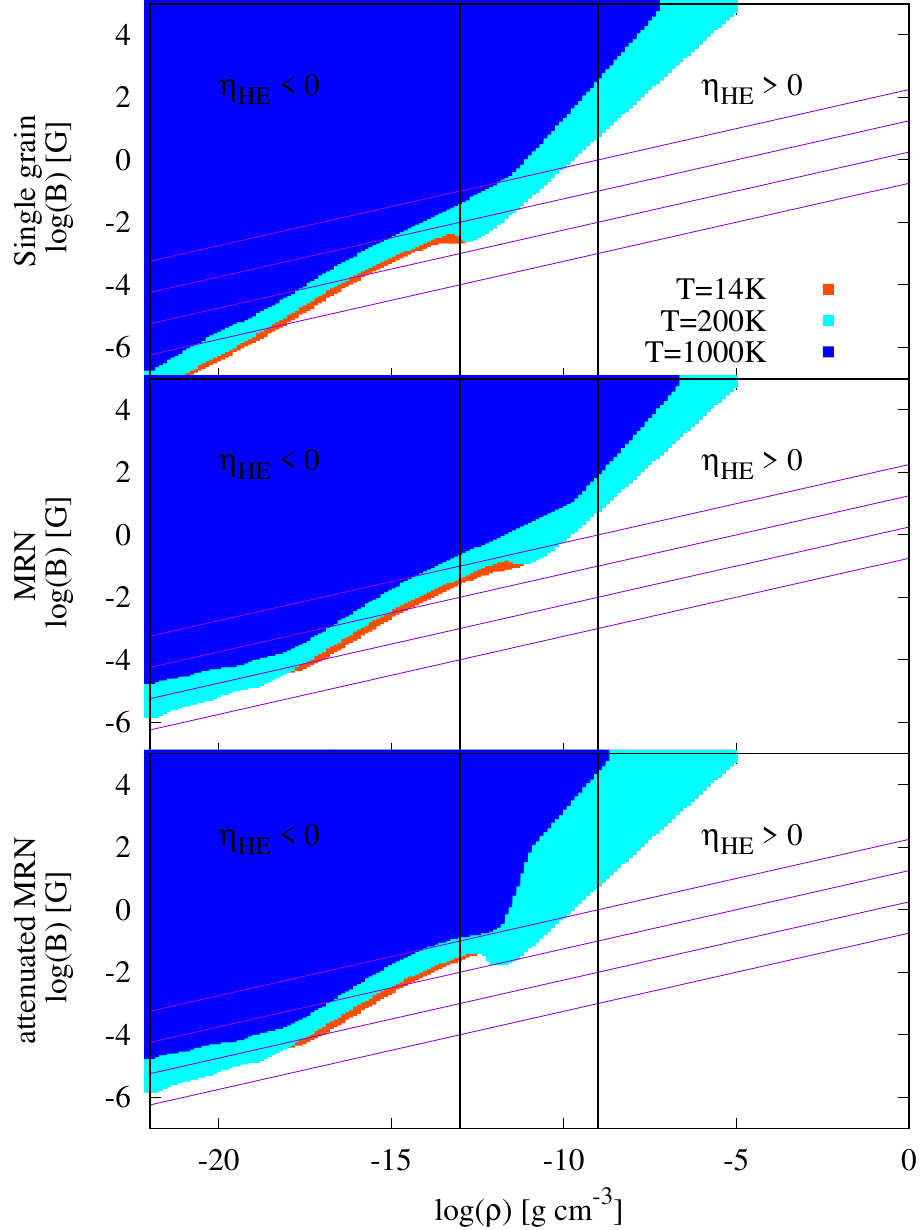}  %Made on Mythos
\caption{The $\rho-B$ phase space where $\eta_\text{HE} < 0$ at $T = 14$, $200$ and $1000$~K (different colours represent different temperatures) and a dust-to-gas ratio of 0.01.   The diagonal lines represent the parameterised magnetic field from Eqn.~\ref{eq:discB} using the four coefficients of $B_0 = 0.001, 0.01, 0.1$ and $1$~G.  For $T \gtrsim 2000$~K,  $\eta_\text{HE} > 0$, while for cooler temperatures, the sign is weakly dependent on temperature; once $\eta_\text{HE} < 0$, then $\eta_\text{HE} < 0$ remains true for all cooler temperatures.  For all grain distributions, $\eta_\text{HE}$ can take either sign at gas densities typically found in discs (densities between the two vertical lines), where reasonably strong magnetic field strengths are required for $\eta_\text{HE} < 0$ .}
\label{fig:nicil:etahall}
\end{figure}

The sign of $\eta_\text{HE}$ is dependent on the gas density, temperature, magnetic field strength, grain distribution and dust-to-gas ratio.  For $T \gtrsim 2000$~K and a dust-to-gas ratio of 0.01,  $\eta_\text{HE} > 0$, independent of the remaining properties; for $T < 2000$~K, the sign depends only weakly on temperature.  \figref{fig:nicil:etahall}  shows the sign of $\eta_\text{HE}$ in the $\rho-B$ phase at $T = 14$, $200$ and $1000$~K for three dust distributions.  

At gas densities typically found in protostellar discs, \rhorange{-13}{-9}, $\eta_\text{HE}$ can take either sign, depending on the magnetic field strength and the dust distribution; this is consistent with our results in Sections~\ref{sec:id} and \ref{sec:rd}.  In lower density regions surrounding protostellar discs, $\eta_\text{HE} < 0$, while in the higher density regions where the non-ideal terms are unimportant, $\eta_\text{HE} > 0$.  Thus, we can expect the sign to change as the gas evolves through the star formation process \citep[in agreement with, e.g.,][]{Wardle2007,Marchand+2016,Wurster2016,XuBai2016}.

For additional analysis regarding the effect the magnetic field strength, dust-to-gas ratio and grain populations have on the sign of $\eta_\text{HE}$, see \citet{XuBai2016}.
%----------------------------------------------------------------------------------------------------------------
\section{Dust settling}
\label{app:settled}
To determine the dust distribution for the settled distribution in \secref{sec:id}, we perform \textsc{Phantom}'s dust settling test \citep{Phantom2018}.  See fig.~10 and associated text of \citet{PriceLaibe2015} for further details and results; see also \citet{Krapp+2018,RiolsLesur2018,Riols+2020,RiolsLesurMenard2020} for further discussion on modelling dust settling.  

Briefly, we simulate the vertical settling of dust in a rectangular box, where the vertical gas density profile mimics that of a disc at a selected radius and the dust is initialised with an MRN distribution that is initially coupled to the gas.  We then allow the dust to settle over the equivalent of 150 orbits at that radius.  Then, for each grain size $i$, we fit the dust-to-gas ratio with one or two Gaussian functions, where the final functions are scaleable to any disc mid-plane gas density and scale-height, via,
\begin{equation}
\label{eq:discratio}
f^i(r,z) = \frac{\rho^i_\text{dust}(r,z)}{\rho_\text{gas}(r,z)}
\end{equation}
where 
\begin{eqnarray}
\label{eq:dustdisc}
\rho^i_\text{dust}(r,z)   &=& \left\{ \begin{array}{l l}  \delta^{i,\text{c}}_\text{dust}(r) \exp\left(\frac{-z^2}{2(Hh^{i,\text{c}})^2}\right) & \text{if } |z| < Hz^i_\text{ce}\\  
                                                                                \delta^{i,\text{e}}_\text{dust}(r) \exp\left(\frac{-z^2}{2(Hh^{i,\text{e}})^2}\right)    &  \text{else}  \end{array}\right. 
\end{eqnarray}
where $\delta^i_\text{dust}(r) = \sigma^i\rho_\text{gas}(r,0)$ is the mid-plane dust mass density, $Hh^i$ is the scale-height of the dust disc, and $Hz^i_\text{ce}$ is the vertical height at which a steeper function is required.  Recall $H \equiv H(r)$ is the scale height of the gas disc.  The grain radii and best-fit parameters for $\sigma^i$ and $h^i$ are given in \tabref{table:dustsettle}.
\begin{table}
\begin{center}
\begin{tabular}{c c c c c c}
\hline\hline
radius (cm)             & $z_\text{ce}$     &                 $\sigma^\text{c}$  & $h^\text{c}$   & $\sigma^\text{e}$           &  $h^\text{e}$ \\
\hline
$1.78\times 10^{-6}$  &     $\infty$        &            $2.48\times 10^{-5}$  &      1.00          &          -                            &        -            \\
$5.62\times 10^{-6}$  &     $\infty$        &            $4.41\times 10^{-5}$  &      1.00          &          -                            &        -            \\
$1.78\times 10^{-5}$  &     $\infty$        &            $7.73\times 10^{-5}$  &      1.00          &          -                            &        -            \\
$5.62\times 10^{-5}$  &     $\infty$        &            $1.39\times 10^{-4}$  &      1.00          &          -                            &        -            \\
$1.78\times 10^{-4}$  &     $\infty$        &            $2.48\times 10^{-4}$  &      1.00          &          -                            &        -            \\
$5.62\times 10^{-4}$  &     $2.78$         &            $4.41\times 10^{-4}$  &      1.00          &    $6.23\times 10^{3}$   &     0.437       \\
$1.78\times 10^{-3}$  &     $2.40$         &            $7.88\times 10^{-4}$  &      1.01          &    $2.66\times 10^{5}$   &     0.358       \\
$5.62\times 10^{-3}$  &     $2.05$         &            $1.43\times 10^{-3}$  &      1.01          &    $8.35\times 10^{14}$ &     0.222       \\
$1.78\times 10^{-2}$  &     $1.51$         &            $2.70\times 10^{-3}$  &      1.06          &    $1.09\times 10^{4}$   &     0.264       \\
$5.62\times 10^{-2}$  &     $0.972$       &            $5.73\times 10^{-3}$  &      1.39          &    $9.98\times 10^{1}$   &     0.217       \\
\hline\hline
\end{tabular}
\caption{The parameters to fit the dust density from the dust settling test; see equation \ref{eq:dustdisc}.  For the smaller grain sizes, only one exponential function is required, while the larger grains require separate functions to fit the centre and edges of the discs.  For the lager grains, the density drops off rapidly, yielding large uncertainties and the lack of trend in their parameters; however, a visual inspection suggests that these fits are reasonable.}
\label{table:dustsettle}
\end{center}
\end{table}

This settled dust distribution is currently in one of \textsc{Nicil}'s example programmes and was used in \secref{sec:id}, however, there are caveats with using this distribution.  First, it assumes that the vertical dust profile is the same at all radii, \emph{modulo} a scaling factor (which it clearly is not in reality; \citealp{Pinte+2016}).  Next, it assumes a constant $H/R = 0.05$, while in the discs in \secref{sec:id}, $H/R \in (0.11,0.17)$; these ratios have been tested in \textsc{Phantom}'s dust settling routine, but yielded unstable results due to the high temperature of the disc.  Although the dust profiles are not realistic (or at least as realistic as the other distributions we present), we have discussed this distribution to show the effect of crude dust settling and as a proof-of-concept that \textsc{Nicil} can function well with reading in local dust-to-gas ratios.
%----------------------------------------------------------------------------------------------------------------
\label{lastpage}
\end{document}